\documentclass[journal]{IEEEtran}

\usepackage[colorlinks,urlcolor=blue,linkcolor=blue,citecolor=blue]{hyperref}
\usepackage{cite}
\usepackage{amsmath,amssymb,amsfonts}
\usepackage{graphicx}
\usepackage{textcomp}
\usepackage{xcolor}
\usepackage{amsmath}
\usepackage{float} % package needed to keep figure at exact position [H]
\usepackage{tikz}
\usetikzlibrary{%
    shapes,%
    calc%
}

\newcommand{\todo}[1]{\textcolor{blue}{\textbf{ TODO:} #1}}
\renewcommand\todo[1]{} %to remove todo

\usepackage{booktabs}
\usepackage{makecell}
\usepackage{adjustbox}

\usepackage[ruled,linesnumbered]{algorithm2e}
 \usepackage{algpseudocode}
\usepackage[normalem]{ulem}
\usepackage{multirow}
\usepackage{subcaption}

\def\BibTeX{{\rm B\kern-.05em{\sc i\kern-.025em b}\kern-.08em
    T\kern-.1667em\lower.7ex\hbox{E}\kern-.125emX}}
    
\makeatletter
\let\original@algocf@latexcaption\algocf@latexcaption
\long\def\algocf@latexcaption#1[#2]{%
	\@ifundefined{NR@gettitle}{%
		\def\@currentlabelname{#2}%
	}{%
		\NR@gettitle{#2}%
	}%
	\original@algocf@latexcaption{#1}[{#2}]%
}
\makeatother

\begin{document}

% \title{Pipe-based Route Switching with Topology Control in Autonomous UAV Networks for Search and Surveillance Missions}
\title{Pipe Routing with Topology Control for UAV Networks}
% \title{Pipe-based Routing with Topology Control Scheme for Autonomous UAV Networks in Search and Surveillance Missions}

\author{Shreyas~Devaraju\thanks{S. Devaraju is with Computational Science Research Center, San Diego State University \& University of California, Irvine, USA, sdevaraju@sdsu.edu},~\IEEEmembership{Student~Member,~IEEE},
\and Shivam~Garg\thanks{S. Garg is with Electrical \& Computer Engineering Department, San Diego State University, San Diego, California, USA, sgarg@sdsu.edu},~\IEEEmembership{Member,~IEEE},
\and Alexander~Ihler\thanks{A. Ihler is with School of Information \& Computer Science, University of California, Irvine, USA, ihler@ics.uci.edu},~\IEEEmembership{Member,~IEEE},
\and Sunil~Kumar\thanks{S. Kumar is with Electrical \& Computer Engineering Department, San Diego State University, San Diego, California, USA, skumar@sdsu.edu},~\IEEEmembership{Senior~Member,~IEEE} 
}

% \authoraddress{S. Devaraju is with Computational Science Research Center, San Diego State University, CA 92182 USA (e-mail: sdevaraju@sdsu.edu). A. Ihler. is with School of Information \& Computer Science, University of California, Irvine, CA 92697, USA (e-mail: ihler@ics.uci.edu). S. Kumar is with Electrical \& Computer Engineering Department, San Diego State University, San Diego, CA 92182 USA (e-mail: skumar@sdsu.edu). \emph{(Corresponding author: Sunil Kumar.)}}

% The paper headers
% \markboth{IEEE Transactions xxxx}
% {S. Devaraju \MakeLowercase{{et al.}}: ...................}
% \markboth{DEVARAJU ET AL.}{DQN - CAP .......X XX XXX XXXX....}

\maketitle

\begin{abstract}
%A hybrid mobility and congestion-aware routing protocol with route maintenance using topology control is designed for autonomous and decentralized UAV networks used in target search and surveillance applications.
%
Routing protocols help in transmitting the sensed data from UAVs monitoring the targets (called target UAVs) to the BS. However, the highly dynamic nature of an autonomous, decentralized UAV network leads to frequent route breaks or traffic disruptions. Traditional routing schemes cannot quickly adapt to dynamic UAV networks and/or incur large control overhead and delays.
To establish stable, high-quality routes from target UAVs to the BS, we design a hybrid reactive routing scheme called pipe routing that is mobility, congestion, and energy-aware. 
The pipe routing scheme discovers routes on-demand and proactively switches to alternate high-quality routes within a limited region around the active routes (called the pipe) when needed, reducing the number of route breaks and increasing data throughput.
We then design a novel topology control-based pipe routing scheme to maintain robust connectivity in the pipe region around the active routes, leading to improved route stability and increased throughput with minimal impact on the coverage performance of the UAV network.
\end{abstract}


% \chapter{Pipe Routing with Topology Control for UAV Networks} \label{chap-TC}

\section{Introduction}\label{Intro-TC}
%\textcolor{red}{
Unmanned aerial vehicles (UAVs) equipped with self-localization and sensing capabilities 
have become popular for applications such as 
area monitoring, surveillance, search-and-rescue, and target tracking \cite{SAR,bsCon,b2}.
We focus on low SWaP (size, weight, and power), fixed-wing UAVs, which offer a balance of portability and area coverage (with higher speeds and longer lifetimes than rotor-based UAVs).
However, low SWaP UAVs face some practical difficulties: they have limited communication range and are more prone to failure than larger UAVs.  These issues motivate a \textbf{decentralized} and \textbf{autonomous} network of UAVs, in which the nodes explore an area, perform local sensing, and communicate with their neighbors without any global knowledge of network topology.
%} 

We consider a network of low SWaP fixed-wing UAVs 
deployed in an area without any fixed communication infrastructure, such as a cellular network.
Example targets might be buildings on fire or a damaged bridge in a disaster area, or enemy troop movements on a battlefield, so the targets need constant monitoring.
When one or more targets of interest are found, a UAV (called the \emph{target UAV}) continuously monitors each target and transmits the sensed information back to the base station (BS) in real time. 
The remaining UAVs continue to collaboratively monitor the area to find new targets while maintaining connectivity and 
communication routes to the BS.
% removed this: problem statement does not require BS-CAP?
%, using the BS-CAP mobility model \cite{BS-CAP_ref}.

\begin{figure*}[htbp]
\centering
    \includegraphics[width=0.9\textwidth]{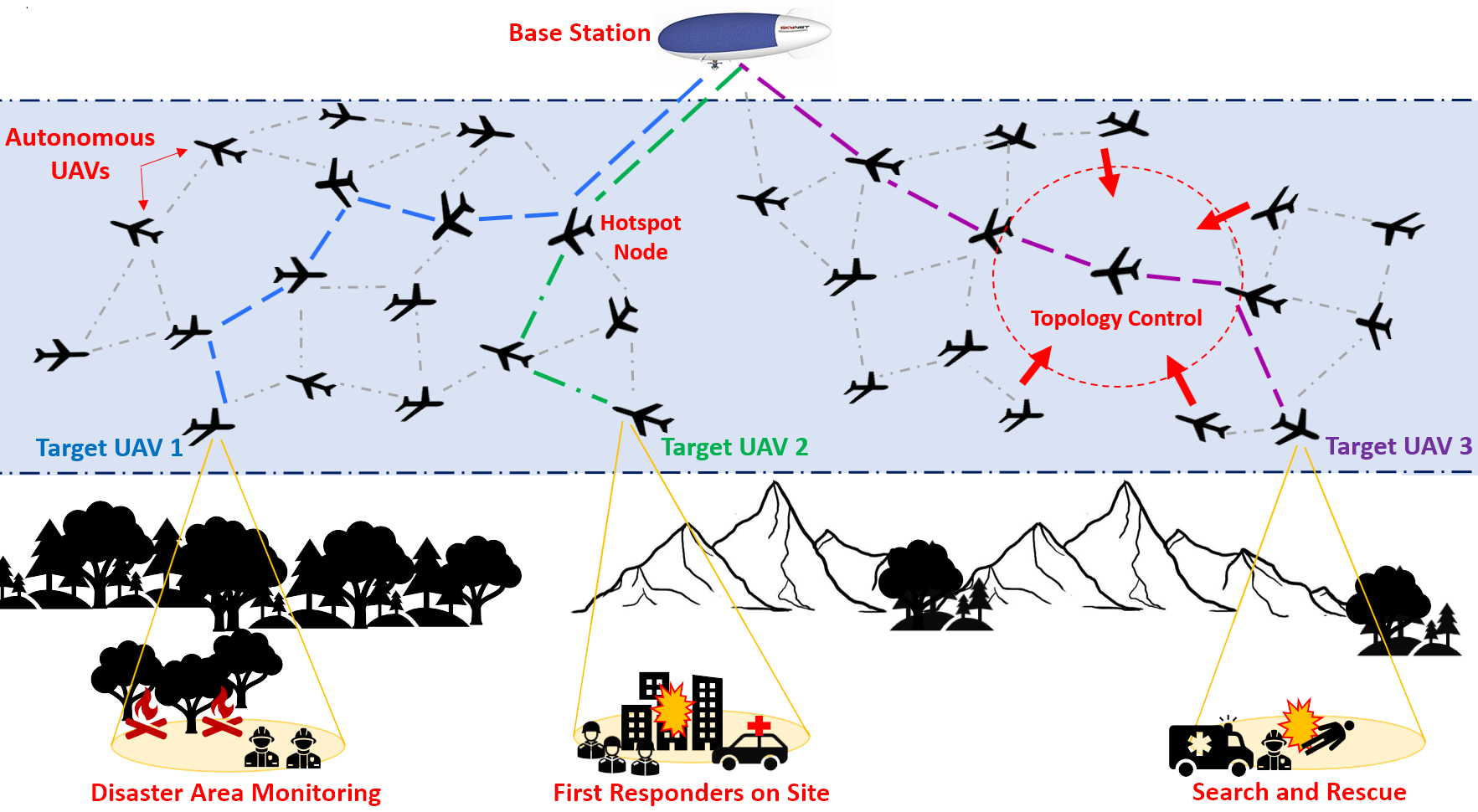}
    \caption{Decentralized, autonomous UAV network performing monitoring, search and surveillance in inaccessible disaster areas.}
    \label{fig_tcpipe_applications}
\end{figure*}

The goal of a routing protocol is to help establish communication routes along which the sensed data can be transmitted from each target UAV to the BS. However, the highly dynamic nature of an autonomous decentralized UAV network can result in frequent changes in the network topology, causing existing routes to break or creating traffic disruptions. The disrupted network may experience loss of data (packet drops), increased control and computational overhead, as more resources are devoted to the frequent route rediscoveries, and corresponding increased delays in the transmission of the target data.
Most existing routing schemes cannot update the routing table or discover a new route before the current route breaks.
\emph{Reactive} routing schemes then create significant delays, since they perform new route discovery while the current routes are broken. On the other hand, \emph{proactive} routing schemes introduce large overheads, as each node periodically transmits control messages to maintain and update the routes in the network \cite{mca-olsr-shivam}. 
Another concern in our application, in which all the target UAVs need to relay their information to a single base station, is that the routes may converge and create high resource usage (and congestion) at nodes along any shared routes, especially at high data rates. 
These ``hot-spot'' nodes along shared routes can become critical points of failure, causing communication disruptions.
Low SWaP UAVs have limited battery capacity, and their residual energy levels may vary depending on usage (flight duration and trajectories). 
Selecting UAVs with low energy reserves to participate in routes to the BS inevitably results in disruption of the data flows when these low-energy UAVs fail. 

To mitigate these issues, a good routing protocol should be mobility, congestion, and energy-aware. 
We say a routing protocol is ``mobility-aware'' if its selection of routes in a network is based on the trajectories of nodes within the route; it can select long-lasting routes by considering the estimated lifespan of a chosen route 
\cite{khudayer2020efficient, garg2021adaptive, sirmollo2021mobility}. 
% \cite{lin2018shortest, khudayer2020efficient, garg2021adaptive, sirmollo2021mobility, tiwari2008mobility}.
``Congestion-aware'' routing protocols select routes that minimize congestion by avoiding congested nodes \cite{mca-olsr-shivam, zhang2022adaptive, pu2018link, song2018mobility, P[5]}.
% \cite{congest-aware1, congest-aware2, congest-aware3, congest-aware4, mca-olsr-shivam}. 
Finally, ``energy-aware'' routing involves selecting nodes above a minimum energy level required to provide stable and long-lasting routes. 

%\textcolor{red}{
Of course, the ability of a routing protocol to find or switch routes depends on the availability of routes between the source and destination. Hence, routing protocols alone cannot guarantee the stability and quality of the routes; the nodes' mobility model plays a crucial role. 
For this reason, we use the BS-CAP mobility \cite{BS-CAP_ref}, a pheromone-based connectivity-aware mobility model, to improve overall network and BS connectivity.
Our focus is on enhancing the stability and quality of routes between target UAVs and the BS; to this end, we propose a pheromone-based topology control mechanism to increase the density of UAVs around an active route. This improves the availability of high-quality routes and enables the routing protocol to choose the best route (long-lasting and less congested) from the available options.

We build our approach on a hybrid reactive routing protocol called MCA-AODV \cite{mca-aodv-shivam}, and design a novel energy-aware routing protocol, called \textbf{Pipe routing}, as well as a complementary topology control scheme, together called \textbf{TC-Pipe routing}. These schemes select a stable route between each target UAV and the BS, while also considering the energy level of the nodes along the selected route. 
A relay based routing scheme is also discussed for comparison against the performance of both pipe based routing schemes. 

\begin{itemize}
    \item We design a mobility, congestion, and energy-aware hybrid reactive routing protocol, called \textbf{Pipe routing}, by modifying the MCA-AODV \cite{mca-aodv-shivam} routing protocol. This routing protocol is able to adapt to highly dynamic UAV networks and proactively switch or select stable and high-quality routes.
    \item Energy awareness prevents the selection of routes with low-energy UAVs; this avoids disruption of the data flows when these low-energy UAVs fail. 
    \item Congestion-aware routes reduce the congestion at hot-spot nodes when multiple flows to the BS are present in the network (especially at high data rates).
    \item A novel pheromone-based topology control scheme is designed; it works in conjunction with the Pipe routing scheme, and is called \textbf{TC-Pipe routing}. The TC-Pipe scheme improves the availability of high-quality routes between the target UAVs and BS. 
    % 
    %\item We study and analyze the TC-Pipe routing scheme's improvement in routing performance (PDR, route breaks, route up time) and its impact on  the coverage performance (percentage cells covered and coverage fairness).
    % 
    \item The TC-Pipe routing scheme maintains robust connectivity around the active routes, leading to improved route stability and increased throughput with minimal impact on the coverage performance of the UAV network.
    % 
    %\item TC-Pipe and Pipe routing schemes reduce the number of route breaks and achieved better PDR than the AODV and ConCov-Pipe routing schemes.
    % 
    \item Both the TC-Pipe and Pipe schemes' PDR and coverage performance decrease gracefully in proportion to the \% UAVs failing in the network.
    \item A relay based routing scheme that uses dedicated relay nodes is also discussed for comparison against the performance of the Pipe and TC-Pipe routing schemes
    % 
    % \item The proposed pipe routing schemes achieve reduced control and computational overhead compared to other proactive routing schemes \textcolor{red}{- Have you compared with proactive routing?}. 
\end{itemize}

\emph{Paper Organization:} 
We first review several existing routing and topology control schemes that focus on providing robust routes
(Section \ref{Related_Work_section-TC}). Next, we discuss our proposed pipe routing and topology control schemes in Sections \ref{sect_pipe_routing} and \ref{toplogy_control}, respectively, followed by the relay based routing scheme in Section \ref{Relay_scheme}. We evaluate the performance of our approach in simulations and analyze the results in Section \ref{Simulation-TC}.

\section{Related Work}\label{Related_Work_section-TC}

Designing a resilient, robust routing protocol for autonomous, decentralized UAV networks is challenging: the network topology is highly dynamic, and traffic undergoes frequent fluctuations \cite{topologyrouting-survey, P[2], P[5], P[7]}.  Proactive routing schemes, such as OLSR and its variants \cite{ P[10], mca-olsr-shivam}, introduce large control and computational overhead \cite{mca-olsr-shivam} and may increase congestion in the network. 

On the other hand, reactive routing schemes 
\cite{AODV, LEPR, BAOMDV-LR-paper, P[7], P[12],P[19],P[20],P[21],P[22],P[23],P[24]} discover new routes only on demand, and thus incur lower control and computational overhead. 
However, reactive methods create higher route discovery delays.
Route re-discovery can be especially problematic when routes break frequently, as in a dynamic UAV network. 
Many reactive routing schemes \cite{AODV, P[2]} suffer from control packet (RREQ) flooding problems, known as ``broadcast storms" \cite{P[7]}.
% \cite{mca-aodv-shivam}. 
Although some schemes \cite{P[7], P[19],P[12]} explicitly try to reduce the overhead and delay caused by frequent route discoveries, this often comes at the cost of using the best information. For example, in \cite{P[19]}, when a route breaks, the intermediate nodes search for an alternate route only in their respective k-hop neighborhood; this 
reduces control overhead by avoiding network-wide route discovery by the source node. 
% AI: Clarify? Not sure I understand from the text how this is different/worse than the pipe search?
In \cite{P[12]}, RREQ flooding is limited by only forwarding the RREQ to certain nodes identified in different zones based on the neighboring nodes' location and connectivity. However, since the full network topology is not used for route discovery, these schemes can result in the selection of poor routes. 

Many applications are time sensitive (e.g., video streaming) and demand a certain bandwidth level from the network. Reactive routing schemes that are QoS-aware \cite{mca-aodv-shivam, P[20], P[21], P[112]} use network statistics such as node mobility, link stability, hop count, signal strength, node energy level, congestion, and traffic load to select the best route. In one such scheme \cite{P[20]}, RREQ is rebroadcast by only those nodes that meet specific QoS thresholds like time-to-live (TTL), bandwidth, and flow jitter. However, this protocol does not estimate or track node mobility and link stability, leading to frequent route breaks.
In \cite{P[21]}, an intermediate node in the route can adapt to the changes in the network topology and traffic conditions by selecting a new downstream node (towards the destination). This selection is based on neighbor distance, latency, traffic load, and reliability. However, this 
%AI: why does it result in long/inefficient routes? 
scheme can sometimes result in longer and less efficient routes. An energy-aware variant of AODV \cite{P[112]} uses the link stability and residual energy of the nodes to select stable routes.

Multipath reactive routing protocols \cite{BAOMDV-LR-paper, P[22]} find multiple node-disjoint or link-disjoint paths during route discovery and cache them as backup routes. When the current route quality degrades, the backup routes from the cache are used instead of performing a new route discovery. However, like the current route, the quality of the backup routes also typically decline over time. To avoid selecting bad routes, \cite{P[23], P[24]} periodically monitor the quality of the backup routes. Whereas \cite{P[7]} performs local route repair using route recovery messages and uses fuzzy logic route selection based on node mobility, residual energy, and link quality. 
%\todo{AI: transition? also passive}
Link stability based reactive routing protocol \cite{LEPR} uses a semi-proactive route switching mechanism, which reduces route discovery overhead incurred in \cite{P[7],P[23],P[24]}. Here, the intermediate nodes notify the destination node of the route whenever the quality of the link to their downstream nodes degrades. Then the destination node sends a new RREP packet (containing the updated route stability value) to the source node along each cached route.
The source node then switches to an alternate route, which avoids the high overhead and delay of new route discovery. However, this scheme is unable to adjust to traffic changes and congestion and also lacks a route repair mechanism. 

%\todo{AI: this whole section reads like an obligatory list: X does A, but lacks B. Y does C. Z does D, but doesn't have E.}

A scalable hybrid routing protocol called mobility and congestion-aware AODV (MCA-AODV) \cite{mca-aodv-shivam} combines reactive and proactive routing methods to adapt to dynamic UAV networks with lower control overhead than traditional proactive networks. Here, the source node maintains knowledge of alternate routes and their quality within a 2-hop region around the selected route (called a ``pipe") to perform proactive route switching when the quality of the current route degrades. MCA-AODV is both mobility and congestion-aware. It selects the best route based on various factors such as hop count, interference, route lifetime, and estimated latency to provide long-lasting and less congested routes.
% vv AI: doesn't sound very different / novel? explain?
%\textcolor{blue}{In this paper, we propose an energy-aware hybrid routing protocol by modifying MCA-AODV. The proposed protocol selects the best route based on hop count, interference, route lifetime, and node residual energy. Then, topology control is added to improve the availability and stability of routes in the pipe region improving its network performance.}

\noindent\textbf{Topology Control for Robust 
Routes:}  Some schemes (e.g., \cite{TC-survey, CAP_ref, BS-CAP_ref, bsConCov, survey_multi-robots3, sensor14}) modify the mobility models to provide stable network connectivity. In this section, we review topology control and routing schemes 
that maintain a stable active route between the source and destination UAVs.
% 
% AI: discussion vvv is a bit list-y; little context
In \cite{BR_AODV}, the AODV routing protocol \cite{AODV} and a modified Boids mobility model \cite{boids} are used to maintain the links over the route in a highly dynamic autonomous UAV network. 
%The movements of UAVs participating in the route are restricted to maintain links while data is being transmitted.
%In the standard AODV protocol, only a single source is assumed, and the congestion and energy resources of the nodes are not considered when establishing the route. %Moreover, the impact on the area coverage performance is not studied. 
%\todo{AI: mixing critiques of method with paper? impact of what on coverage?}
%This scheme is similar to the Relay scheme (discussed in Section \ref{Relay_scheme}) implemented for comparison.
% 
%\todo{transition?}
In \cite{rahmati2021dynamic}, a global joint 3D trajectory and power allocation scheme was designed using spectral graph theory and convex optimization techniques to form interference-aware routes between a single source and destination (BS) using rotary wing UAVs. %This approach is not suitable for decentralized UAV networks.
%\todo{why? (strong statement with no obvious support?)}
% 
In \cite{virtspring-TC}, a mobility control system for FANETs based on virtual springs maintains an aerial mesh network to ensure ground-user coverage and QoS for disaster recovery.
In \cite{TC-routing-crowd-control}, a virtual force-based UAV mobility model and energy-efficient mobility-aware cluster head (CH) UAV selection is used to improve ground-user coverage and network connectivity. Authors also propose a topology-aware routing protocol that uses Q-learning to find stable, low-delay and energy-efficient routes between CH UAVs and BS. 

Our proposed routing protocol for autonomous UAV networks combines the advantages of both proactive and reactive schemes, along with topology control, to establish robust and low-overhead routes that are also mobility, congestion and energy aware. This protocol is based on our connectivity-aware pheromone mobility model that balances maintaining connectivity with providing coverage \cite{BS-CAP_ref}.

\subsection{Overview of UAV Mobility Models}\label{overview_mobility_models}
Pheromone mobility models use digital repel pheromones to promote exploration and fast coverage of an area with no prior information \cite{CAP_ref, b2}.
We consider a decentralized connectivity-aware pheromone-based UAV mobility such as BS-CAP \cite{BS-CAP_ref} in our experiments, to provide fast area coverage, strong node degree, and base station connectivity. In BS-CAP, the UAVs move and deposit repel pheromones in virtual maps divided into grid cells, and the trajectory of a UAV (controlled by its next-waypoint) is decided based on the repel pheromone in cells and the node degree at a set of candidate waypoints, while ensuring that the next-waypoint maintains a route to the BS. The coverage and connectivity trade-off is controlled by adjusting the weight given to the UAVs' node degree at the waypoints.
In the sequel, we also modify the BS-CAP model to provide topology control in the pipe, and increase the density of UAVs around an active route (see Sections \ref{sect_pipe_routing} and \ref{toplogy_control}).

Another connectivity-aware UAV mobility model used in our experiments is the \textit{ConCov model} \cite{bsConCov}, which uses a modified flocking behavior to provide trade-offs between coverage and BS connectivity. The ConCov model uses artificial potential fields, along with 1-hop neighbor locations and their routing information, to minimize overlap in the area sensed by neighboring UAVs but maintain BS connectivity.
The heading of a UAV is the weighted sum of repelling forces from neighboring UAVs and an attractive force to a neighboring UAV with a route to the BS. The coverage/connectivity trade-off can be tuned by adjusting the relative weight of these two forces.

\section{Pipe Routing Scheme} 
\label{sect_pipe_routing}
The Pipe routing scheme selects and maintains a stable route between each target UAV and the BS, known as an \emph{active route}. It is based on a mobility and congestion-aware, hybrid reactive routing scheme \cite{mca-aodv-shivam}.
Since the UAVs participating on an active route also participate in area coverage and exploration tasks (e.g., following the BS-CAP mobility model \cite{BS-CAP_ref}), the resulting network topology changes over time, which can cause link breaks and/or changes in route quality. 
Pipe routing proactively maintains the network information of a narrow, 2-hop region of nodes around the active route (called the ``pipe"; see Section \ref{pipe_form}), and the source node performs a ``multi-metric'' route selection (Sections \ref{routing_metrics} and \ref{active_route_select}) to switch to a new, higher-quality route when required.
Because it takes into account the node mobility, available energy, and congestion, the Pipe routing scheme selects less congested and long-lasting routes with fewer route breaks, while also incurring little route discovery overhead. When the quality of the current route drops below a pre-defined threshold, the source node proactively switches to a new route within the pipe without triggering a new route discovery. 

\subsection{Desctiption of Routing Metrics}\label{routing_metrics}
To choose a stable, low congestion route with a long lifetime, we use several route information metrics, capturing route length, interference, anticipated route lifetime, and available energy.  In particular, we use the hop count (HC, the number of links between the target UAV and the BS) along with the total number of interfering links (IL) across all UAVs in the route.
The route lifetime (RLT) is the estimated time duration after which the route is likely to break; it is the minimum estimated link lifetime (LLT) over all the links that form the route. A UAV estimates the LLT of a link with its neighbors using the GPS and current trajectory information found in the Hello message from its neighbors. Additional details on computing these metrics can be found in \cite{mca-aodv-shivam}.
% \cite{section 2 in [11] from mca-aodv}.

In addition to the given route metrics, the residual energy of the route ($En_R$) is computed as the minimum energy across all nodes in the route. If $En_i$ is the energy level of node $i$ in the route (R), and $N_R$ is the set of nodes, then
\begin{equation} \label{route_energy}
En_R = \min_{i \in N_R}(En_i)
\end{equation}
UAVs with energy levels below a threshold ($En_{th}$) are likely to run out of battery soon, resulting in node failure and causing a route break. Therefore, these UAVs should safely disengage from the route.
For this reason, routes with $En_R \leq En_{th} + \delta_e$ are not considered during route selection in our scheme. 
Since the source node receives updated topology information (node energy) every 2 s, we select the value of the tolerance factor $\delta_e$ to be twice the per-second energy depletion rate 
% AI: is this what you meant? ^^
%as 2$\times$Energy\_Depletion\_Rate 
of a UAV to account for the delay.
Moreover, preemptive route switching is performed when the currently active route's $En_R$ falls below the threshold $En_{th}$.

\subsection{Active Route Selection}\label{active_route_select}
During the route discovery process, if the target UAV (source node) does not already have a route to the destination (in our setting, the BS), it broadcasts an RREQ message towards the BS. 
Unlike MCA-AODV \cite{mca-aodv-shivam}, which floods RREQ messages in the entire network, our scheme takes advantage of the UAV's knowledge about the BS's GPS location and broadcasts the RREQ messages only in the direction of the BS.
In response, the BS sends back RREP messages to the target UAV over the reverse routes, identified during the RREQ forwarding phase. 
To find stable routes, with a low control packet overhead, 
RREQ forwarding is limited to intermediate nodes that meet both the link stability condition (as in Eq. (1) of \cite{mca-aodv-shivam}) and the energy threshold condition, $En_i > En_{th}$.
% . This ensures that the RREQs and RREPs find stable routes to the BS that last longer.
% AI: one example of the repeated sentence vvv
%\textcolor{blue}{These two conditions serve to ensure that the RREP process results in stable and long-lasting routes to the BS.}

The target UAV receives RREPs containing the HC and IL values of the respective routes.
The target UAV then computes the best route $R^*$ (route with least cost $C_R$, defined in \eqref{route_cost})
by selecting among all routes that satisfy a lifetime condition \eqref{rlt_cond} and an energy condition \eqref{en_cond}.
The selected route is long-lasting, less-congested, and energy-sufficient.
Here, the lifetime condition is,
\begin{equation} \label{rlt_cond}
RLT_R > TTL,
\end{equation}
where $TTL$ is the time-to-live value of the data packets.
The energy condition is,
\begin{equation} \label{en_cond}
% new energy cond. - 
En_R > En_{th} + \delta_e,
\end{equation}
For each satisfying route $R$, we compute the cost, 
\begin{equation} \label{route_cost}
C_R = w_1 (\frac{HC_R}{HC_{min} }) + w_2 (\frac{IL_R}{IL_{min} + \alpha\, IL_R}),
\end{equation}
% where the route length of the active route $R^*$ is $HC_{min}$, 
% 
where $(\cdot)_{min} = \min_{R \in X}(\cdot)_R $ and 
$X$ is the set of all routes known by the target UAV that satisfy the constraints \eqref{rlt_cond} and \eqref{en_cond}.
The weights $w_1$, $w_2$ determine the balance between short paths (low HC) and low-interference paths (low IL), with $w_1 + w_2 = 1$; we set $w_1 = w_2 = 0.5$. 
To reduce the route computation delay, we only consider routes with $HC_R \leq HC_{min}+3$. We use $\alpha = 0.3$ as the scaling factor to convert the IL cost values to a similar range as the HC cost values 
\footnote{Since $HC_{min} \geq$ 2, the range of $(\frac{HC_R}{HC_{min} })$ is (1, 2.5]. Similarly,  since $IL_{min} \geq$ 2, $(\frac{IL_R}{IL_{min} + \alpha\, IL_R})$ can vary from 0.77 to 3.33. However, in our simulation settings, the range of this ratio is [1, 2.5) because $IL_R$ is generally greater than $IL_{min}$.}.

\subsection{Pipe Formation} \label{pipe_form}

Since the UAV network topology changes frequently, the quality of the current active route may deteriorate over time and new, better-quality routes may become available.  Ideally, the target UAV (source node) should proactively switch routes when the current route deteriorates.  In practice, we accomplish this by switching to a new route when the quality of the current active route drops below a certain threshold. This proactive switching can improve throughput, reduce interruptions in data flow, and prevent packet drops due to route breaks.

\begin{figure}[htbp]
\centerline{\includegraphics[width=0.6\linewidth]
{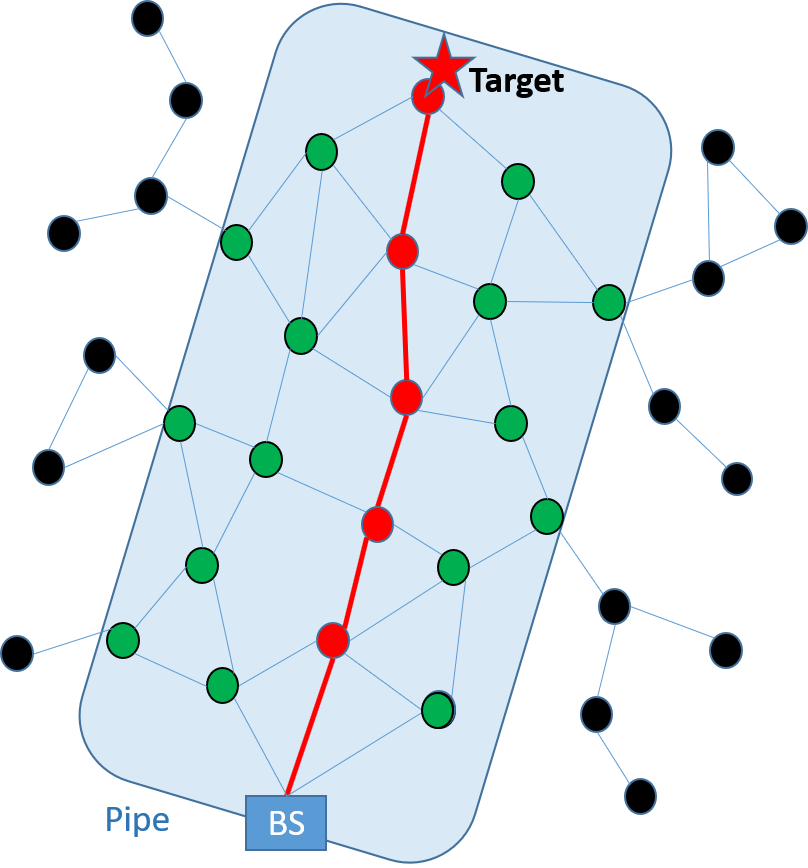}}
% {images/pipe_formation.png}}
\caption[Illustration of a pipe around the active route]{Illustration of a pipe around the active route (red links) from the target UAV to BS. The pipe consists of nodes (green nodes) that are up to 2-hop from the nodes along the active route (red nodes). }
\label{fig_pip_formation}
\end{figure}

Since network topology information is not centralized and no node knows the entire topology, the nodes must periodically broadcast their information via the protocol's Hello packets.
As in MCA-AODV \cite{mca-aodv-shivam}, the target UAV (source node) collects and builds limited network topology information around the currently used active route to form a ``pipe" and enable proactive route switching.
When a target is discovered, the target UAV performs route discovery and selects the best route to BS. Also, a pipe (see Figure \ref{fig_pip_formation}) is established along the active route from the target UAV to the BS \cite{mca-aodv-shivam}. The pipe comprises the nodes along the active route (red nodes) and their 2-hop neighbors (green nodes).
The topology information (including LLT, En, and IL values) of the nodes within the pipe is transmitted periodically to the source node as discussed in \cite{mca-aodv-shivam}.
A pipe width of 2-hop neighbors around the active route appears to give a good trade-off between the usefulness of the neighborhood information and its incurred overhead \cite{mca-aodv-shivam}.
Increasing the pipe width correspondingly increases the amount of the known network topology; while this may help identify more alternate routes, it comes at the cost of more control packets and computational overhead.

\subsection{Proactive Route Switching}\label{route_switching}
Proactive route switching is triggered when the quality of the active route ($R^*$) degrades below a threshold, given by the conditions \eqref{rlt_cond} and \eqref{en_cond}, or when a shorter alternate route becomes available in the pipe. In such a situation, the source node initiates a breadth-first search (BFS) algorithm to explore all the available routes within the pipe to the destination node and selects the route with least cost (calculated using \eqref{route_cost}).
When the source node switches to a new active route, the existing pipe is abandoned and a new pipe is formed around the new active route. If no suitable alternate route is available, a new route discovery is triggered. 
% \textcolor{red}{The route switching process is described in the flowchart in Figure [?].} 
%
By using the local pipe topology when possible, proactive route switching reduces the number of new route discoveries, and thus their overhead and delays, while minimizing the interruptions in data throughput.

\section{Topology Control} \label{toplogy_control}
The UAVs in the network follow the BS-CAP mobility scheme for improved coverage and connectivity \cite{BS-CAP_ref}. In this section, we discuss a topology control scheme to alter the trajectory of UAVs near the pipe region in order to maintain a robust pipe and a stable route. The proposed scheme, which combines both pipe-based routing (Section \ref{sect_pipe_routing}) and topology control, is called \emph{pipe routing with topology control} (TC-Pipe).

\subsection{Need for Topology Control}
Stable and high quality routes, along with proactively identified alternate routes in the pipe, can increase data throughput and reduce the number of route rediscoveries, thereby reducing packet loss, overhead, and delay.
Although the BS-CAP mobility model tries to provide a strong node degree as well as BS connectivity, it can be difficult to achieve both fast area coverage and stable connectivity for low to medium density UAV swarms. As a result, the route between the target UAV and BS may not be stable or have sufficient quality to meet the QoS requirements.  Additionally, the advantages of our pipe-based routing and proactive switching critically depend on alternate routes being available in the pipe, which is also less likely at lower densities. 
%When the UAV density is low in the pipe, the likelihood of finding alternate routes decreases (see Figure \ref{fig_pip_thin}).
% 
% The BS-CAP connectivity condition is satisfied even if any path exists between a UAV and BS. This alone is not sufficient to ensure stable data communication between the target UAV and the BS. The dynamic nature of the network causes the routes between the target UAV and the BS to constantly change, leading to route breaks, data flow interruptions, and packet drops.
% 
%In addition to connectivity to the BS, we are interested in providing high-quality routes to meet the QoS requirements.

\begin{figure}[htbp]
 \centerline{\includegraphics[width=0.5\linewidth]{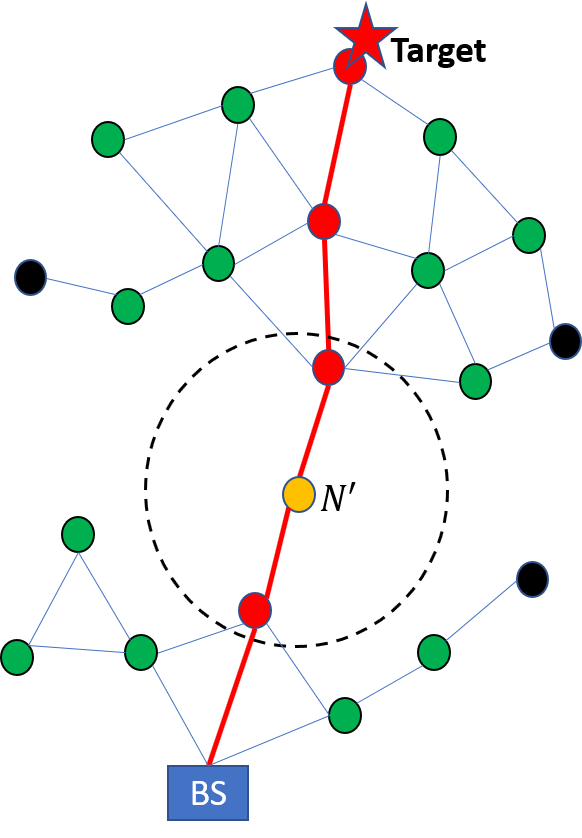}}
\caption[Illustration of pipe thinning problem]{Illustration of pipe thinning problem, where the node $N'$ has no 1-hop neighbors except the upstream and downstream nodes on the active route.}
\label{fig_pip_thin}
\end{figure}

The BS-CAP mobility model uses a repel pheromone, in which UAVs fly away from already-covered areas to explore the entire map, without taking into account the active route(s) being used for data communication. This can lead to sections of the pipe becoming thin, where some intermediate node(s) in the active route have very few or no neighbors. We call this effect the `pipe thinning' problem, and illustrate it in Figure \ref{fig_pip_thin}. 
%Pipe thinning can also occur as UAVs fly away from each other (due to repel pheromones) to cover all the areas of the map. 

Typically, existing node mobility schemes do not consider how to facilitate a robust set of alternative routes to improve the reliability and throughput of ongoing communication to BS \cite{BR_AODV, TC-survey, topologyrouting-survey}.
Our proposed topology control scheme uses the digital pheromone map and node-degree information of the nodes along the active route between the target UAV and BS to maintain a sufficient number of neighboring UAVs in the pipe region.  Our topology control is performed by 
applying a pheromone mask that negates the repel pheromone (effectively acting like attract pheromone)
to bring UAVs toward regions of each active route's pipe where the UAV density is low.  Combined with pipe-based routing, this reduces data flow interruptions and improves throughput. It can be particularly helpful when the targets are located far from the BS (a long route length) in a network of low node density. 

%To address these issues, we design a topology control scheme to work with the pipe-based routing which can reduce data flow interruptions and improve the throughput.
% 

% 
%Utilization of the pipe depends on the stable routes (that satisfy the route quality thresholds) it can provide (as described in Section \ref{route_switching}). 
% 

%This would improve the stability of the ongoing data communication to BS along the active route in the pipe. 

% \textcolor{red}{We study the improvements in routing performance and throughput and its trade-off with coverage performance.}\\
%

\subsection{Topology Control Scheme}
% %
% \begin{figure}[htbp]
% \centerline{\includegraphics[width=0.7\linewidth]{images/pipe_thinning.PNG}}
% \caption{Pipe thinning problem }
% \label{fig_pip_thin}
% \end{figure}
% %
\begin{figure}[htbp]
\centerline{\includegraphics[width=0.7\linewidth]{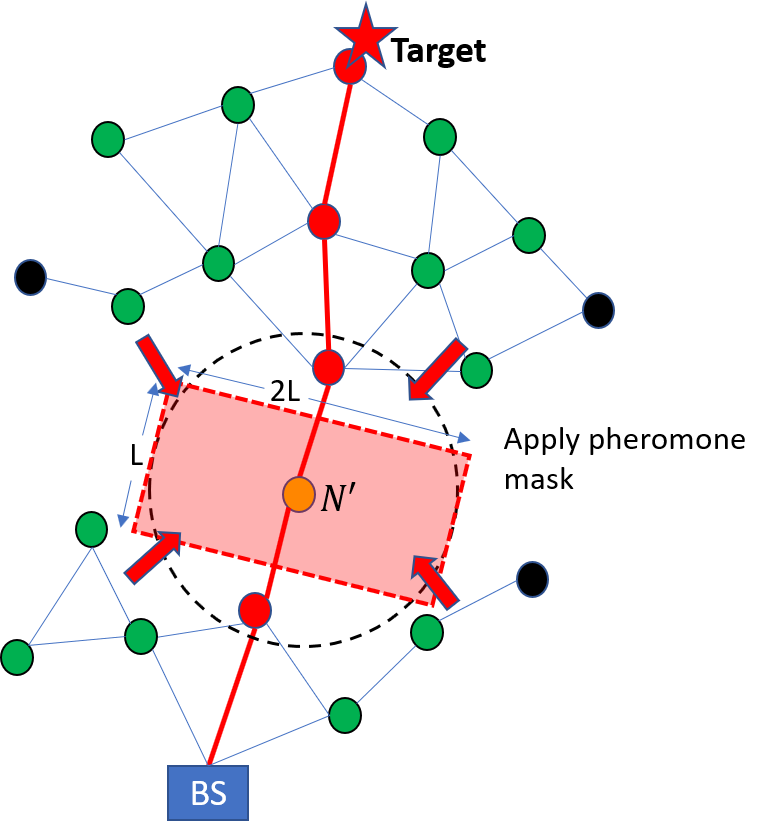}}
\caption{Applying a pheromone mask to attract UAVs.}
% \centerline{\includegraphics[width=0.45\linewidth]{images/Attract_pheromone.PNG}}
% \caption{Illustration of depositing the \textcolor{red}{attract} pheromones.}
\label{fig_pip_add_attract}
\end{figure}
%
%\begin{figure}[htbp]
%\centerline{\includegraphics[width=0.99\linewidth]{images/remove_attract_pher.PNG}}
%\caption{Remove attract pheromone }
%\label{fig_pip_remove_attract}
%\end{figure}
%

In Figure \ref{fig_pip_thin}, the node $N'$ (shown in yellow) on the active route has a low node degree, below a user-selected threshold $Th$, due to pipe thinning. This low degree reduces the probability of finding an alternate route in the pipe region from the target UAV to the BS. 
%If the quality of the current active route becomes worse or if the active route is about to break and no alternate routes are available to switch to, then data flow from the UAV to BS is more likely to break. Th
%is interruption in the data flow leads to decreased throughput, increased route discoveries, overhead, delays, and packet loss.
In this section, we propose a topology control scheme that uses pheromone-based mobility control to resolve the pipe thinning problem. 

Since all UAVs in the network follow the BS-CAP mobility model (which is a repel pheromone-based mobility scheme), their mobility is influenced by the pheromone values of the cells \cite{BS-CAP_ref}. The digital pheromone information is propagated to the neighbors using the Hello messages.
In our updated approach, nodes on an active route 
% deposit \textcolor{red}{attract} pheromones 
apply a pheromone mask to nullify the nearby repel pheromone 
if their node degree is below $Th$. 
As a result, neighboring nodes that are not a part of the active route are induced to move into the region;
see Figure \ref{fig_pip_add_attract} for an illustration.

\begin{algorithm}[htbp]
\SetAlgorithmName{Algorithm}{}
\LinesNumbered 
\setcounter{AlgoLine}{0}
\begin{small}
    // For an active route\\
    \For{each node $N'$ in active route; except BS}{
        Get $d_{N'}$ = 1-hop node degree excluding upstream and downstream nodes\;
        //Node degree threshold, $Th=2$\\
        \uIf{$d_{N'} \leq Th$}{
            // Pipe node density low in region\\
            % Add \textcolor{red}{attract} pheromone to 
            % pheromone map, $P = P_R \odot A$ from \eqref{P_map_attarct}\;
            Apply pheromone mask, $P = P_R \odot A$ from \eqref{P_map_attarct}\;            
            }
        \Else{// Pipe node density not low in region\\
            % Remove \textcolor{red}{attract} pheromone in pheromone map, $P = P_R$; 
            Remove pheromone mask, $P = P_R$;
        }
        }
\end{small}
\caption{Topology Control}
\label{pseudo:topology_control}
\end{algorithm}

% We assume that node $N'$ is in cell $(x,y)$. If no attract pheromone is deposited by node $N'$ in cell $(x,y)$, then its pheromone map ($P$) is given by,
% \begin{equation} \label{P_map_repel}
% P = P_R
% \end{equation}
% where $P_R$ is the repel pheromone map. 

% If an attract pheromone is deposited by node $N'$ in cell $(x,y)$, then its resulting pheromone map ($P$) is given by,
% \begin{equation} \label{P_map_attarct}
% P = P_R \odot A
% \end{equation}
% where $\odot$ notation signifies an element-wise product and $A$ is the attract pheromone mask. Here mask $A$ is given by,

% \begin{equation} \label{A_horizontal}
% A_{i,j}=
% \scalebox{1.0}{$
%     \begin{cases}
%         0 & \mbox{$(x-10 \leq i \leq x+10), (y-5 \leq j \leq y+5)$ }\\
%         1 & \mbox{otherwise.}\\
% \end{cases}$}
% \end{equation}

% Orientation of the attract pheromone mask ($A$) applied at cell (x,y) varies based on the angle $\theta$ of the vector connecting the upstream and downstream nodes of $N'$. We apply the attract mask perpendicular to this vector so that the nodes could be attracted to move towards node $N'$ (see Figure \ref{fig_pip_add_attract}). This facilitates the formation of alternate routes between the upstream and downstream nodes of the active route.
% % 
% $A$ is defined as \eqref{A_horizontal} if the $135\leq |\theta| \leq45$. Else, A is defined as \eqref{A_vertical},

% \begin{equation} \label{A_vertical}
% A_{i,j}=
% \scalebox{1.0}{$
%     \begin{cases}
%         0 & \mbox{$(x-5 \leq i \leq x+5), (y-10 \leq j \leq y+10)$ }\\
%         1 & \mbox{otherwise.}\\
% \end{cases}$}
% \end{equation}

Consider a node $N'$ located in cell $(x,y)$, with repel pheromone map $P_R$ \cite{BS-CAP_ref}.
% (see Section \ref{Overview_Pheromone_Mobility_section} for details).
% If an \textcolor{red}{attract} pheromone is deposited by 
If a pheromone mask is applied by
node $N'$, its resulting pheromone map $P$ becomes
\begin{equation} \label{P_map_attarct}
P = P_R \odot A,
\end{equation}
where $A$ is the 
pheromone mask and $\odot$ represents an element-wise product. 
$A$ is a rectangular region of dimension $2L$ x $L$ centered at node $N'$'s cell $(x,y)$, and oriented perpendicular to the vector connecting the upstream and downstream nodes of $N'$ (see Figure \ref{fig_pip_add_attract}). The size $L$ is taken to be the UAV's signal transmission range, i.e., the maximum 1-hop distance between UAVs. The pheromone values of cells in A are set to zero. The mask update can be equivalently viewed as depositing an attract pheromone value in each cell equivalent to its current repel pheromone value, resulting in a net pheromone value of zero. This causes the cells under the $2L$ x $L$  region 
%centered at node $N'$'s cell $(x,y)$ will 
to attract UAVs relative to cells outside the masked region, whose repel pheromone values are generally positive. 

Nodes arriving in the newly masked region provide potential alternate links and improve the robustness of the pipe region around the active route node(s). The 
% \textcolor{red}{attract} 
pheromone mask is then retracted by the node(s) on the active route once its number of 1-hop neighbors exceeds $Th$.  
The topology control algorithm is described in Algorithm \ref{pseudo:topology_control}.
We use the threshold $Th$ = 2 in our simulations. While higher $Th$ attracts more nodes to the 1-hop regions of the active route, doing so does not necessarily improve the routing performance and can degrade the network's area coverage at low node densities. 
% We have also observed that a higher node degree can form redundant links between upstream and downstream nodes.
% \todo{AI: don't follow purpose of the last sentence}

\section{Control Overhead and Computational Complexity}

\emph{Control Overhead:} In addition to the information sent in the RREQ, RREP, RERR and Hello packets in the standard AODV protocol, each node in the propsed TC-Pipe routing scheme includes the following information in its Hello packet:
\begin{itemize}
    \item Its UAV ID (7 bits), its current location (18 bits with 10 m resolution), next-waypoint cell (12 bits for the cell ID, assuming 60 x 60 cells of 100 m x 100 m resolution in a 6 km x 6 km area), and a local pheromone map (6-bit pheromone values in the 5 x 5 cells centered at the UAV’s current cell, for 150 bits total). It also includes the hop count of the shortest route to BS (4 bits), and the pheromone mask flag/cell ID (12 bits).
    % \item Its GPS location (uses 6 bytes for (x,y,z) coordinates) and trajectory information, which includes the center coordinates (6 bytes), and radius and node movement direction in ST mobility model (2 bytes) [29]. Note that a node sends its trajectory information only when it forms a new link or changes its current trajectory.

    \item Its $En$ and IL values (1 to 2 bytes).

    \item If the node is a 1-hop neighbor of an active node (that is participating in data packet forwarding), it broadcasts the LLT, $En$, IL, and isLinkActive values for each of its 1-hop neighbor nodes (variable size depending on neighbor density).
    
\end{itemize} 

The destination node periodically sends the Notify\_Source packet to the source node, which carries the 2-hop neighborhood information (i.e., UAV ID and
their $En$, IL and LLT values) of all intermediate nodes on the route (variable size depending on 2-hop neighbor density along the route).

% \emph{Computational Complexity}: In TC-Pipe routing scheme, a source node finds all routes to the destination node within the pipe using the BFS algorithm, which has a worst time complexity of $O(TV_p^2(E_p + V_p))$, where T is the number of times routes are recomputed, and $V_p$ and $E_p$ are the number of nodes and links, respectively, within the pipe. In addition, the intermediate nodes compute the cliques of their 2-hop neighborhood using the Bron-Kerbosch algorithm [30], which has the worst time complexity of $O(3^{V_p/3})$. Note that the clique computation is optional and helps in reducing the control overhead.
\emph{Computational Complexity}: In the TC-Pipe routing scheme, the source node employs the BFS algorithm to discover all possible routes to the destination node within the pipe. This algorithm's worst-case time complexity is $O(TV_p^2(E_p + V_p))$, where T represents the number of route recomputations, and $V_p$ and $E_p$ denote the number of nodes and links within the pipe, respectively. Additionally, intermediate nodes use the Bron-Kerbosch algorithm to compute cliques within their 2-hop neighborhood. This algorithm's worst-case time complexity is $O(3^{V_p/3})$. It is worth noting that the clique computation is optional but aids in reducing control overhead.

\section{Relay Routing}\label{Relay_scheme}
%
% \todo{AI: kind of agree that these results are of a different character in the results section.  Maybe we could put this text, and a comparison of maybe TC-Pipe by itself to Relay, in a separate section after the main experiments?  Again, the purpose is just to explicitly quantify this tradeoff -- around 0 network breaks and good PDRs, but at the cost of a pretty significant coverage reduction in 30, and less in 50. SHREY: Dr. Kumar decided to leave it as it is for the theis}
%
Our approach to mobility and routing in UAV swarms is explicitly homogeneous: each node in the network is, individually, balancing the two needs of exploration (area coverage) and communication (providing routes within the network and to the BS).  A possible alternative, of course, is to assign \emph{roles} to nodes within the network, allowing some nodes to become solely responsible for providing routing (``relays''), while others continue to explore the environment.
As one example, relay UAVs might be deployed as a separate layer to provide continuous BS connectivity in a hierarchical network \cite{bs24, Steiner4}.
% \cite{[13][25]in_bscap}.

We believe that a homogeneous approach has a number of important advantages.
While a relay-based scheme prioritizes ensuring a stable communication path, by focusing on maintaining a specific route, it may create bottleneck nodes that suffer from congestion, or that can become critical points of failure to the communications structure.  Moreover, these nodes may be more vulnerable to antagonistic interference (e.g., on a battlefield) \cite{heir_net_challenges2}, since differences in the nodes' behavior may indicate their current role.
Finally, in low-density networks, the loss of some nodes to communication infrastructure may significantly degrade area coverage, without an obvious mechanism to trade off between the two.

%In a relay-based routing scheme,
% is a hierarchical system where 
%some UAVs act as dedicated relays, forming a fixed, stable route for communication between the target UAVs and the BS. The relay node architecture prioritizes stable communication but it utilizes more resources and some relay nodes can become the bottleneck nodes, leading to delays and congestion. These nodes can also be vulnerable to attacks \cite{heir_net_challenges2}. % \cite{[15]in_bscap}.
% 
%Relay UAVs may be deployed as a separate layer to provide continuous BS connectivity in a hierarchical network \cite{bs24, Steiner4}. % \cite{[13][25]in_bscap}.

In order to explore the scale of these tradeoffs to our approach,
we implement a simple relay-based routing scheme (called Relay) and compare its performance with our proposed schemes. Initially, all UAVs in the network follow the BS-CAP mobility model. When a UAV finds a target, it finds the shortest route to the BS using AODV routing. The nodes along this route to BS are reassigned as relay nodes. The relay UAVs in this established route start flying in a nearly circular trajectory within or around their current grid cell to maintain the established route. Thus, a stable and permanent route between each target UAV and BS is established for data communication. When a relay node fails, the above process is repeated to establish a new shortest route.
Of course, in the presence of multiple targets, this process of selecting the shortest routes to the BS may lead to multiple active routes sharing links in common, which will result in congestion, especially at high data rates.
The relay UAVs act as a separate layer from the rest of the UAVs, which continue performing area search and coverage while maintaining connectivity using the BS-CAP mobility model.  Again, the more nodes are selected to act as relays, the fewer nodes are available for search area coverage, degrading the swarm's coverage performance.

\section{Simulation Results and Discussion}\label{Simulation-TC}

\subsection{Simulation Setup}
%As discussed in Section \ref{from BSCAP}, 
Our experimental evaluations simulate the behavior of a swarm
of 30 to 50 low SWaP fixed-wing UAVs (e.g., \cite{locust, switchblade})  monitoring a 6 km $\times$ 6 km area in which fixed communication infrastructure, such as a cellular network, is not available.
The aerial BS is located at the bottom center of the map, and the UAVs are launched from the vicinity of the BS. Each UAV is equipped with GPS and has a transmission range of 1 km (e.g., \cite{mca-olsr-shivam, bsConCov, ieee80211ah}). We assume a multihop topology with free space propagation (e.g., \cite{ieee80211ah}).
For simplicity, the UAVs are assumed to be point masses, and their mobility is limited to the X-Y plane flying at a constant altitude (typically from 200 m to 1 km above the ground \cite{switchblade}). 
UAVs perform collision avoidance through trajectory modifications \cite{collision_avoidance}.
To facilitate representing the pheromone map, waypoints, and coverage statistics, the area is divided into grid cells of 100 m $\times$ 100 m each, consistent with other literature \cite{CAP_ref, b2, bsConCov, camera_spec1}.
A UAV scans the cell in which it currently resides and deposits a repel pheromone of magnitude $1$. We use pheromone evaporation $\lambda$ and diffusion $\psi$ rates of 0.006 each. 
Table \ref{sim_table-TC} shows the simulation parameters.

\begin{figure*}[htbp]
\centering
\begin{subfigure}{0.3\linewidth}
 \centering
\begin{tikzpicture}[scale=0.6]
    \draw[darkgray, very thick] (0,0) rectangle (6,6);
  
    \node[star,draw,inner sep=0,fill=black!30, minimum size=.7em] (T) at (1.5,5){};
    
    \node[above, font=\footnotesize] at (1.5,5){[1.5,5]km};

    \node[rectangle,draw,inner sep=0,fill=black!30, minimum size=.7em] (T) at (3,0.2){BS};
\end{tikzpicture}
\caption{Target Location: C1}
\label{fig:caseC1}
\end{subfigure}
\begin{subfigure}{0.3\linewidth}
 \centering
\begin{tikzpicture}[scale=0.6]
    \draw[darkgray, very thick] (0,0) rectangle (6,6);
  
    \node[star,draw,inner sep=0,fill=black!30, minimum size=.7em] (T) at (4,4){}; 

    \node[above, font=\footnotesize] at (4,4){[4,4]km};

    \node[rectangle,draw,inner sep=0,fill=black!30, minimum size=.7em] (T) at (3,0.2){BS};
\end{tikzpicture}
\caption{Target Location: C2}
\label{fig:caseC2}
\end{subfigure}
% % 
% \par\medskip % force a bit of vertical whitespace
% % 
\begin{subfigure}{0.3\linewidth}
 \centering
\begin{tikzpicture}[scale=0.6]
    \draw[darkgray, very thick] (0,0) rectangle (6,6);
  
    \node[star,draw,inner sep=0,fill=black!30, minimum size=.7em] (T) at (2,2){};
    
    \node[above, font=\footnotesize] at (2,2){[2,2]km};

    \node[rectangle,draw,inner sep=0,fill=black!30, minimum size=.7em] (T) at (3,0.2){BS};
\end{tikzpicture}
\caption{Target Location: C3}
\label{fig:caseC3}
\end{subfigure}
\par\medskip % force a bit of vertical whitespace
\begin{subfigure}{0.3\linewidth}
 \centering
\begin{tikzpicture}[scale=0.6]
    \draw[darkgray, very thick] (0,0) rectangle (6,6);
  
    \node[star,draw,inner sep=0,fill=black!30, minimum size=.7em] (T) at (1,4.5){}; 
    \node[star,draw,inner sep=0,fill=black!30, minimum size=.7em] (T) at (3,5){}; 
    \node[star,draw,inner sep=0,fill=black!30, minimum size=.7em] (T) at (5,4.5){}; 
    \node[rectangle,draw,inner sep=0,fill=black!30, minimum size=.7em] (T) at (3,0.2){BS};

    \node[above, font=\footnotesize] at (1,4.5){[1,4.5]km};
    \node[above, font=\footnotesize] at (3,5){[3,5]km};
    \node[above, font=\footnotesize] at (5,4.5){[5,4.5]km};    
\end{tikzpicture}
\caption{Target Location: C4}
\label{fig:caseC4}
\end{subfigure}
% % 
% \par\medskip % force a bit of vertical whitespace
% %
\begin{subfigure}{0.3\linewidth}
 \centering
\begin{tikzpicture}[scale=0.6]
    \draw[darkgray, very thick] (0,0) rectangle (6,6);
  
    \node[star,draw,inner sep=0,fill=black!30, minimum size=.7em] (T) at (1,4){}; 
    \node[star,draw,inner sep=0,fill=black!30, minimum size=.7em] (T) at (3.6,2.5){}; 
    \node[star,draw,inner sep=0,fill=black!30, minimum size=.7em] (T) at (5,4.8){}; 
    \node[rectangle,draw,inner sep=0,fill=black!30, minimum size=.7em] (T) at (3,0.2){BS};

    \node[above, font=\footnotesize] at (1,4){[1,4]km};
    \node[above, font=\footnotesize] at (3.6,2.5){[3.6,2.5]km};
    \node[above, font=\footnotesize] at (5,4.8){[5,4.8]km};
\end{tikzpicture}
\caption{Target Location: C5}
\label{fig:caseC5}
\end{subfigure}
\begin{subfigure}{0.3\linewidth}
 \centering
\begin{tikzpicture}[scale=0.6]
    \draw[darkgray, very thick] (0,0) rectangle (6,6);
  
    \node[star,draw,inner sep=0,fill=black!30, minimum size=.7em] (T) at (1,1.5){}; 
    \node[star,draw,inner sep=0,fill=black!30, minimum size=.7em] (T) at (3,2){}; 
    \node[star,draw,inner sep=0,fill=black!30, minimum size=.7em] (T) at (5,1.5){}; 
    \node[rectangle,draw,inner sep=0,fill=black!30, minimum size=.7em] (T) at (3,0.2){BS};

    \node[above, font=\footnotesize] at (1,1.5){[1,1.5]km};
    \node[above, font=\footnotesize] at (3,2){[3,2]km};
    \node[above, font=\footnotesize] at (5,1.5){[5,1.5]km};
\end{tikzpicture}
\caption{Target Location: C6}
\label{fig:caseC6}
\end{subfigure}
\caption[Target locations]{The target locations in a 6x6 $km^2$ map. Three different target locations are shown for a single target in (a), (b) and (c) and for a group of 3-targets in (d), (e) and (f).}
\label{fig:Targets_loc}
\end{figure*}
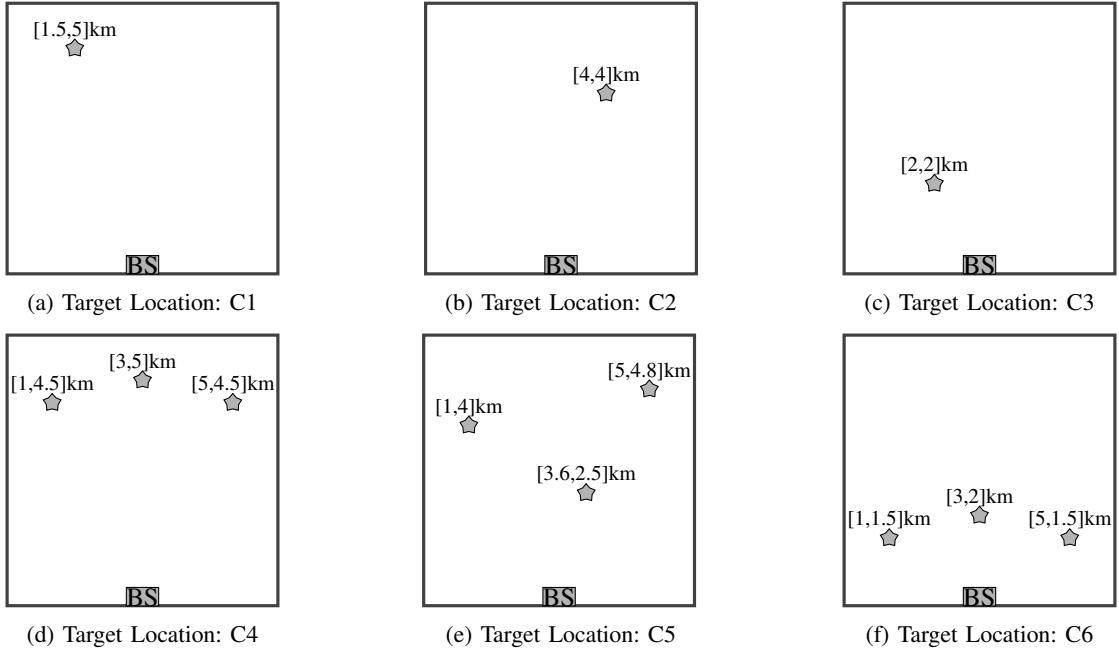

We evaluate monitoring both single and multi-target settings, across various locations; see Figure \ref{fig:Targets_loc}.
%We examine monitoring either single or multiple (3) targets in our simulations. These targets are placed at several location sets, including near and far ends of the map, as shown in Figure \ref{fig:Targets_loc}. 
The UAVs use the BS-CAP mobility model ($\beta=1.5$) to compute their flight trajectories. All the UAVs start out from the region near the base station. If a UAV discovers a target, it becomes a ``target UAV'' and continuously monitors the target by circling around it. Meanwhile, other UAVs continue searching and scanning for new targets or events. The target UAV establishes a route and transmits data packets continuously to the BS.

% AI: some repetition in these sections?
In our simulations, we take 
the channel rate to be 11 Mbps and the UAV transmission range to 1 km. We consider a line-of-sight communication and ignore channel fading and noise. The packet size and TTL values are 1.5 kB and 3 s, respectively.
We consider a  distributed TDMA MAC protocol  \cite{distributed-tdma-mac} and a large MAC queue size. Packets with the smallest time-to-expiry in the queue are pushed to the head-of-line.

%Each simulation duration is 2000 s, and performance metrics are averaged over 30 simulation runs. We study the network's throughput, routing, coverage, and connectivity performances for 2000 s after all the targets in the map are found (between 1000 and 3000 s). 
% AI: needs a better explanation of start/duration?
In order to study the network behavior under traffic, we
first simulate the UAV swarm for 1000 s, after which time all the targets have been found.  Then we evaluate the throughput, routing and coverage performance for 2000 s (i.e., from 1000 to 3000 s) as the swarm monitors the targets and continues to search. We report performance metrics averaged over 30 simulation runs. %Note that all the targets are found within the first 1000 s in our simulations but we show performance plots after 1000 s even when some targets could be found earlier than that. 
%Occasionally, finding the far-away targets could take up to 1500 s for a node density of 30; we still observe the performance for 2000 s after the targets are found. However, for the sake of uniformity, all the performance plots assume that the targets are found 

Low SWaP UAVs are prone to failure due to mechanical malfunctioning or energy depletion. To understand the impact of these issues on our routing scheme, we study the method's performance when a fraction of nodes randomly fail during the simulation time. In particular, we assign a fixed fraction of nodes (20\% or 30\%) a failure time during the simulation, with the time selected uniformly between 1000 and 3000s. This simulates nodes failing progressively over the course of the deployment.
% \textcolor{red}{Define $En_{th}$ value.}
% 
\begin{table}[htbp]
	\small
	\centering 
	\caption{Simulation Parameters}
	\label{sim_table-TC}
	\begin{tabular}{l l}
	\toprule
		Parameters & Values  \\
	\midrule
		Simulation Time &2000 s\\
		Map Area 		&6 km $\times$ 6 km\\
		% Cell Size & 100 m $\times$ 100 m \\
		%\textcolor{red}{Sensor Coverage Area} &100 m $\times$ 100 m (1 cell) \\
		Transmission Range 	&1 km	\\
		Number of UAVs		&30, 50\\
		UAV Speed 	&20 m/s, 40 m/s	\\
        Number of Targets &1, 3\\

        Channel Range &11 Mbps\\
        Packet size &1.5 kB\\
        Data Rate &1 - 3.5Mbps\\
        TTL &3 s\\
        
	\bottomrule
	\end{tabular}\\
\end{table}

\subsection{Performance Metrics} \label{sec:perfmetrics-TC}
We measure the performance of our UAV network using several routing and coverage metrics.

The \emph{routing} properties of the network are measured by:
\begin{itemize}
    \item \textit{Packet Delivery Ratio ($PDR$)}:  The ratio of total data packets received at the BS to the total data packets generated at the target UAVs. PDR is a normalized throughput and can be used to calculate other metrics such as flow throughput (PDR x data rate) and packet loss ratio (1 - PDR).
    
    \item \textit{Route Breaks ($R_b$)}:  The number of
    route breaks per data flow during the 2000 s simulation duration. Since a route break leads to new route discovery, a lower $R_b$ value signifies more stable routes and a lower control overhead.

    \item \textit{Route Up Time ($R_{u}$)}:  The percentage of time a target UAV has a valid route to the BS throughout the simulation (2000 s duration). A higher value is desired as it signifies a more stable route to BS with fewer data flow interruptions.

    % \item \textit{Route Up Time ($R_{u}$)}: is the percentage of time a target UAV has a valid route to the BS after communication starts. A higher value is desired as it signifies a more stable route to BS with fewer data flow interruptions.    
    
\end{itemize}

The \emph{coverage} properties of the network are measured by:
\begin{itemize}
    \item \textit{Coverage ($C_v$)}: The average percentage of cells in the map visited by UAVs at least once within a given duration of time; we prefer a higher $C_v$ in a given duration of time.
    % \item \textit{Coverage Time ($Tc$)}: The average time taken to scan 90\% of cells in the map. A lower value of $Tc$ indicates faster area coverage.
    \item \textit{Coverage Fairness ($F$)}: Represents how equally all the cells of the map are visited during a given time period, as measured by Jain’s fairness index \cite{Jains_Index},
    \begin{equation} \label{fairness_eq-TC}
        F=\frac{(\sum_i x_i)^2}{n\sum_i x_i^2}
    \end{equation}
    where $x_i$ is the number of scans of cell $i$, and $n$ is the total number of cells in the map. A higher value of $F$ is desired. 

    \item \textit{Cell Visitation Frequency ($V_f$)}: is the average number of times a grid cell of the area map is visited by UAVs over a given duration of time (2000 s of simulation).
    
\end{itemize}
%
% The \emph{connectivity} properties are measured by:
% \begin{itemize}
%     \item \textit{Number of Connected Components ($NCC$)}: Average number of disjoint components in the UAV network (with no path between them), sampled every 10 s. NCC measures how disconnected the network is;  its optimal (minimal) value is 1.
%     \item \textit{Average Node Degree ($AND$)}: The node degree of a node $u$ ($N_D(u)$) is its number of links (or 1-hop neighbors). $AND$ is the average node degree of all $V$ nodes in the network, computed by sampling the network every 10 s.
%     \begin{equation} \label{AND_eq}
%         AND =\frac{\sum_u N_D(u)}{V}
%     \end{equation}
    
%     % \item \textit{Percentage of Time Connected to BS ($T_{bs}$)}: Average percentage of time a UAV is connected, directly or through a multihop path, to the BS throughout the simulation.  
%     % \item \textit{Giant Component ($G$)}: Average size of the connected subgraph (component) with the largest number of nodes in the network; larger $G$ indicates that fewer UAVs are isolated.
% \end{itemize}

Our proposed pipe routing with topology control scheme (Section \ref{toplogy_control}) is referred to as `TC-Pipe'. Pipe routing without topology control (Section \ref{sect_pipe_routing}) is referred to as `Pipe'. The `AODV' routing scheme does not use a pipe region and topology control. The TC-Pipe, Pipe and AODV routing schemes all follow the BS-CAP mobility model. 
Pipe routing under the ConCov mobility scheme \cite{bsConCov} is referred to as `ConCov-Pipe'.
Finally, the relay-based routing scheme uses the BS-CAP mobility model and is referred to as `Relay' (see Section \ref{Relay_scheme}).
Suffix ``-20'' (e.g., TC-Pipe-20, AODV-20) indicates UAVs at 20 m/s, while ``-40'' indicates UAVs at 40 m/s speeds.

We first discuss the routing and coverage performances of these schemes for different single and 3-targets settings with no node failure in Sections \ref{1T-results} and \ref{3T-results}, respectively. Next, we discuss the effect of 20\% and 30\% node failure on the performance of these schemes in Section \ref{Effects_node_failure}. Lastly, we summarize the results in Section \ref{Results_summary}.

\subsection{Performance for Single Target} \label{1T-results}

We first consider the routing and coverage performance for a single flow from a target (source) to BS. The performance of a routing scheme depends on the number and location of the targets (sources).  
In our simulations, we considered various target locations; however, for the sake of convenience, we detail the results for three particular settings, designated $C_1$, $C_2$ and $C_3$, in which the target is located at [1.5 km, 5 km], [4 km, 4 km] and [2 km, 2 km], respectively (see Figures \ref{fig:caseC1}, \ref{fig:caseC2}, and \ref{fig:caseC3}). Here $C_1$ (target located far from BS) and $C_3$ (target located close to BS) represent the two extremes of the map. We see that the performance of the network for other target locations (such as $C_2$) generally tends to fall within the performance range of the two extreme settings ($C_1$ and $C_3$).

\subsubsection{Routing Performance:}
\label{sect:routing_performance-1T}

\begin{figure*}[htbp]
\centerline{\includegraphics[width=0.9\textwidth]{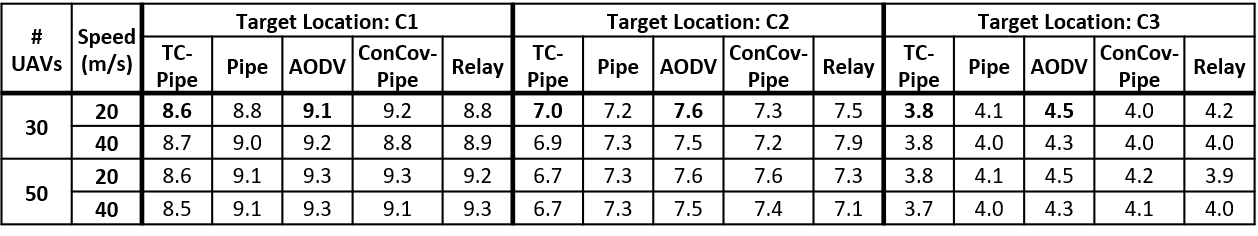}}
\caption[Average route length for single target locations]{Average route length for single target locations ($C_1$, $C_2$, and $C_3$).}
\label{fig:hops}
\end{figure*}

\begin{figure*}[htbp]
\centering
\begin{subfigure}[b]{0.325\textwidth}
\centering
\includegraphics[width=\textwidth]{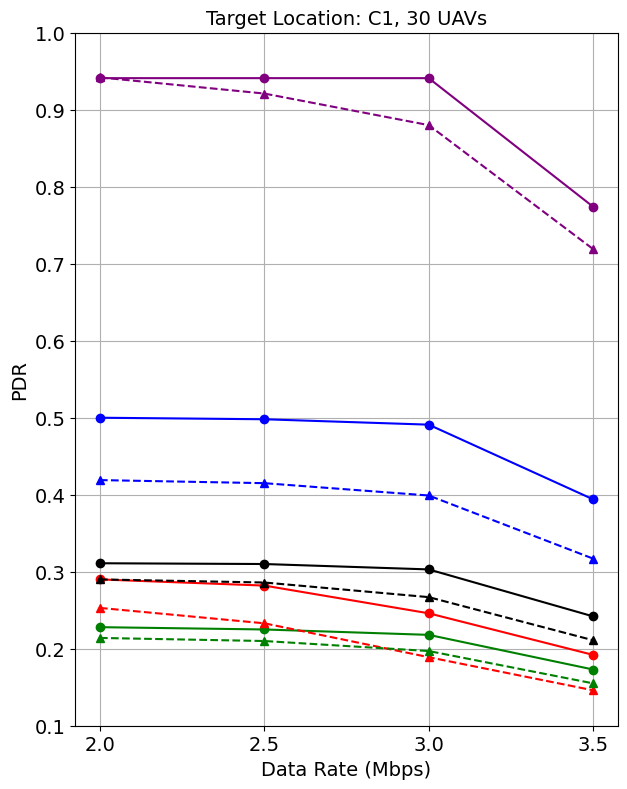} 
\caption{PDR (Target C1, 30 UAVs)}
\label{fig:15-1target-pdr-30n}
\end{subfigure}
\begin{subfigure}[b]{0.325\textwidth}
\centering
\includegraphics[width=\textwidth]{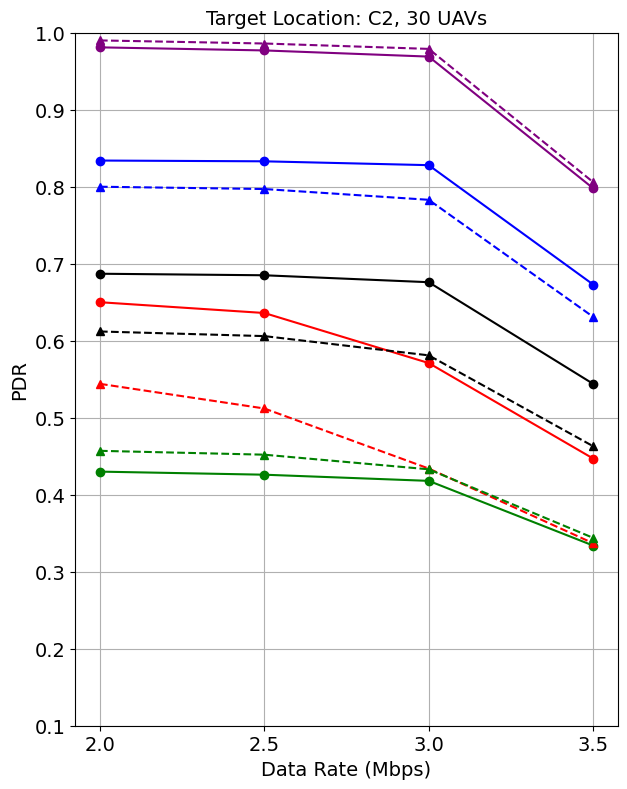} 
\caption{PDR (Target C2, 30 UAVs)}
\label{fig:44-1target-pdr-30n}
\end{subfigure}
\begin{subfigure}[b]{0.325\textwidth}
\centering
\includegraphics[width=\textwidth]{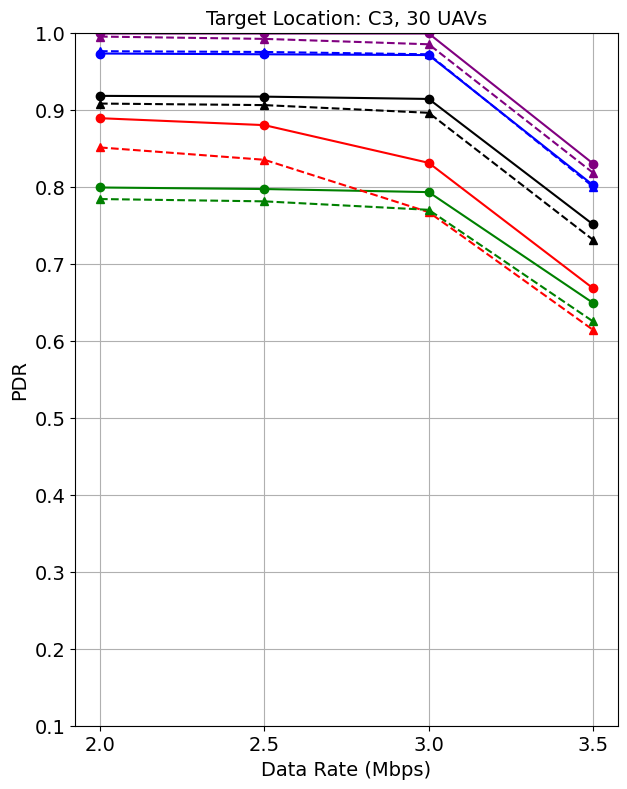} 
\caption{PDR (Target C3, 30 UAVs)}
\label{fig:22-1target-pdr-30n}
\end{subfigure}

\par\medskip % force a bit of vertical whitespace
\begin{subfigure}[b]{0.325\textwidth}
\centering
\includegraphics[width=\textwidth]{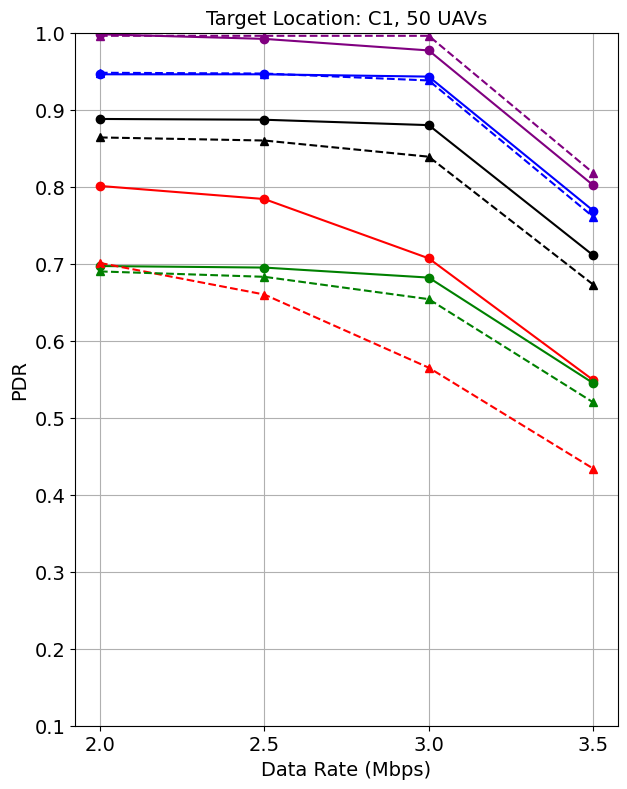}
\caption{PDR (Target C1, 50 UAVs)}
\label{fig:15-1target-pdr-50n}
\end{subfigure}
\begin{subfigure}[b]{0.325\textwidth}
\centering
\includegraphics[width=\textwidth]{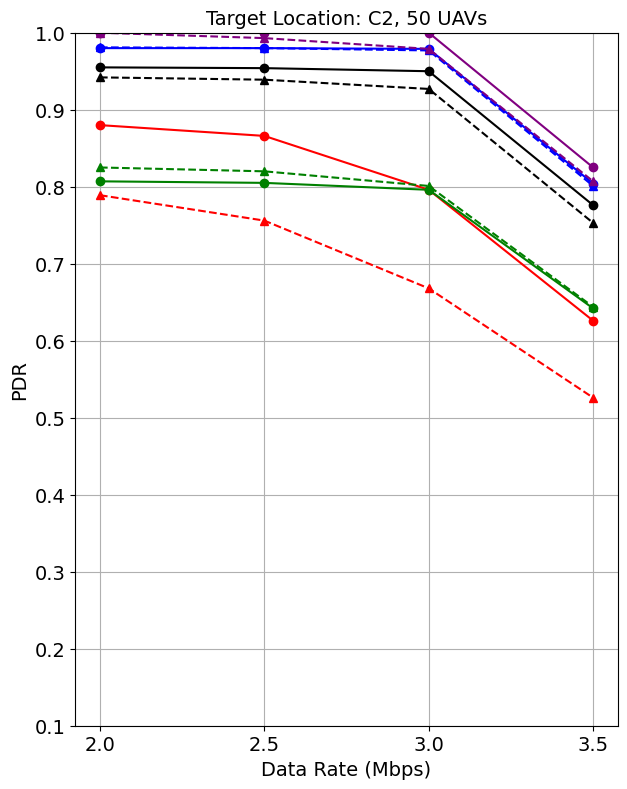}
\caption{PDR (Target C2, 50 UAVs)}
\label{fig:44-1target-pdr-50n}
\end{subfigure}
\hspace{-1.2em}
\raisebox{-.33em}{
\begin{subfigure}[b]{0.325\textwidth}\centering
% \includegraphics[width=\textwidth]{RESULTS/1target/22-PDR-50UAVs-drop0.png}
% Position the legend image on top of the plot image
\begin{tikzpicture}
\node at (0,0) {\includegraphics[width=\textwidth]{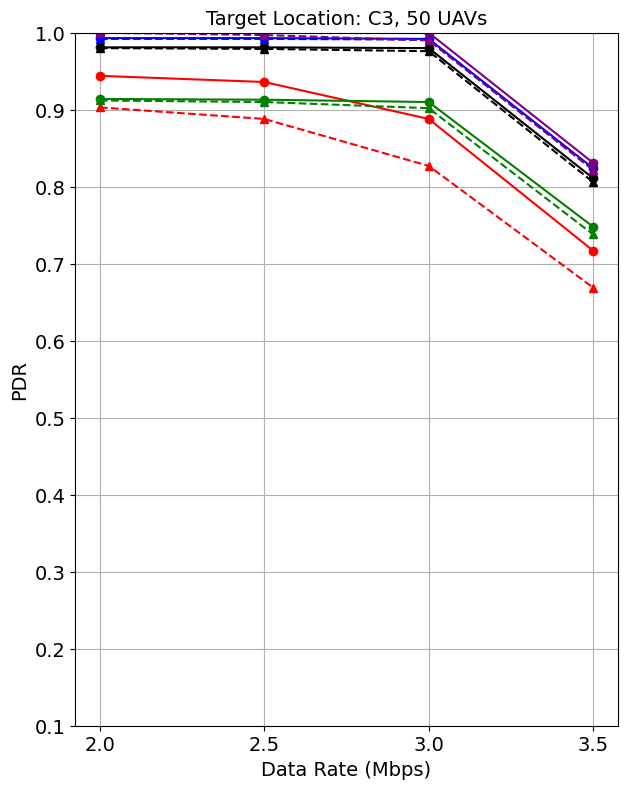}};
\node at (0.4,-2) {\includegraphics[width=0.74\textwidth]{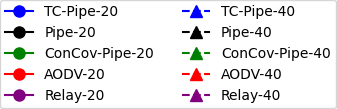}};
\end{tikzpicture}
\caption{PDR (Target C3, 50 UAVs)}
\label{fig:22-1target-pdr-50n}
\end{subfigure}}
\caption[PDR for single target settings]{Routing Performance: PDR for single target settings $C_1$, $C_2$ and $C_3$. Suffix“-20” (e.g., TC-Pipe-20, AODV-20) indicate UAVs at 20
m/s, while “-40” indicates UAVs at 40 m/s speeds.}
\label{fig:1target-PDR}
\end{figure*}

\begin{figure*}[htbp]
\centering
\begin{subfigure}{0.8\textwidth}
\includegraphics[width=\textwidth]{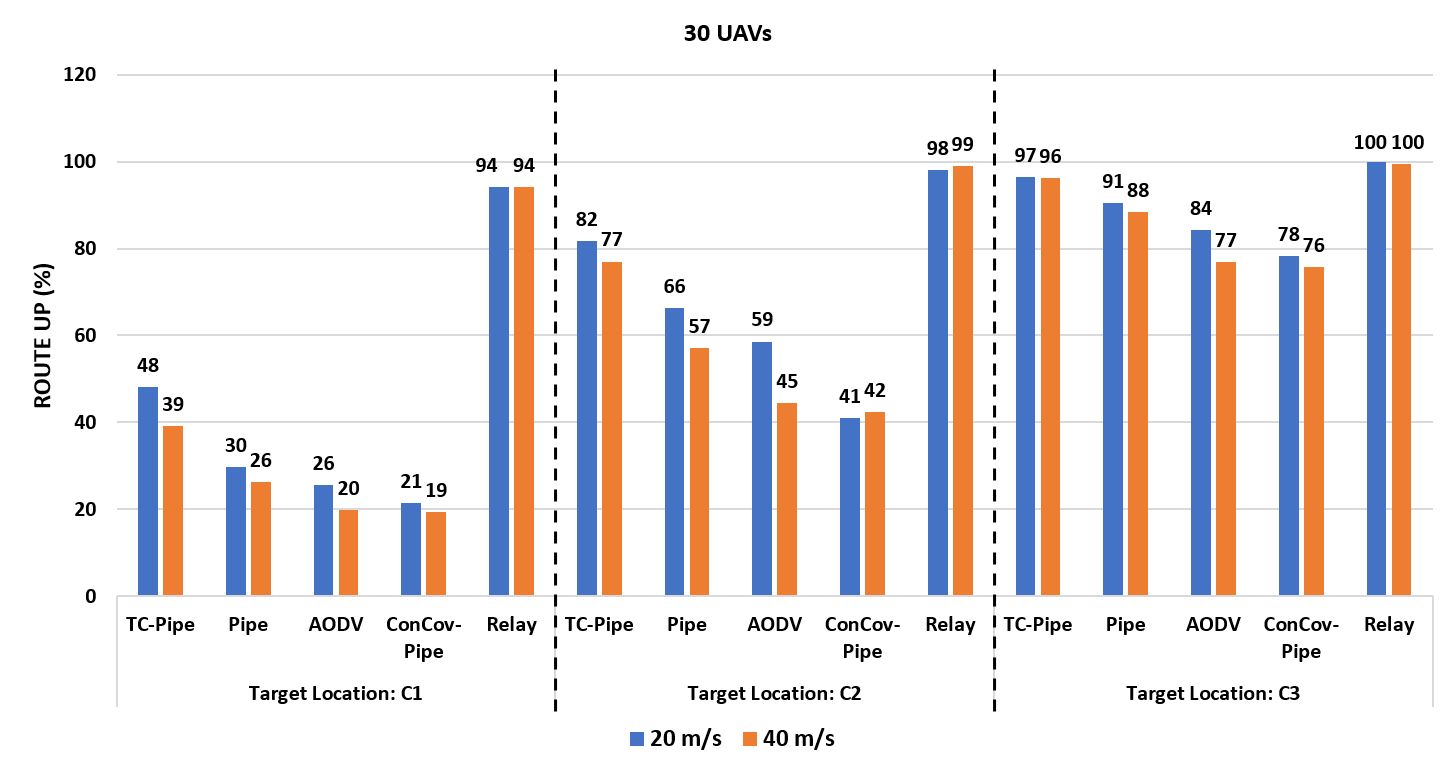}
\caption{ Route Up (30 UAVs)}
\label{fig:1target-routeUP-30n}
\end{subfigure}

\par\medskip % force a bit of vertical whitespace

\begin{subfigure}{0.8\textwidth}
\includegraphics[width=\textwidth]{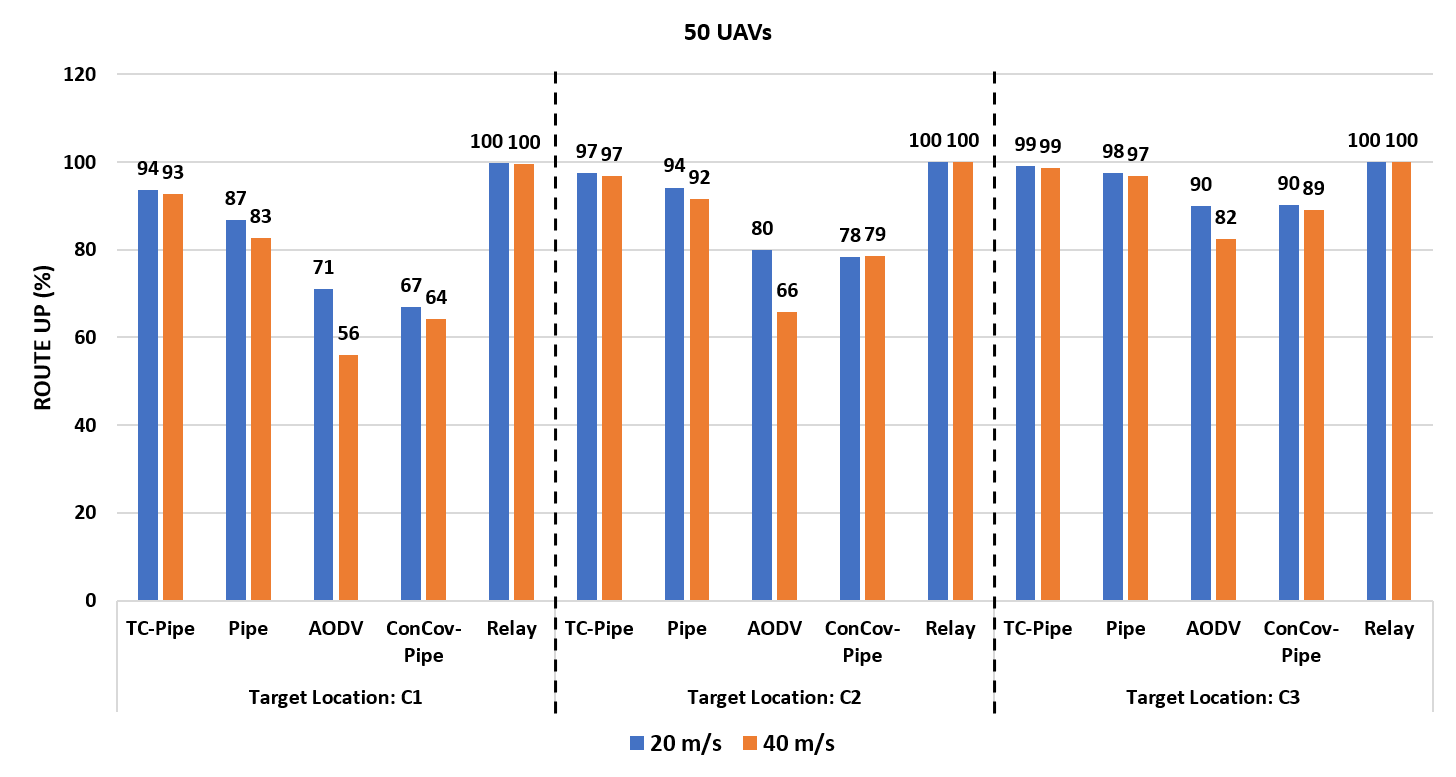}
\caption{Route Up (50 UAVs)}
\label{fig:1target-routeUP-50n}
\end{subfigure}
\caption[Route Up for single target settings]{Routing Performance: Route Up for single target settings $C_1$, $C_2$ and $C_3$.}
\label{fig:1target-RouteUP}
\end{figure*}

\begin{figure*}[htbp]
\centering
\begin{subfigure}{0.9\textwidth}
\includegraphics[width=\textwidth]{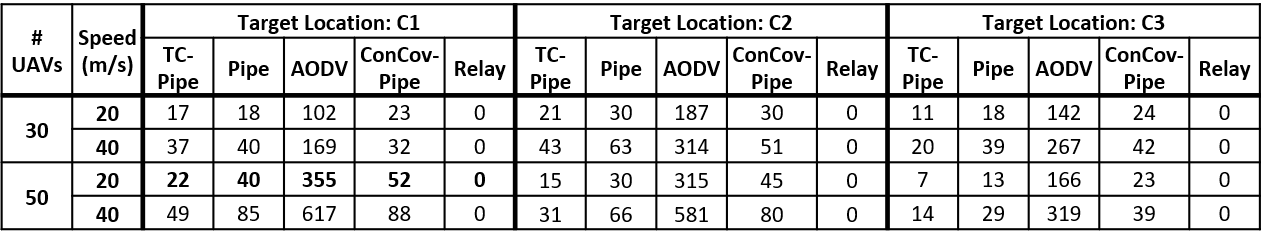}
% \caption{Route Breaks}
\end{subfigure}
\caption[Route Breaks for single target settings]{Routing Performance: Route Breaks for single target settings $C_1$, $C_2$ and $C_3$.}
\label{fig:1target-RouteBreaks}
\end{figure*}

Figures \ref{fig:hops}-\ref{fig:1target-RouteBreaks} show the routing performance (measured by average route length, PDR, route up time and number of route breaks) of different schemes (TC-Pipe, Pipe, ConCov-Pipe, AODV, and Relay) for a single data flow from a target to BS for three different target locations, represented by $C_1$, $C_2$ and $C_3$. Here, the UAV network consists of two different node densities (30 and 50 UAVs) and two node speeds (20 and 40 m/s).

\textbf{Average Route Length:} Figure \ref{fig:hops} shows the average route length achieved by all the routing schemes for both densities at both speeds. 
We observe that the route length increases with distance between the target and BS nodes, but it does not vary noticeably when the node density and speed are changed.
A shorter route length typically provides better data throughput and smaller delays. For $C_1$ where the target is far from the BS, our TC-Pipe scheme achieves the average route length of $\approx$8.6, while AODV achieves $\approx$9.2. On the other hand, for $C_3$ where the target is relatively close to the BS, our TC-Pipe scheme achieves the average route length of $\approx$3.8, while AODV achieves $\approx$4.4. %We observe similar average route lengths at node speeds of 40 m/s. 
Overall, the TC-Pipe scheme achieves the shortest average route length, while AODV has the longest route length, with other routing schemes (Pipe, Relay and ConCov-Pipe) providing average route lengths in between the two. This indicates the ability of topology control with pipe routing to establish stable, shorter routes using the pipe region. 

\textbf{Average PDR:} Figure \ref{fig:1target-PDR} shows the average PDR achieved by all the routing schemes for both node densities at both speeds. The average PDR achieved by a routing scheme increases as the distance between the target and BS decreases; this is because longer routes are more likely to break due to node mobility. Further, the average PDR increases for a higher density because route discovery and maintenance of pipe region is more efficient. Also, the average PDR decreases with node speed because more route breaks occur.
For different network settings and target locations ($C_1$, $C_2$ and $C_3$), our proposed TC-Pipe scheme achieves higher average PDR than the Pipe, AODV, and ConCov-Pipe, except the Relay scheme, as discussed below (see Figure \ref{fig:1target-PDR}). 

For setting $C_1$, the PDR at different data rates for both UAV densities and speeds are shown in Figures \ref{fig:15-1target-pdr-30n} and \ref{fig:15-1target-pdr-50n}. For longer routes in Figure \ref{fig:15-1target-pdr-30n}, the TC-Pipe scheme provides up to 63\%, 105\% and 129\% increase in PDR compared to Pipe, AODV and ConCov-Pipe schemes, respectively, for 30 UAVs at 20 m/s. 
%Similarly, for 50 UAVs at 20 m/s, the TC-Pipe scheme achieves up to 8\%, 40\% and 40\% increase in PDR compared to Pipe, AODV and ConCov-Pipe schemes, respectively.
For higher UAV speed of 40 m/s, the TC-Pipe scheme achieves up to 52\%, 113\% and 100\% increase in PDR compared to Pipe, AODV and ConCov-Pipe schemes, respectively. We observed similar trends for 50 nodes in Figure \ref{fig:15-1target-pdr-50n}. 
%Similarly, for 50 UAVs at 40 m/s, the TC-Pipe scheme achieves up to 13\%, 77\% and 46\% increase in PDR compared to Pipe, AODV and ConCov-Pipe schemes, respectively. 
Even at higher UAV speeds where route breaks are more frequent, the TC-Pipe scheme achieves significant improvement in PDR compared to the Pipe and other schemes because the use of topology control in the pipe region makes the route more robust. 
For settings $C_2$ and $C_3$ too, TC-Pipe scheme provides higher PDR than Pipe, AODV and ConCov-Pipe, but the relative PDR improvement is reduced with the distance of the target from BS. 

Further, the Pipe routing scheme performs better than AODV for both densities at both speeds for all three target locations because the use of pipe region facilitates proactive route switching instead of route rediscovery when the route quality degrades. Also, the Pipe routing scheme (which uses BS-CAP mobility model with pipe routing) performs better than ConCov-Pipe (which uses ConCov mobility model \cite{bsConCov} with pipe routing); this demonstrates the advantage of using the BS-CAP mobility model which provides a better node degree and BS-connectivity than the ConCov model \cite{BS-CAP_ref}. 
The Relay scheme achieves higher PDR performance than all other schemes, as it uses dedicated relay UAVs to establish the active route. However, as a trade-off, its coverage performance is decreased, as discussed in Section \ref{sect:cover_performance-1T}. 

% abc
% def

%Greater route stability is achieved with higher $R_u$ and lower $R_b$.
\textbf{Route Up ($R_u$):} A Higher $R_u$ value indicates that the route from target UAV to BS exists for a longer duration of time, which generally results in a more stable route providing a higher PDR. Figure \ref{fig:1target-RouteUP} shows the $R_u$ value of all the considered schemes for both node densities at both speeds.
The value of $R_u$ increases for a higher node density but decreases for a higher node speed. 
Here, the Relay scheme achieves the highest $R_u$ (nearly 100\%) because a stable route is established. The TC-Pipe scheme achieves the highest $R_u$ value after the Relay scheme for both node densities at both speeds.  In Figure \ref{fig:1target-routeUP-50n}, TC-Pipe achieves $R_u$ value of 94\% and 93\%, for the target location $C_1$ at a node density of 50 and speeds of 20 m/s and 40 m/s, respectively, compared to 87\% and 83\% achieved by the Pipe routing scheme.
The TC-Pipe scheme's improved performance over the Pipe scheme is due to its ability to maintain a robust pipe along the active route.
AODV (which uses the BS-CAP mobility model without pipe region) and ConCov-Pipe (which uses the ConCov \cite{bsConCov} mobility model with a lower BS connectivity performance compared to the BS-CAP mobility model \cite{BS-CAP_ref}) provide lower $R_u$ values.
This shows that Pipe routing combined with the BS-CAP mobility model provides better routing performance (as in TC-Pipe, Pipe). 
A similar performance trend is observed at 30 UAV density in Figure \ref{fig:1target-routeUP-30n}; however, the $R_u$ performances decreases considerably, especially for longer route lengths in target setting $C_1$, due to a lower node density.

\textbf{Route Breaks ($R_b)$:} Figure \ref{fig:1target-RouteBreaks} shows the number of route breaks achieved by all the routing schemes for both node densities at both speeds.
Proactive route switching using the alternate route in the pipe reduces data flow interruptions and delays in pipe routing based schemes (i.e., TC-Pipe, Pipe and ConCov-Pipe). Due to better pipe robustness, the TC-Pipe scheme achieves the least route breaks ($R_b$) among the pipe-based schemes for both densities at both speeds and for all three target locations, reducing new route discoveries and associated control overhead. 
The pipe routing based schemes have far less route breaks than the AODV scheme for both densities at both speeds. 
For example, in target setting $C_1$ for 50 UAVs, TC-Pipe achieves lower $R_b$ value of 22 at 20 m/s compared to 40 for the Pipe scheme. The ConCov-Pipe and AODV schemes yield higher $R_b$ of 52 and 355 at 20 m/s, respectively. 
At UAV speed of 40 m/s, more route breaks are observed due to dynamic network topology. 
We find a similar performance trend across all three target locations ($C_1$, $C_2$ and $C_3$) at both UAV densities at both speeds.
Since the Relay scheme establishes stable routes, it experiences zero route breaks.

\subsubsection{Coverage Performance:}
\label{sect:cover_performance-1T}

\begin{figure*}[htbp]
\centering
\begin{subfigure}[b]{0.325\textwidth}
\centering
\includegraphics[width=\textwidth]{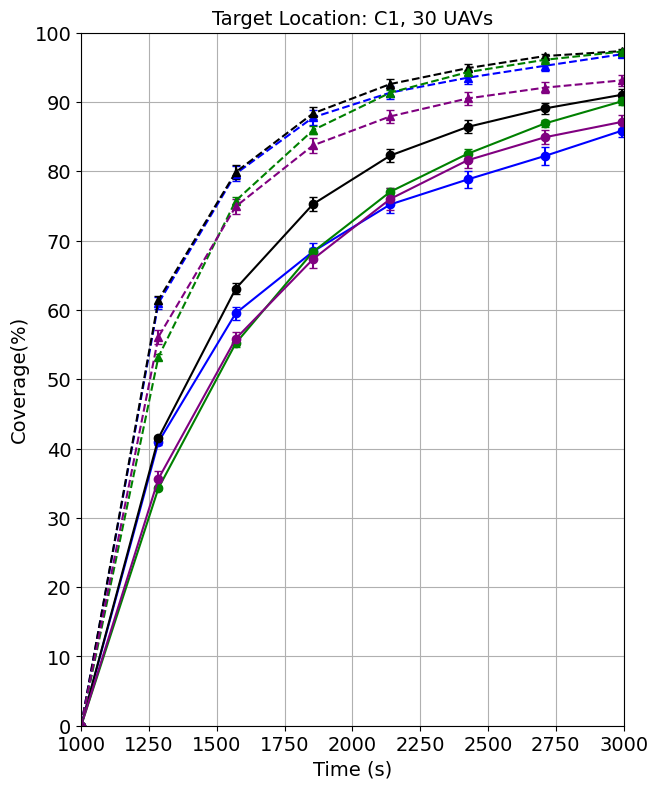} 
\caption{Coverage (Target C1, 30 UAVs)}
\label{fig:1target-cov-30n-CASE15}
\end{subfigure}
\begin{subfigure}[b]{0.325\textwidth}
\centering
\includegraphics[width=\textwidth]{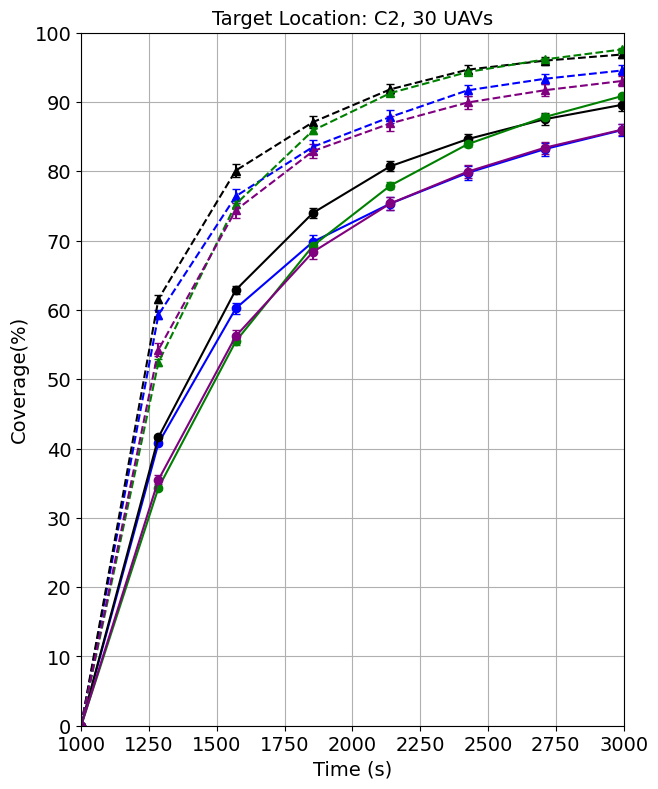} 
\caption{Coverage (Target C2, 30 UAVs)}
\label{fig:1target-cov-30n-CASE44}
\end{subfigure}
\begin{subfigure}[b]{0.325\textwidth}
\centering
\includegraphics[width=\textwidth]{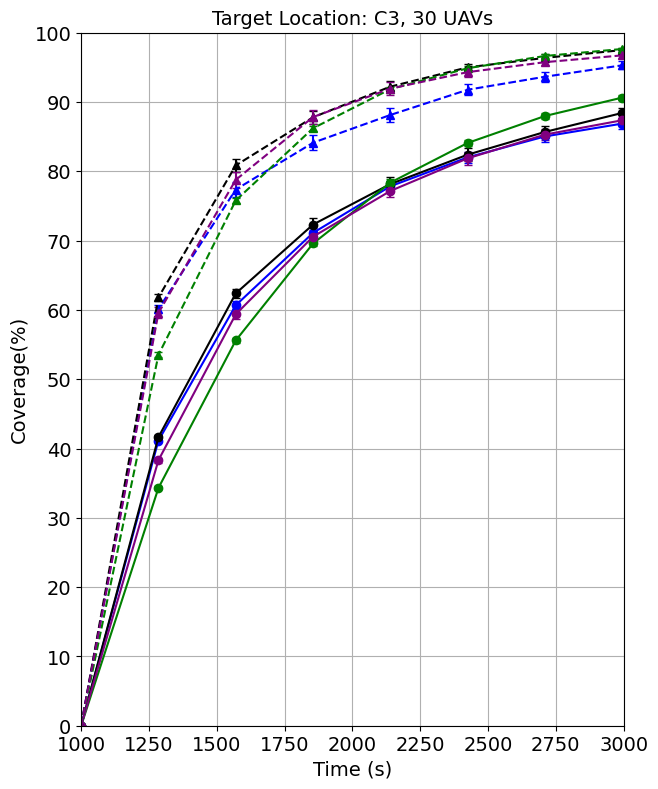} 
\caption{Coverage (Target C3, 30 UAVs)}
\label{fig:1target-cov-30n-CASE22}
\end{subfigure}

\par\medskip % force a bit of vertical whitespace
\begin{subfigure}[b]{0.325\textwidth}
\centering
\includegraphics[width=\textwidth]{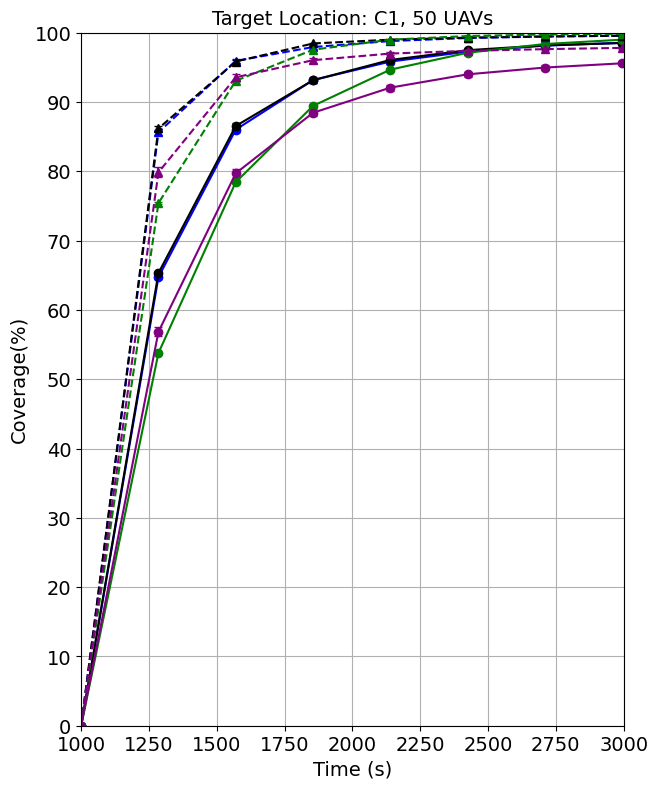}
\caption{Coverage (Target C1, 50 UAVs)}
\label{fig:1target-cov-50n-CASE15}
\end{subfigure}
\begin{subfigure}[b]{0.325\textwidth}
\centering
\includegraphics[width=\textwidth]{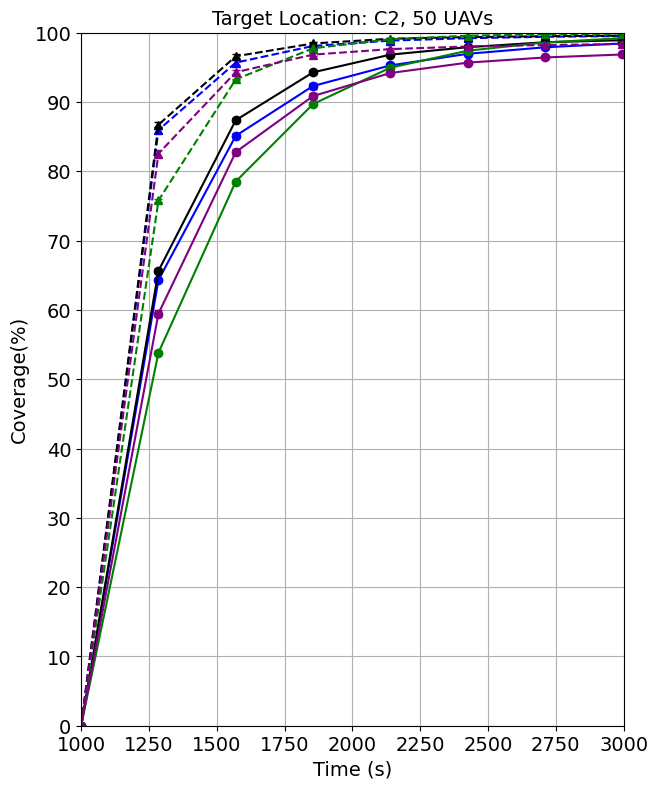}
\caption{Coverage (Target C2, 50 UAVs)}
\label{fig:1target-cov-50n-CASE44}
\end{subfigure}
\hspace{-1em}\raisebox{-.33em}{
\begin{subfigure}[b]{0.325\textwidth}
\centering
% \includegraphics[width=\textwidth]{RESULTS/1target/22-Cov1000_sem-50UAVs-drop0.png}
% Position the legend image on top of the plot image
\begin{tikzpicture}
\node at (0,0) {\includegraphics[width=\textwidth]{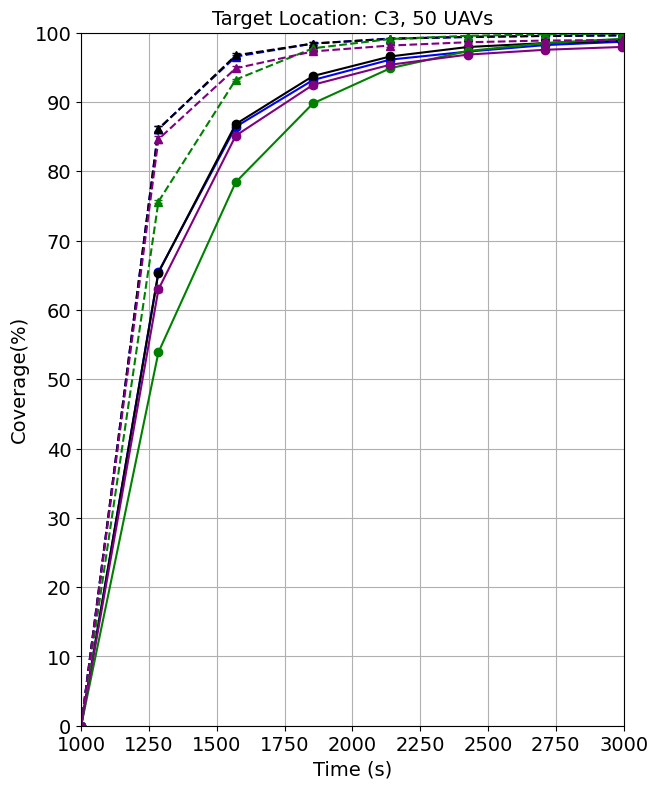}};
\node at (0.4,-2) {\includegraphics[width=0.74\textwidth]{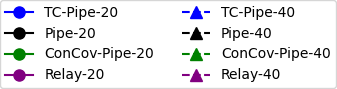}};
\end{tikzpicture}
\caption{Coverage (Target C3, 50 UAVs)}
\label{fig:1target-cov-50n-CASE22}
\end{subfigure}}

\caption[Coverage vs. Time plots for single target settings]{Coverage vs. Time plots for single target settings $C_1$, $C_2$ and $C_3$.}
\label{fig:1target-coveragevstime}
\end{figure*}

\begin{figure*}[htbp]
\centering
\begin{subfigure}{0.9\textwidth}
\includegraphics[width=\textwidth]{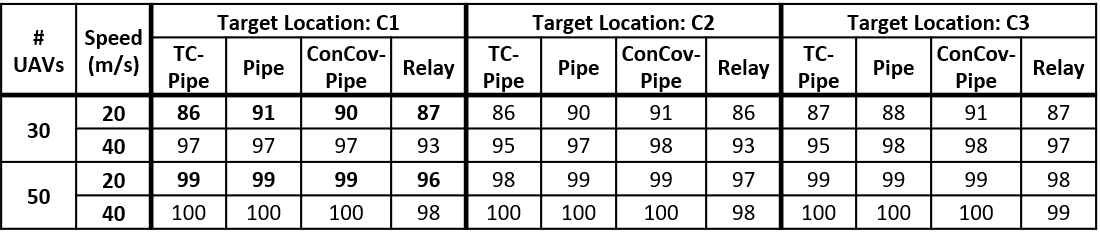}
% \caption{Coverage (\%) from 1000s to 3000s ($C_v$)}
% \label{fig:1target-COV3000s}
\end{subfigure}
\caption[Coverage $C_v$ for single target settings]{Coverage Performance: $C_v$ for target single settings $C_1$, $C_2$ and $C_3$.}
% \label{fig:1target-TotalCoverage}
\label{fig:1target-COV3000s}
\end{figure*}

\begin{figure*}[htbp]
\centering
\begin{subfigure}{0.8\textwidth}
\includegraphics[width=\textwidth]{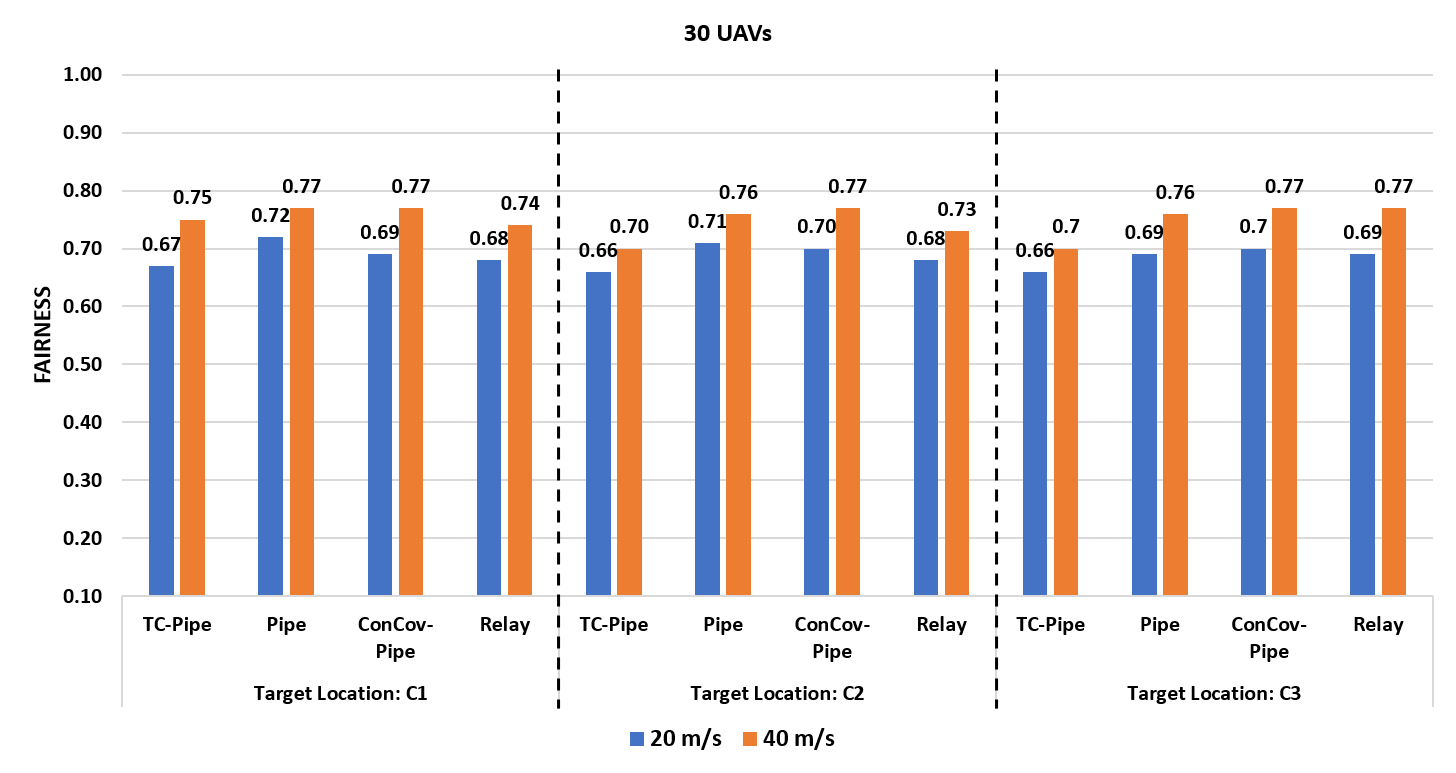}
\caption{ Fairness (30 UAVs)}
\label{fig:1target-Fairness-30n}
\end{subfigure}

\par\medskip % force a bit of vertical whitespace

\begin{subfigure}{0.8\textwidth}
\includegraphics[width=\textwidth]{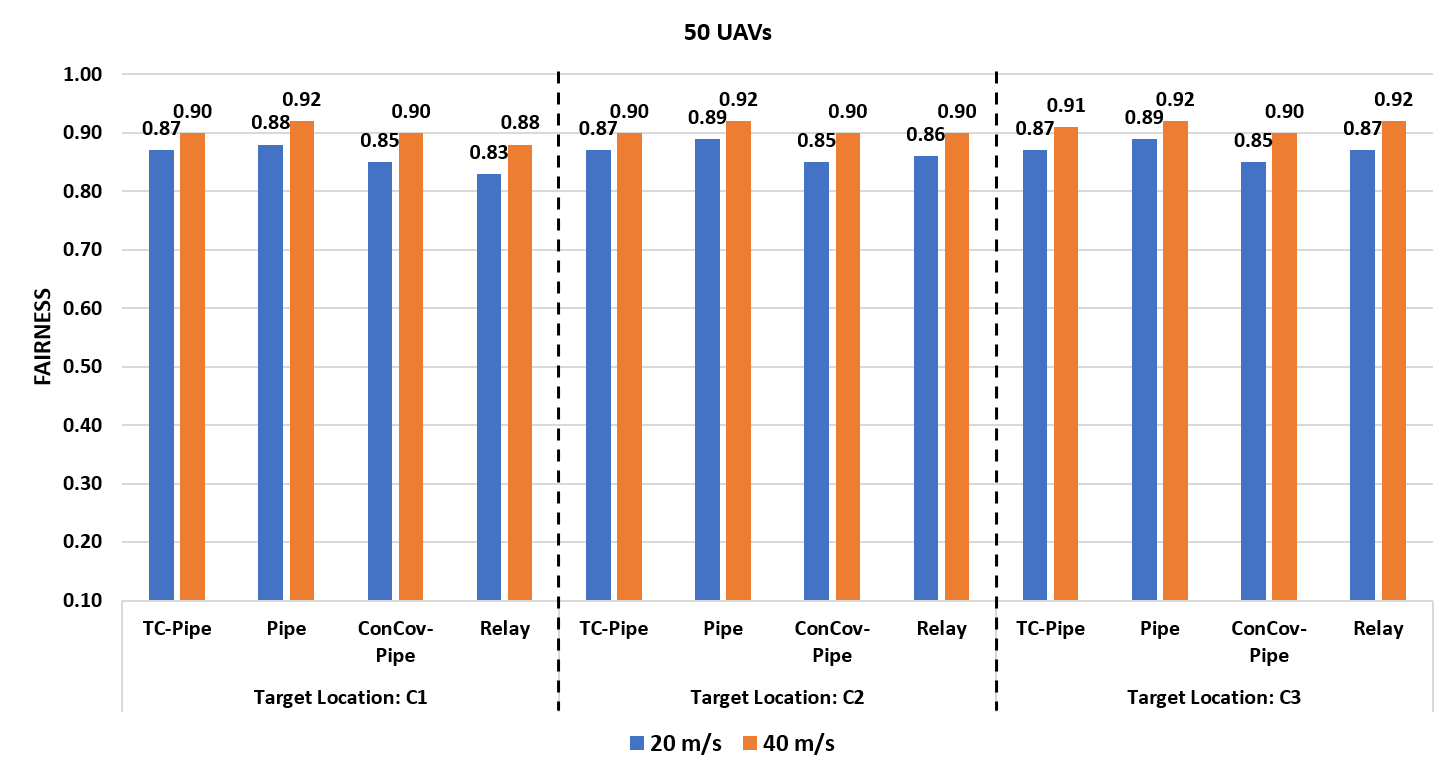}
\caption{Fairness (50 UAVs)}
\label{fig:1target-Fairness-50n}
\end{subfigure}
\caption[Coverage Fairness for single target settings]{Coverage Performance: Fairness for single target settings $C_1$, $C_2$ and $C_3$.}
\label{fig:1target-Fairness}
\end{figure*}

Figures \ref{fig:1target-coveragevstime} - \ref{fig:1target-Fairness} show the coverage performance (measured by coverage vs. time, total coverage, and fairness) of different schemes (TC-Pipe, Pipe, ConCov-Pipe, and Relay) for a single data flow from a target to BS for three different target locations, represented by $C_1$, $C_2$ and $C_3$. Here, the UAV network consists of two different node densities (30 and 50 UAVs) and two node speeds (20 and 40 m/s). The performance of AODV scheme is not shown because it uses the BS-CAP mobility model with coverage performance identical to Pipe scheme. 

We select the BS-CAP mobility model with $\beta=1.5$ (for TC-Pipe, Pipe, and Relay) and ConCov mobility with $\omega=0.5$ (for ConCov-Pipe). Among the three tuning parameter values of $\beta$ and $\omega$ experimented in \cite{BS-CAP_ref}, the aforementioned middle values give a balanced coverage and connectivity trade-off performance.
Also, these values give relatively similar coverage profiles (coverage vs. time curves) for the two mobility models, allowing us to compare differences in their routing performances.

The \textbf{Coverage vs. Time} plots in Figure \ref{fig:1target-coveragevstime} represent the total map area covered in a given time for both node densities at both speeds. The rate of area covered increases with node density as well as speed. 
The Relay scheme achieves a lower coverage performance compared to the Pipe and ConCov-Pipe schemes, as it reassigns some of its UAVs as relay nodes, which do not actively cover the map. This impact is more apparent at low UAV density (30 UAVs) and multiple targets (see Section \ref{sect:cover_performance-3T}). 
Similarly, at low UAV density, the TC-Pipe scheme achieves a lower coverage performance because some of the UAVs are attracted to the pipe region. 
For setting $C_1$ in Figure \ref{fig:1target-COV3000s} at 30 UAVs and 20 m/s, the TC-Pipe and Relay schemes achieve $C_v$ of $\approx$ 86\%, compared to $\approx$ 90\% achieved by Pipe and ConCov schemes. Note that TC-Pipe and Relay coverage performance is comparable to Pipe and ConCov for higher node density and/or speed. For example, for 50 UAVs at 20 m/s, TC-Pipe, Pipe and ConCov achieve $C_v$ of 99\%, compared to 96\% by Relay scheme. %A similar performance trend is observed for node speed of 40 m/s and for $C_2$ and $C_3$ settings.

The \textbf{Fairness ($F$)} performance of all the schemes for both node densities at both speeds is shown in Figure \ref{fig:1target-Fairness}. Each scheme achieves a higher $F$ when the density and/or node speed increase as all the cells in the area under monitoring are visited more frequently.   
The TC-Pipe has slightly lower value of $F$ than the Pipe and ConCov-Pipe schemes for 30 UAVs at both speeds for all three target locations, as some UAVs are attracted to maintain a stable pipe region. Since enough UAVs are available at node density of 50, the fairness performance of the TC-Pipe scheme is comparable to Pipe and ConCov-Pipe at both speeds.
For example, in Figure \ref{fig:1target-Fairness-30n}, for 30 UAVs in setting $C_1$, TC-Pipe achieves an $F$ of 0.67 and 0.75 at 20 m/s and 40 m/s, respectively, compared to $F$ value around 0.71 and 0.77 (resp.) for Pipe and ConCov.
In Figure \ref{fig:1target-Fairness-50n}, for 50 UAVs in setting $C_1$, TC-Pipe achieves an $F$ of 0.87 and 0.90 at 20 m/s and 40 m/s, respectively. Meanwhile, Pipe and ConCov achieve $F$ values of 0.88 and 0.85 at 20 m/s, and 0.92 and 0.90 at 40 m/s, respectively.
The Relay scheme achieves $F$ values of 0.83 and 0.88 for 50 UAVs at 20 m/s and 40 m/s, respectively. We further examine Relay scheme's coverage and fairness performance for three targets in Section \ref{sect:cover_performance-3T}.

\subsection{Performance for Three Targets} \label{3T-results}

In this section, we consider the routing and coverage performance for multiple flows from three targets to the BS.    
In our simulations, we considered various target locations; however, for the sake of convenience, we detail the results for three particular settings, designated $C_4$, $C_5$ and $C_6$, in which the targets are located at ([1 km, 4.5 km], [3 km, 5 km], [5 km, 4.5 km]), ([1 km, 4 km], [3.6 km, 2.5 km], [5 km, 4.8 km]) and ([1 km, 1.5 km], [3 km, 2 km], [5 km, 1.5 km]), respectively (see Figures \ref{fig:caseC4}, \ref{fig:caseC5} and \ref{fig:caseC6}).
Here $C_4$ (targets are located far from BS) and  $C_6$ (targets located close to BS) represent two extremes of the map. We see that the performance of the network for other target locations (such as $C_5$) generally tends to fall within the performance range of the two extreme settings ($C_4$ and $C_6$).

\subsubsection{Routing Performance:}
\label{sect:routing_performance-3T}

\begin{figure*}[htbp]
\centerline{\includegraphics[width=0.9\textwidth]{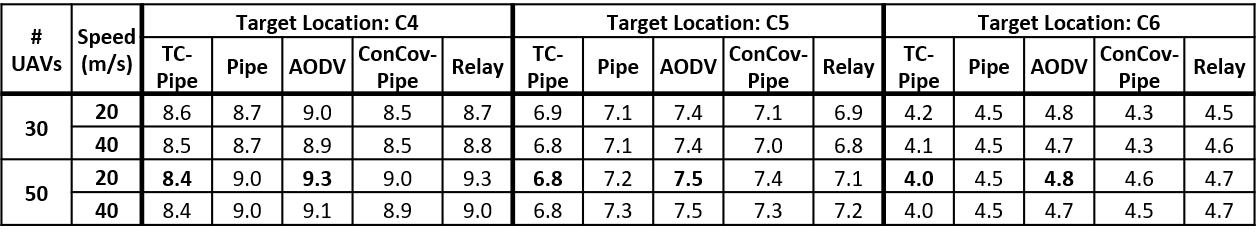}}
\caption[Average route length for 3-target locations]{Average route length for 3-target locations.}
\label{fig:hops-3T}
\end{figure*}

\begin{figure*}[htbp]
\centering
\begin{subfigure}[b]{0.325\textwidth}
\centering
\includegraphics[width=\textwidth]{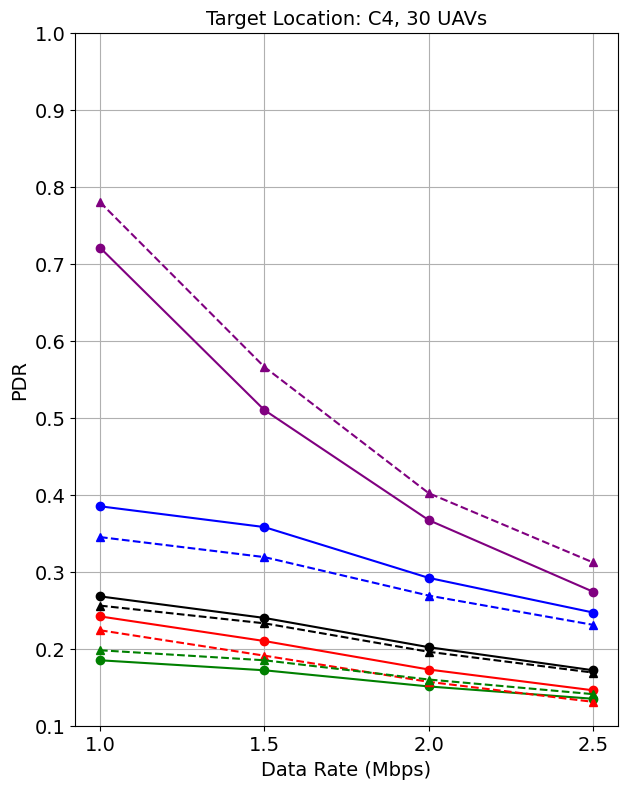} 
\caption{PDR (Targets C4, 30 UAVs)}
\label{fig:6-3targets-pdr-30n}
\end{subfigure}
\begin{subfigure}[b]{0.325\textwidth}
\centering
\includegraphics[width=\textwidth]{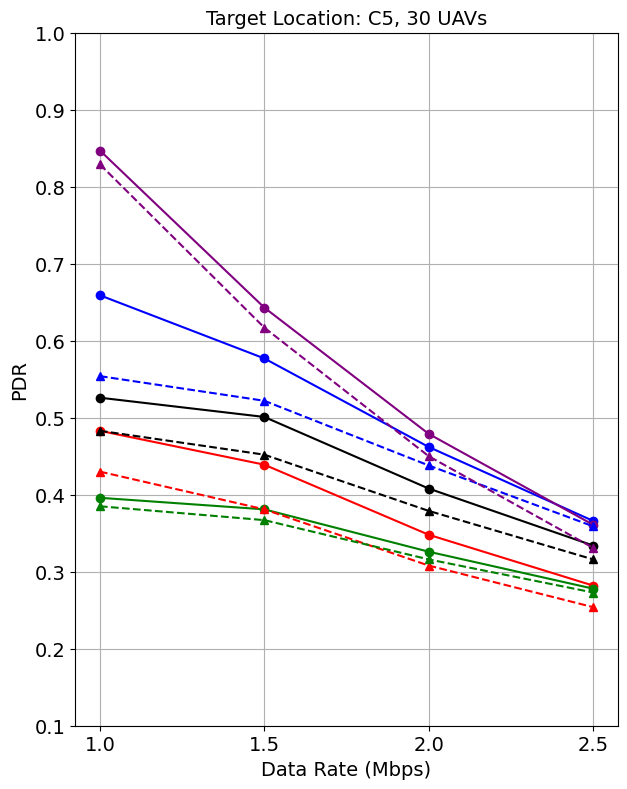} 
\caption{PDR (Targets C5, 30 UAVs)}
\label{fig:5-3targets-pdr-30n}
\end{subfigure}
\begin{subfigure}[b]{0.325\textwidth}
\centering
\includegraphics[width=\textwidth]{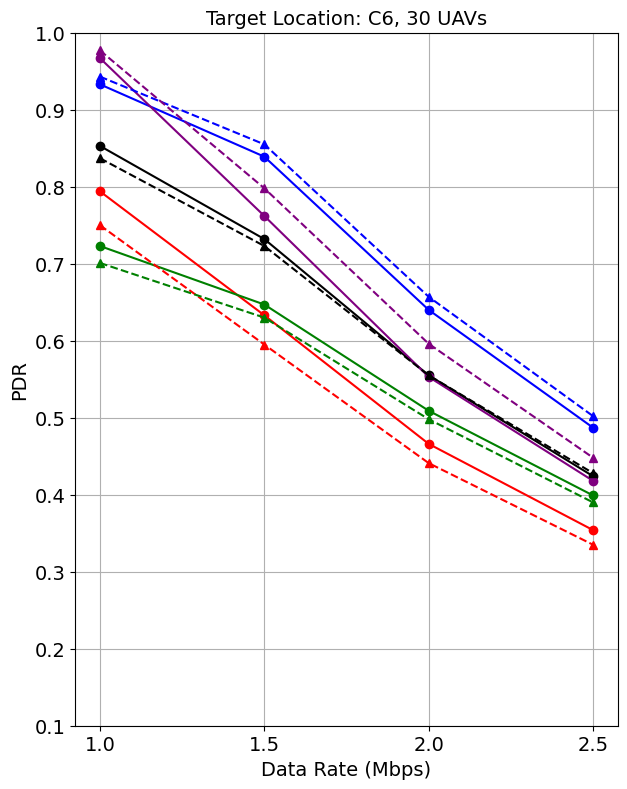} 
\caption{PDR (Targets C6, 30 UAVs)}
\label{fig:7-3targets-pdr-30n}
\end{subfigure}

\par\medskip % force a bit of vertical whitespace
\begin{subfigure}[b]{0.325\textwidth}
\centering
\includegraphics[width=\textwidth]{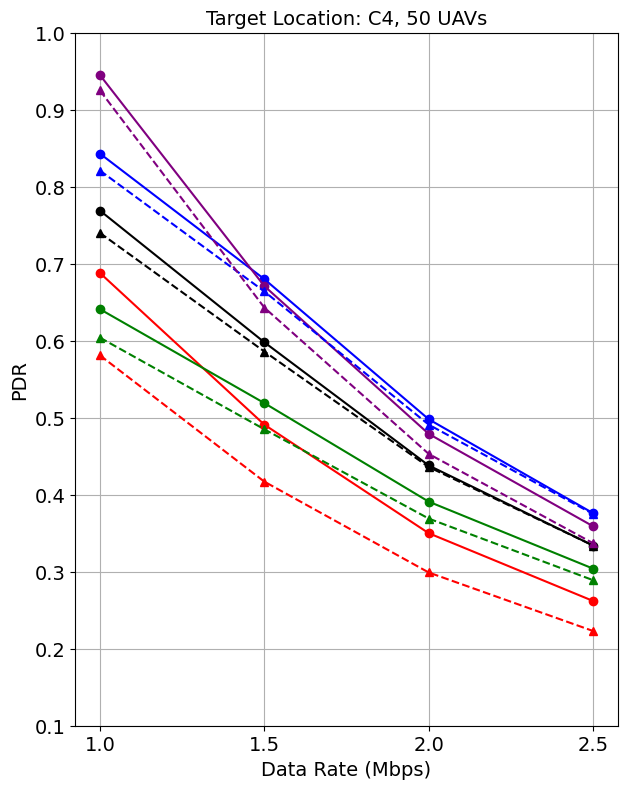}
\caption{PDR (Targets C4, 50 UAVs)}
\label{fig:6-3targets-pdr-50n}
\end{subfigure}
\begin{subfigure}[b]{0.325\textwidth}
\centering
\includegraphics[width=\textwidth]{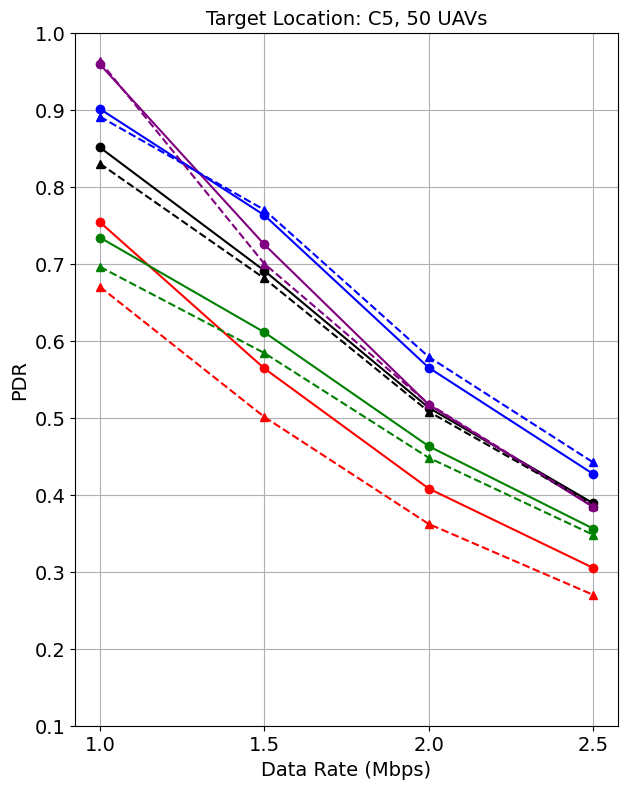}
\caption{PDR (Targets C5, 50 UAVs)}
\label{fig:5-3targets-pdr-50n}
\end{subfigure}
\hspace{-1.2em} \raisebox{-.33em}{
\begin{subfigure}[b]{0.325\textwidth}
\centering
% \includegraphics[width=\textwidth]{RESULTS/3targets/7-PDR-50UAVs-drop0.png}
% Position the legend image on top of the plot image
\begin{tikzpicture}
\node at (0,0) {\includegraphics[width=\textwidth]{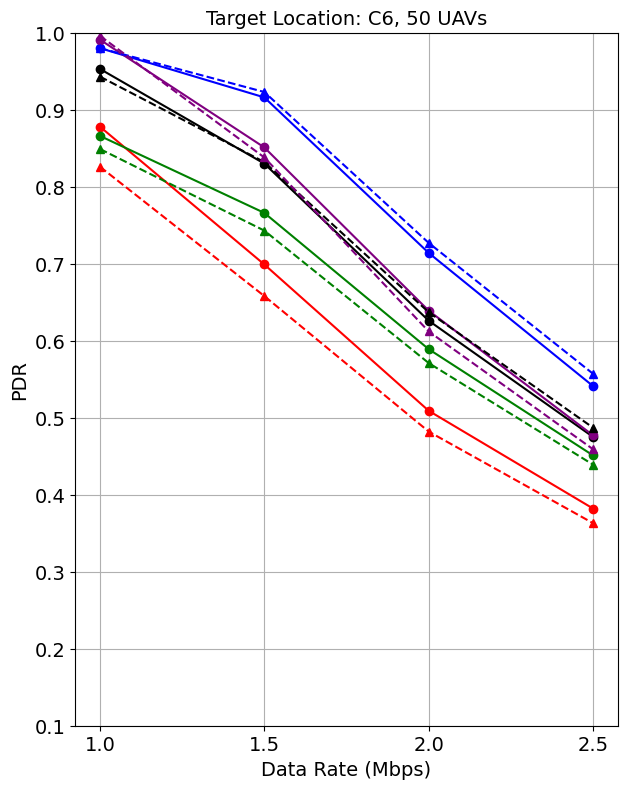}};
\node at (0.4,-2) {\includegraphics[width=0.74\textwidth]{RESULTS/pdr-legend.png}};
\end{tikzpicture}

\caption{PDR (Targets C6, 50 UAVs)}
\label{fig:7-3targets-pdr-50n}
\end{subfigure}}

\caption[PDR for 3-targets settings]{Routing Performance: PDR for 3-targets settings $C_4$, $C_5$ and $C_6$.}
\label{fig:3targets-PDR}
\end{figure*}

\begin{figure*}[htbp]
\centering
\begin{subfigure}{0.8\textwidth}
\includegraphics[width=\textwidth]{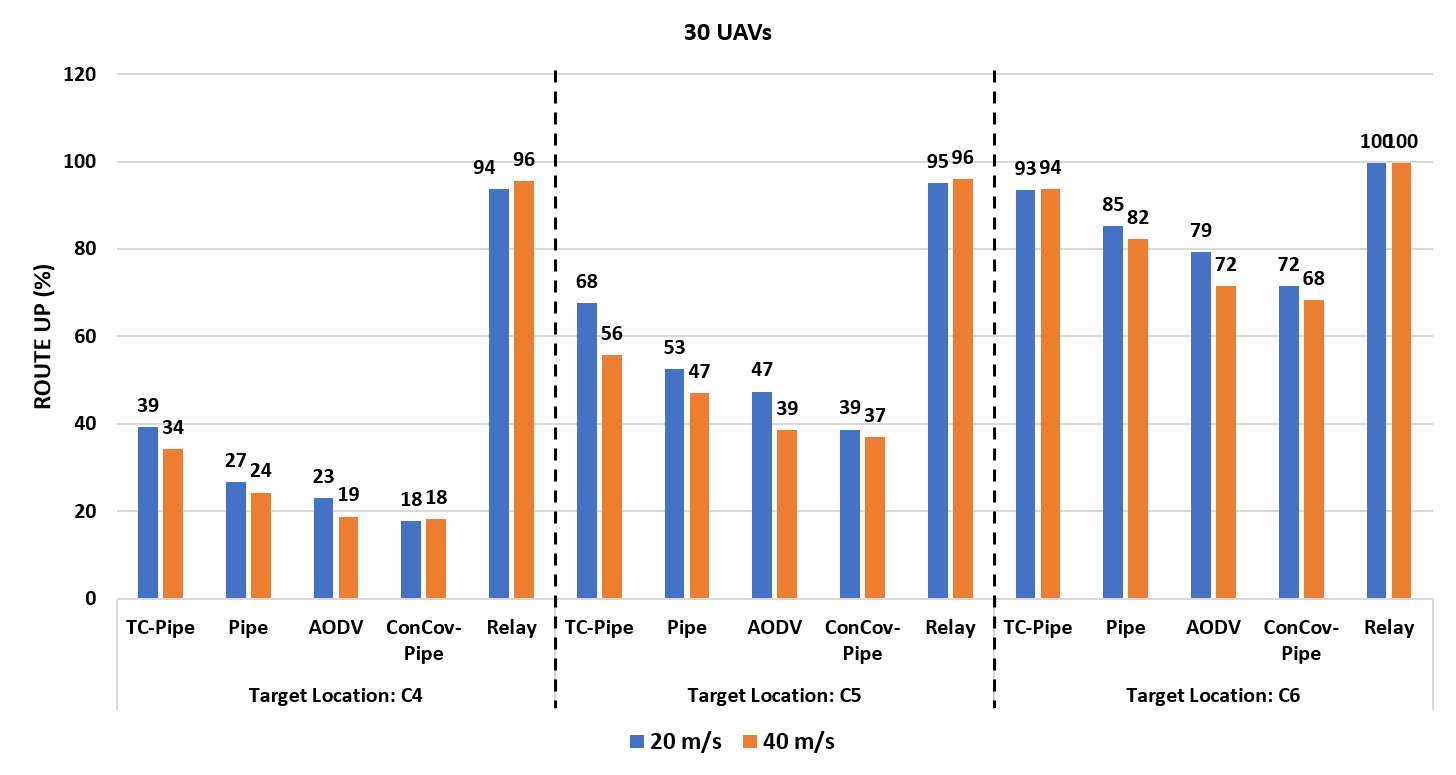}
\caption{ Route Up (30 UAVs)}
\label{fig:3targets-routeUP-30n}
\end{subfigure}

\par\medskip % force a bit of vertical whitespace

\begin{subfigure}{0.8\textwidth}
\includegraphics[width=\textwidth]{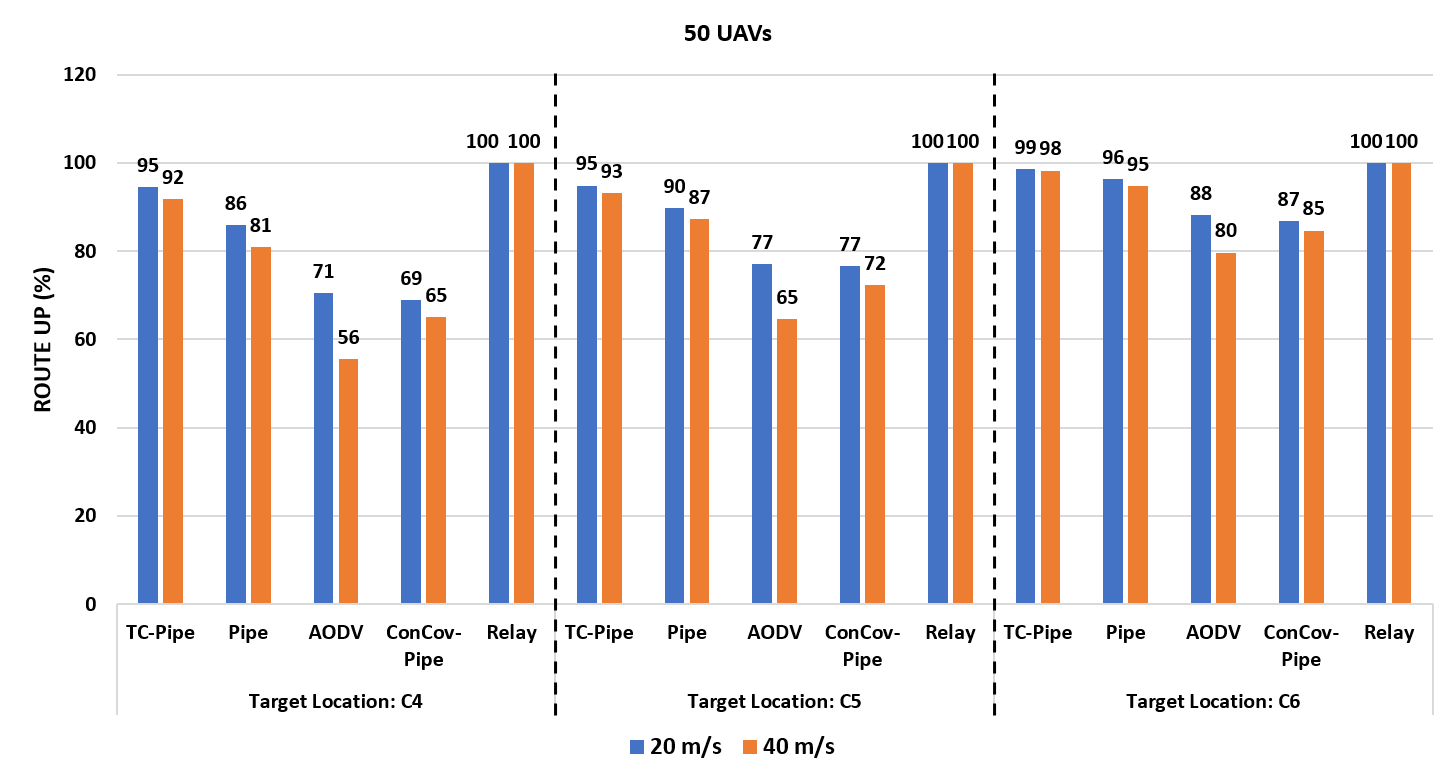}
\caption{Route Up (50 UAVs)}
\label{fig:3targets-routeUP-50n}
\end{subfigure}

\caption[Route Up for 3-targets settings]{Routing Performance: Route Up for 3-targets settings $C_4$, $C_5$ and $C_6$.}
\label{fig:3targets-RouteUP}
\end{figure*}

\begin{figure*}[htbp]
\centering
\begin{subfigure}{0.9\textwidth}
\includegraphics[width=\textwidth]{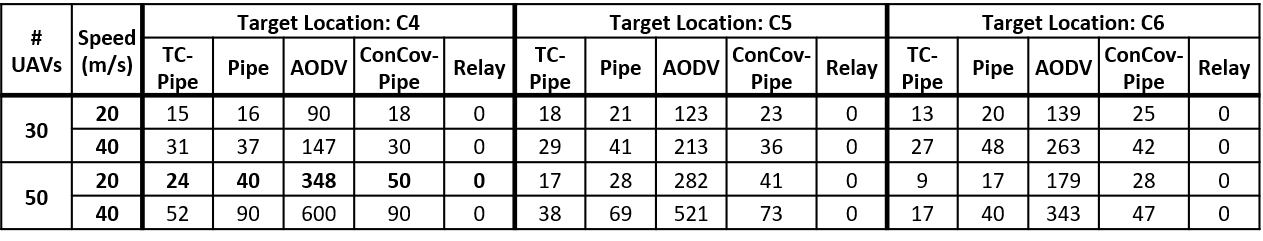}
% \caption{Route Breaks}
\end{subfigure}
\caption[Route Breaks for 3-targets settings]{Routing Performance: Route Breaks for 3-targets settings $C_4$, $C_5$ and $C_6$.}
\label{fig:3targets-RouteBreaks}
\end{figure*}

Figures \ref{fig:hops-3T} - \ref{fig:3targets-RouteBreaks} show the routing performance (measured by average route length, PDR, route up time and number of route breaks) of different schemes (TC-Pipe, Pipe, ConCov-Pipe, AODV, and Relay) for three data flows (one flow from each of the three targets to the BS), for three different settings of target locations, represented by $C_4$, $C_5$ and $C_6$. Here, the UAV network consists of two different node densities (30 and 50 UAVs) and two node speeds (20 and 40 m/s).

Interference and congestion may occur when multiple targets with flows to the BS exist, especially at higher data rates. The TC-Pipe and Pipe routing schemes take into consideration the congestion and interfering links between different routes when selecting an active route or when proactively switching the active routes, leading to better performance.
Using the TC-Pipe scheme to maintain a robust pipe (with alternate routes) around each flow (active route) allows for selecting high-quality routes to BS with less interference and congestion. Whereas, the Pipe routing scheme (without topology control) selects less interfering or less congested routes as and when they become available in the pipe region. The AODV scheme only considers the shortest route and can therefore select routes with congested or interfering links. 

\textbf{Average Route Length:} Figure \ref{fig:hops-3T} shows the average route length achieved by all the routing schemes for both densities at both speeds.
We observe that the route length increases with distance between the target and BS nodes, but it does not vary noticeably when the node density and/or speed change.
A shorter route length is desired because it can provide better data throughput and smaller delays. In setting $C_4$ where the targets are far from BS, our TC-Pipe scheme achieves the average route length of $\approx$8.5, while AODV achieves $\approx$9.1. 
On the other hand, for $C_6$ where the targets are relatively close to the BS, our TC-Pipe model achieves the average route length of $\approx$4.1, while AODV achieves $\approx$4.8. 
Overall, the TC-Pipe scheme achieves the shortest average route length, while AODV has the longest route length, with other routing schemes (Pipe, Relay and ConCov-Pipe) providing average route lengths in between the two. This indicates the ability of topology control with pipe routing to establish
stable, shorter routes using the pipe region.

\textbf{Average PDR:} 
Figure \ref{fig:3targets-PDR} shows the average PDR achieved by all the routing schemes for both node densities at both speeds. The average PDR achieved by a routing scheme increases as the distance between the target and BS decreases; this is because longer routes are more likely to break due to node mobility. Further, the average PDR increases for a higher density because route discovery and maintenance of pipe region is more efficient. Also, the average PDR decreases with node speed because more route breaks occur.
For different network and 3-target settings ($C_4$, $C_5$ and $C_6$), our proposed TC-Pipe scheme achieves higher average PDR than the Pipe, AODV, and ConCov-Pipe schemes due to a more robust pipe which helps in selecting better quality routes (i.e., stable routes with low interference and congestion).

For target setting $C_4$, the PDR at different data rates for both UAV densities and speeds are shown in Figures \ref{fig:6-3targets-pdr-30n} and \ref{fig:6-3targets-pdr-50n}, respectively. For longer routes, the advantage of using TC-Pipe scheme becomes more prominent.
In Figure \ref{fig:6-3targets-pdr-30n}, the TC-Pipe scheme achieves up to 50\%, 71\% and 112\% increase in PDR compared to Pipe, AODV and ConCov-Pipe schemes, respectively, for 30 UAVs at 20 m/s. 
% Similarly, for 50 UAVs at 20 m/s, the TC-Pipe scheme achieves up to 15\%, 46\% and 31\% increase in PDR, respectively.
For higher UAV speeds of 40 m/s and 30 UAVs, the TC-Pipe scheme achieves up to 39\%, 77\% and 75\% increase in PDR compared to Pipe, AODV and ConCov-Pipe schemes, respectively. 
% Similarly, for 50 UAVs at 40 m/s, the TC-Pipe scheme achieves up to 15\%, 73\% and 37\% increase in PDR, respectively. 
We observed similar trends for 50 nodes in Figure \ref{fig:6-3targets-pdr-50n}.
Even at higher UAV speeds where route breaks are more frequent, the TC-Pipe scheme achieves significant improvement in PDR compared to the Pipe scheme because the use of topology control in the pipe region makes the route more robust.
For settings $C_5$ and $C_6$ too, TC-Pipe scheme provides higher PDR than Pipe, AODV and ConCov-Pipe but the relative PDR improvement is reduced with the distance of the target from BS. For all three target settings and both UAV densities and speeds, the Pipe scheme achieves higher PDR than ConCov and AODV schemes.

For all three settings $C_4$, $C_5$ and $C_6$ (for 50 UAVs) and for $C_6$ (for 30 UAVs), the TC-Pipe achieves better PDR than the Relay scheme for higher data rates. This is because forming shortest paths to BS without considering congestion and interference in the Relay scheme leads to poor PDR at higher data rates. 
% In Figure \ref{fig:7-3targets-pdr-30n}, for setting $C_6$ with shorter routes, even with 30 UAV density, the TC-Pipe outperforms the Relay scheme at higher data rates. 
Whereas, in Figure \ref{fig:6-3targets-pdr-30n}, for setting $C_4$ with longer routes for 30 UAVs, the Relay scheme outperforms the TC-Pipe scheme.
Also, the Relay scheme for three targets requires more reassigned relay UAVs to form the routes, resulting in a greater decrease in coverage performance (as discussed in Section \ref{sect:cover_performance-3T}).

\textbf{Route Up ($R_u$):}  
A Higher $R_u$ value indicates that the routes from targets UAV to BS exist for a longer duration of time, which generally provides a higher PDR. 
Figure \ref{fig:3targets-RouteUP} shows the $R_u$ value of all the considered schemes for both node densities at both speeds.
We observe that the value of $R_u$ increases for a higher node density but decreases for a higher node speed. 
The Relay scheme achieves the highest $R_u$ (nearly 100\%) because a stable route is established. The TC-Pipe scheme achieves the second highest $R_u$ value for both node densities at both speeds.
In Figure \ref{fig:3targets-routeUP-50n}, TC-Pipe achieves $R_u$ value of 95\% and 92\%, for the target location setting $C_4$ at a node density of 50 and speeds of 20 m/s and 40 m/s, respectively, compared to 86\% and 81\% achieved by the Pipe routing scheme. 
%The TC-Pipe scheme's improved performance over the Pipe scheme is due to its ability to maintain a robust pipe along the active routes.
% 
AODV (which uses the BS-CAP mobility model without pipe region) and ConCov-Pipe (which uses the ConCov \cite{bsConCov} mobility model with a lower BS connectivity performance compared to the BS-CAP mobility model \cite{BS-CAP_ref}) provide lower $R_u$ values.
% AODV achieves a $R_u$ value of 71\% and 56\%, at 20 m/s and 40 m/s, respectively. Meanwhile, ConCov-Pipe achieves a $R_u$ value of 69\% and 65\% at 20 m/s and 40 m/s, respectively. Though ConCov-Pipe uses Pipe route switching, UAVs follow the ConCov \cite{bsConCov} mobility model (instead of the BS-CAP mobility model), which has a lower BS connectivity performance \cite{BS-CAP_ref}.
This shows that Pipe routing combined with the BS-CAP mobility model provides better routing performance (as in TC-Pipe and Pipe). 
A similar performance trend is observed at 30 UAV density in Figure \ref{fig:3targets-routeUP-30n}. For 30 UAVs, in setting $C_4$ where targets are located far from the BS, all schemes achieve lower $R_u$ values (except Relay) because there are fewer UAVs in the network to maintain the routes.

\textbf{Route Breaks ($R_b$):} Figure \ref{fig:3targets-RouteBreaks} shows the number of route breaks achieved by all the routing schemes for both UAV densities at both speeds.
Proactive route switching using the alternate route in the pipe reduces data flow interruptions and delays in pipe routing based schemes (i.e., TC-Pipe, Pipe and ConCov-Pipe). Due to better pipe robustness, the TC-Pipe scheme achieves the least route breaks ($R_b$) among the pipe-based schemes for both densities at both speeds and for all three target location settings, reducing new route discoveries and associated control overhead.
The pipe routing based schemes have far less route breaks than the AODV scheme for both densities at both speeds.
In Figure \ref{fig:3targets-RouteBreaks}, in setting $C_4$ for 50 UAVs,  TC-Pipe achieves lower $R_b$ values of 24 at 20 m/s compared to 40 in the Pipe scheme. The ConCov-Pipe and AODV schemes yield $R_b$ of 50 and 348 at 20 m/s, respectively. 
At UAV speed of 40 m/s, more route breaks are observed due to dynamic network topology. 
We find a similar performance trend across all three target location settings ($C_4$, $C_5$ and $C_5$) for both UAV densities at both speeds.
Since the Relay scheme establishes stable routes, it experiences zero route breaks.

\subsubsection{Coverage Performance:}
\label{sect:cover_performance-3T}

\begin{figure*}[htbp]
\centering
\begin{subfigure}[b]{0.325\textwidth}
\centering
\includegraphics[width=\textwidth]{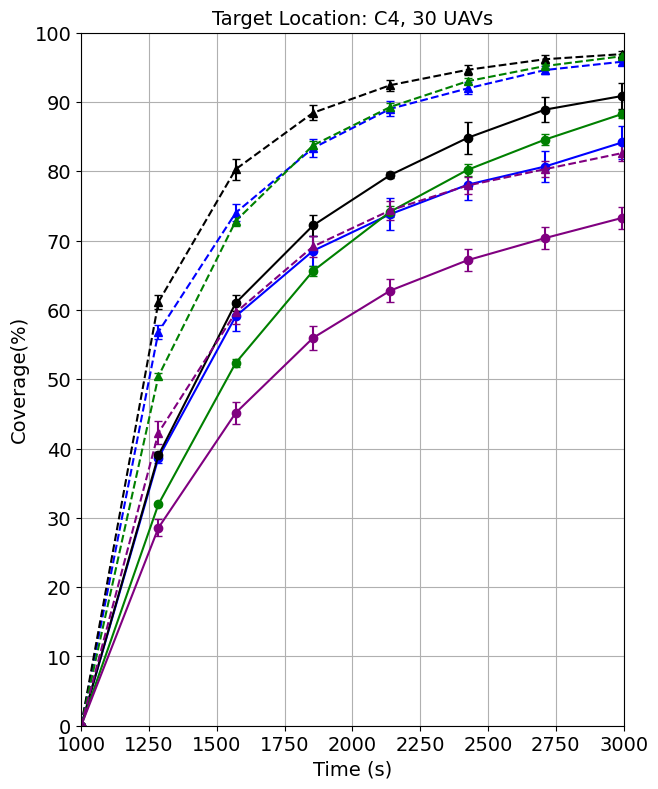} 
\caption{Coverage (Targets C4, 30 UAVs)}
\label{fig:3targets-cov-30n-CASE6}
\end{subfigure}
\begin{subfigure}[b]{0.325\textwidth}
\centering
\includegraphics[width=\textwidth]{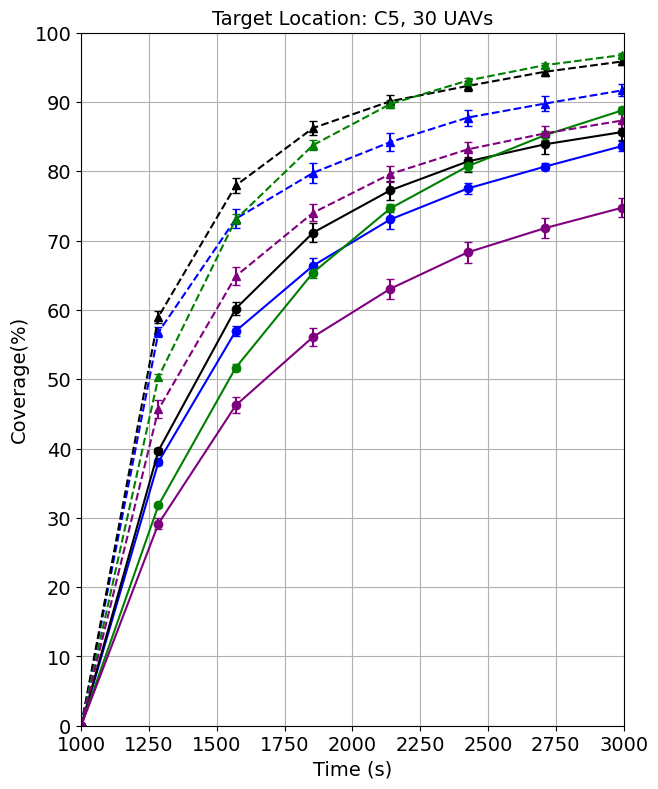} 
\caption{Coverage (Targets C5, 30 UAVs)}
\label{fig:3targets-cov-30n-CASE5}
\end{subfigure}
\begin{subfigure}[b]{0.325\textwidth}
\centering
\includegraphics[width=\textwidth]{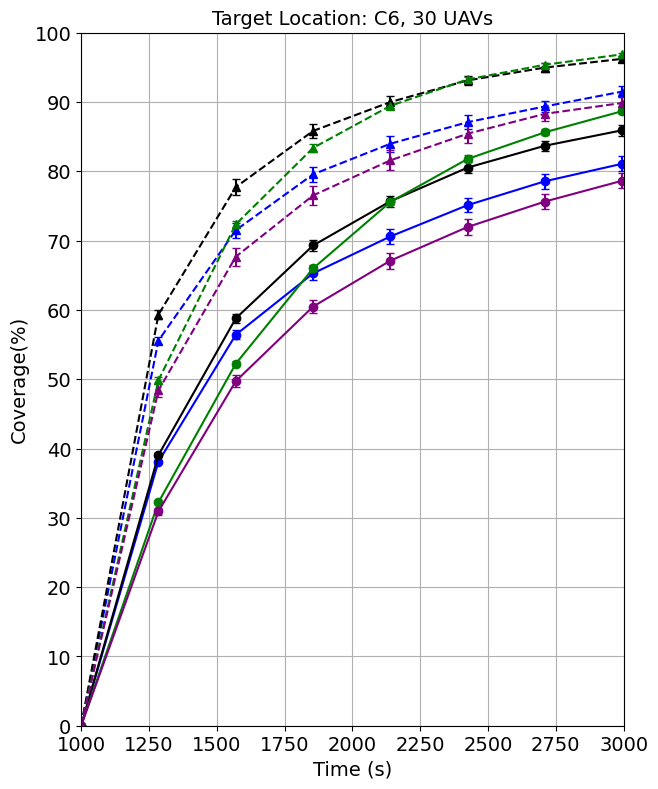} 
\caption{Coverage (Targets C6, 30 UAVs)}
\label{fig:3targets-cov-30n-CASE7}
\end{subfigure}

\par\medskip % force a bit of vertical whitespace
\begin{subfigure}[b]{0.325\textwidth}
\centering
\includegraphics[width=\textwidth]{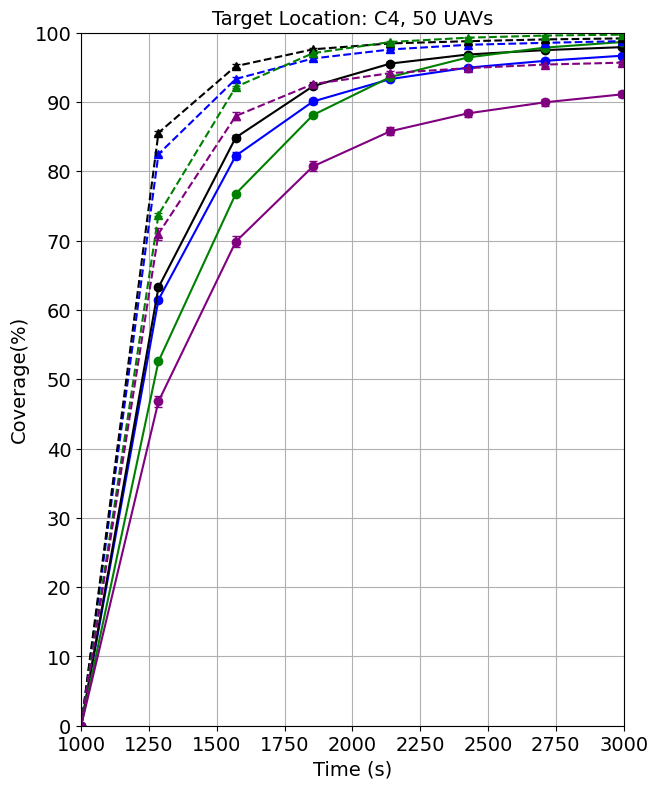}
\caption{Coverage (Targets C4, 50 UAVs)}
\label{fig:3targets-cov-50n-CASE6}
\end{subfigure}
\begin{subfigure}[b]{0.325\textwidth}
\centering
\includegraphics[width=\textwidth]{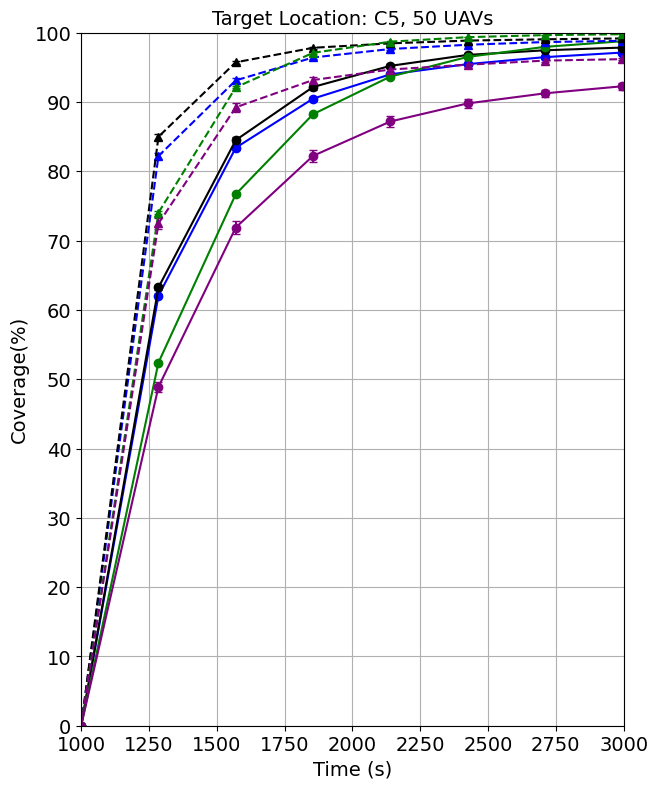}
\caption{Coverage (Targets C5, 50 UAVs)}
\label{fig:3targets-cov-50n-CASE5}
\end{subfigure}
\hspace{-1.1em} \raisebox{-.33em}{
\begin{subfigure}[b]{0.325\textwidth}
\centering
\begin{tikzpicture}
\node at (0,0) {\includegraphics[width=\textwidth]{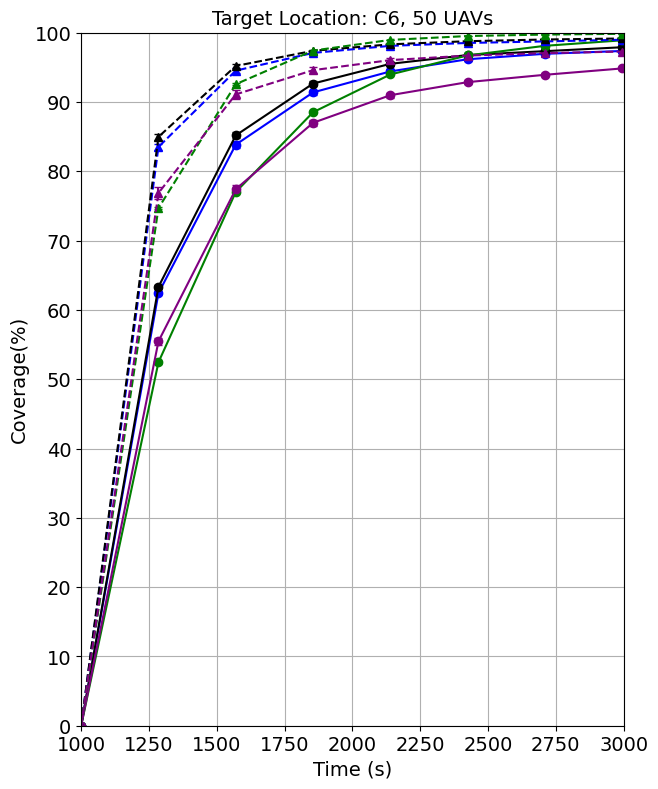}};
\node at (0.4,-2) {\includegraphics[width=0.74\textwidth]{RESULTS/coverage-legend.png}};
\end{tikzpicture}

\caption{Coverage (Targets C6, 50 UAVs)}
\label{fig:3targets-cov-50n-CASE7}
\end{subfigure}}

\caption[Coverage vs. Time plots for 3-targets settings]{Coverage vs. Time plots for 3-targets settings $C_4$, $C_5$ and $C_5$.}
\label{fig:3targets-coveragevstime}
\end{figure*}

\begin{figure*}[htbp]
\centering
\begin{subfigure}{0.9\textwidth}
\includegraphics[width=\textwidth]{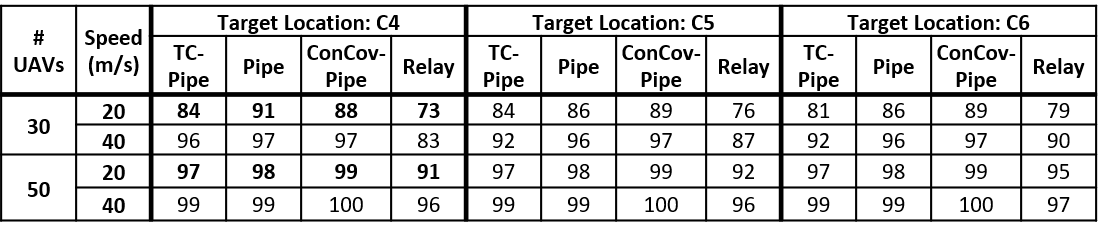}
% \caption{Coverage (\%) from 1000s to 3000s ($C_v$)}
% \label{fig:3targets-COV3000s}
\end{subfigure}
\caption[Coverage $C_v$ for 3-targets settings]{Coverage Performance: $C_v$ for 3-targets settings $C_4$, $C_5$ and $C_6$}
% \label{fig:3targets-TotalCoverage}
\label{fig:3targets-COV3000s}
\end{figure*}

\begin{figure*}[htbp]
\centering

\begin{subfigure}{0.8\textwidth}
\par\medskip % force a bit of vertical whitespace
\includegraphics[width=\textwidth]{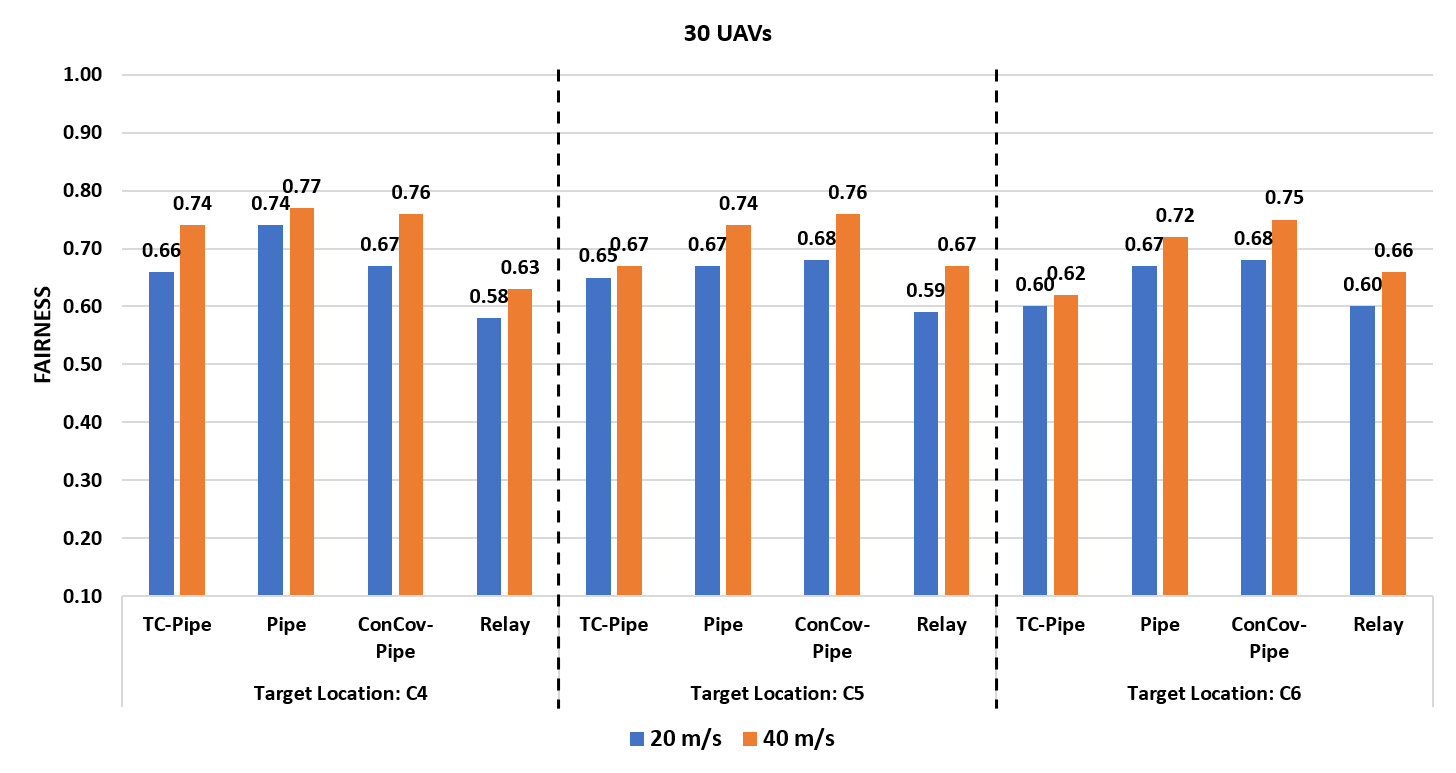}
\caption{ Fairness (30 UAVs)}
\label{fig:3targets-Fairness-30n}
\end{subfigure}

\par\medskip % force a bit of vertical whitespace
\begin{subfigure}{0.8\textwidth}
\includegraphics[width=\textwidth]{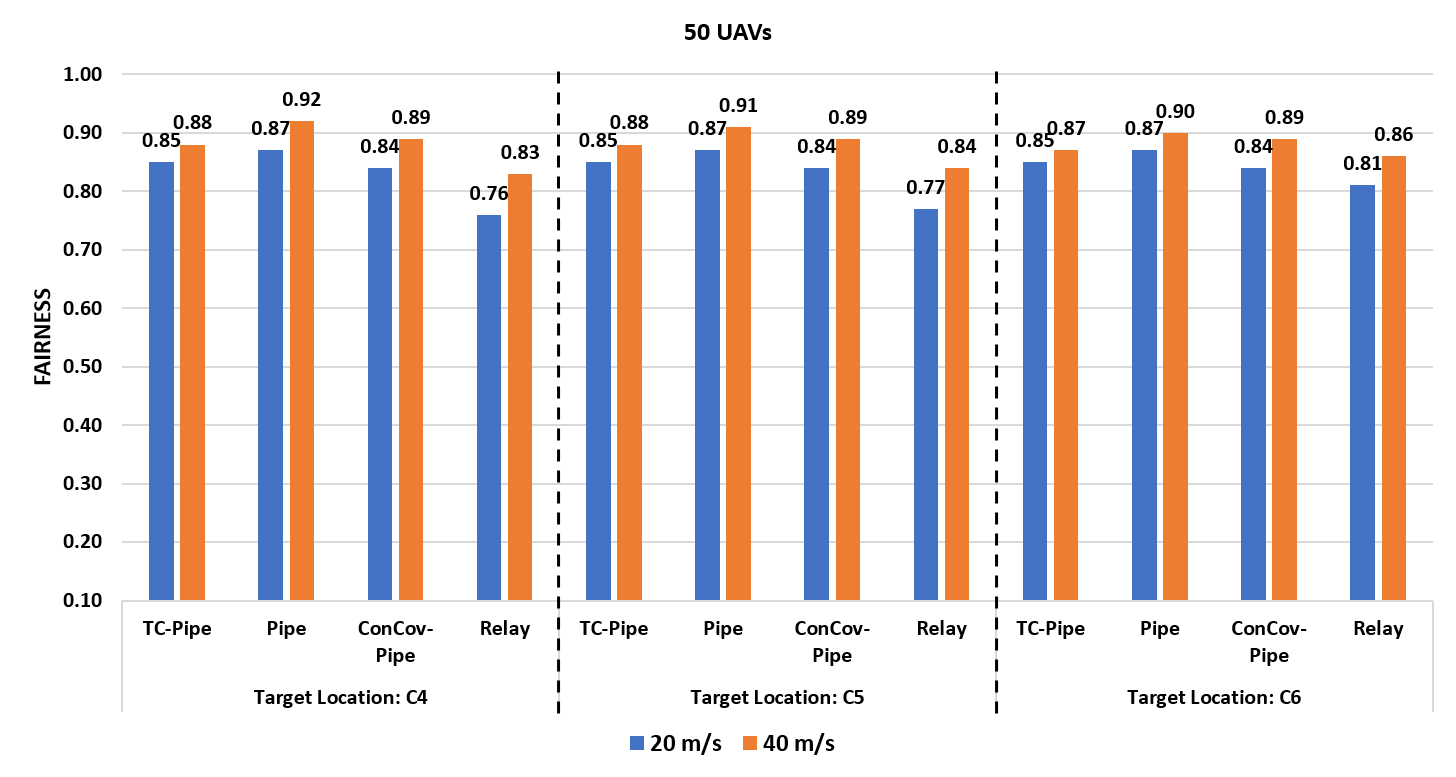}
\caption{Fairness (50 UAVs)}
\label{fig:3targets-Fairness-50n}
\end{subfigure}

\caption[Coverage Fairness for 3-targets settings]{Coverage performance: Fairness for 3-targets settings $C_4$, $C_5$ and $C_6$}
\label{fig:3targets-Fairness}
\end{figure*}

Figures \ref{fig:3targets-coveragevstime} - \ref{fig:3targets-Fairness} show the coverage performance (measured by coverage vs. time, total coverage, and fairness) of different schemes (TC-Pipe, Pipe, ConCov-Pipe, and Relay) for three data flows from targets to BS for three different target location settings, represented by $C_4$, $C_5$ and $C_6$. Here, the UAV network consists of two different node densities (30 and 50 UAVs) at two node speeds (20 and 40 m/s). The performance of AODV scheme is not shown because it uses the BS-CAP mobility model with coverage performance identical to Pipe scheme.
As explained in Section \ref{sect:cover_performance-1T}, we select the BS-CAP mobility model with $\beta=1.5$ (for TC-Pipe, Pipe, AODV, and Relay) and ConCov mobility model with $\omega=0.5$ (for ConCov-Pipe). 
%The coverage performance of AODV scheme is not shown in the plots because it uses the BS-CAP mobility model (like Pipe) with coverage performance identical to the Pipe scheme. 
%Note that use of topology control in the TC-Pipe scheme reduces its coverage performance compared to the Pipe scheme where the UAVs follow the BS-CAP mobility model with no topology control.

The \textbf{Coverage vs. Time} plots in Figures \ref{fig:3targets-coveragevstime} represent the total map area covered in a given time for both node densities at both speeds. The rate of area covered increases with node density as well as speed.
The Relay scheme achieves a lower coverage performance compared to the Pipe and ConCov-Pipe schemes, as it reassigns some of its UAVs as relay nodes in the three routes to BS; these relay nodes do not actively cover the map. This impact is more apparent at low UAV density (30 UAVs). 
%In the settings of three targets, three routes to BS need to be established; thus, more UAVs are reassigned as relay nodes. 
The longer the route length of these routes, the greater the number of UAVs reassigned as relay nodes, and the greater the decrease in coverage performance. This is observed in Figures \ref{fig:3targets-coveragevstime} and \ref{fig:3targets-COV3000s}, where the Relay scheme's coverage performance is much worse in target setting $C_4$ than in $C_6$.
At low UAV density (30 UAVs), the TC-Pipe scheme achieves a slightly lower coverage performance because some of the UAVs are attracted to the pipe region to maintain its node density. However, the TC-Pipe scheme allows the UAVs participating in the route to move and cover the surrounding area, hence giving better coverage than the Relay scheme.
For setting $C_4$ in Figure \ref{fig:3targets-COV3000s} at 30 UAVs and 20 m/s, the TC-Pipe and Relay schemes achieve $C_v$ of 84\% and 73\%, respectively, compared to $\approx$ 90\% achieved by Pipe and ConCov schemes. 
TC-Pipe’s coverage performance is comparable to Pipe and ConCov for higher node density, whereas Relay has a lower performance, especially for $C_4$ and $C_5$ (which correspond to medium and long route lengths). For example, for 50 UAVs at 20 m/s, TC-Pipe, Pipe and ConCov achieve $C_v$ of $\approx$ 98\%, compared to $\approx$ 91\% achieved by Relay for $C_4$ and $C_5$. 
%similar performance trend is observed for node speed of 40 m/s and for $C_6$ and $C_5$ target settings.

The \textbf{Fairness ($F$)} performance of all the schemes for both node densities at both speeds is shown in Figure \ref{fig:3targets-Fairness}. Each scheme achieves a higher $F$ when the density and/or node speed increase as all the cells in the area under monitoring are visited more frequently.
$F$ value of the TC-Pipe and Relay schemes is lower than Pipe and ConCov-Pipe schemes for 30 UAVs at both speeds for all three target location settings. In TC-Pipe, some UAVs cover the region near the pipes of three flows (due to pheromone masking), leading to a decrease in coverage fairness performance. Whereas, in Relay, more UAVs are reassigned as relay nodes to form the routes of the three flows, leading to a bigger decrease in the number of UAVs performing coverage and thus decreasing coverage fairness.
For example, in Figure \ref{fig:3targets-Fairness-30n}, for 30 UAVs and setting $C_4$, TC-Pipe achieves $F$ of 0.66 and 0.74 at 20 m/s and 40 m/s, respectively, compared to $F$ values of 0.74-0.67 and 0.77-0.76 (resp.) for Pipe and ConCov-Pipe. The Relay scheme achieves lower $F$ values of 0.58 and 0.63 for 30 UAVs at 20 m/s and 40 m/s, respectively. 
Since enough UAVs are available at a density of 50 UAVs, the TC-Pipe scheme's fairness performance is comparable to Pipe and ConCov-Pipe, whereas the Relay scheme achieves lower $F$ values (see  Figure \ref{fig:1target-Fairness-50n}).

\subsection{Effect of Node Failures on Performance} \label{Effects_node_failure}
We study the impact of node failures on the routing and coverage performance of the TC-Pipe, Pipe and Relay schemes when a fraction of nodes randomly fail during the simulation time. We consider 20\% and 30\% of the UAVs progressively failing during the simulation due to energy depletion and mechanical failures, and observe that the PDR and coverage performances decrease gracefully in proportion to the \% UAVs failing in the network.

\subsubsection{Effect of Node Failures for Single Target:} \label{Effects_node_failure-1T}

\begin{figure*}[htbp]
\centering
\includegraphics[width=0.9\textwidth]{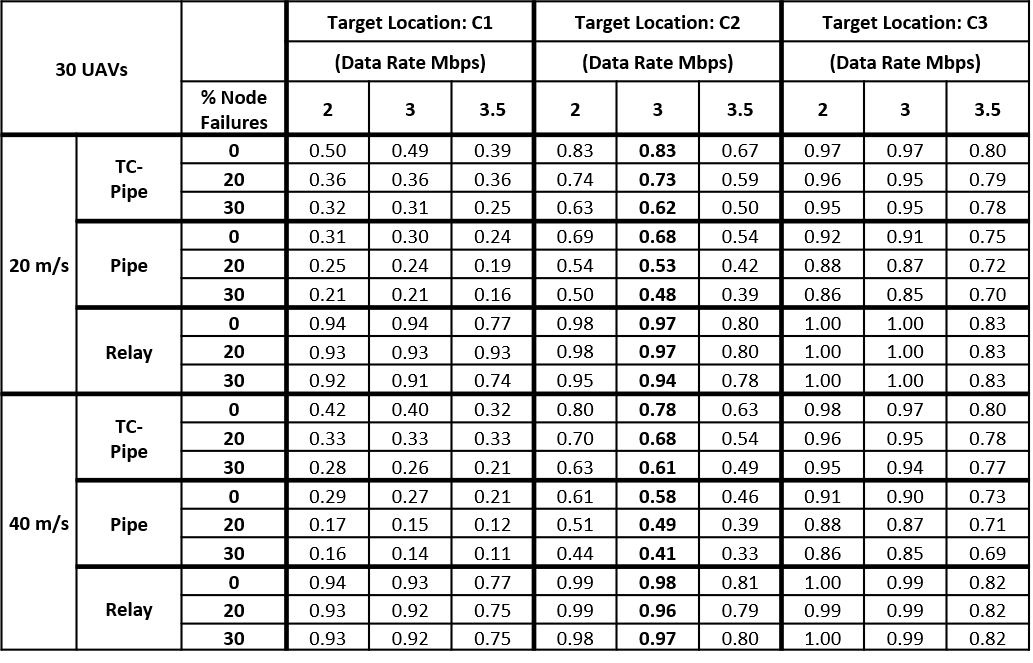}
\caption[Impact of node failures on PDR for 30 UAVs for single-target settings]{PDR values for 30 UAVs for single-target settings $C_1$, $C_2$ and $C_3$.}
\label{fig:1T-PDR-30n}
\end{figure*}

\begin{figure*}[htbp]
\centering
\includegraphics[width=0.9\textwidth]{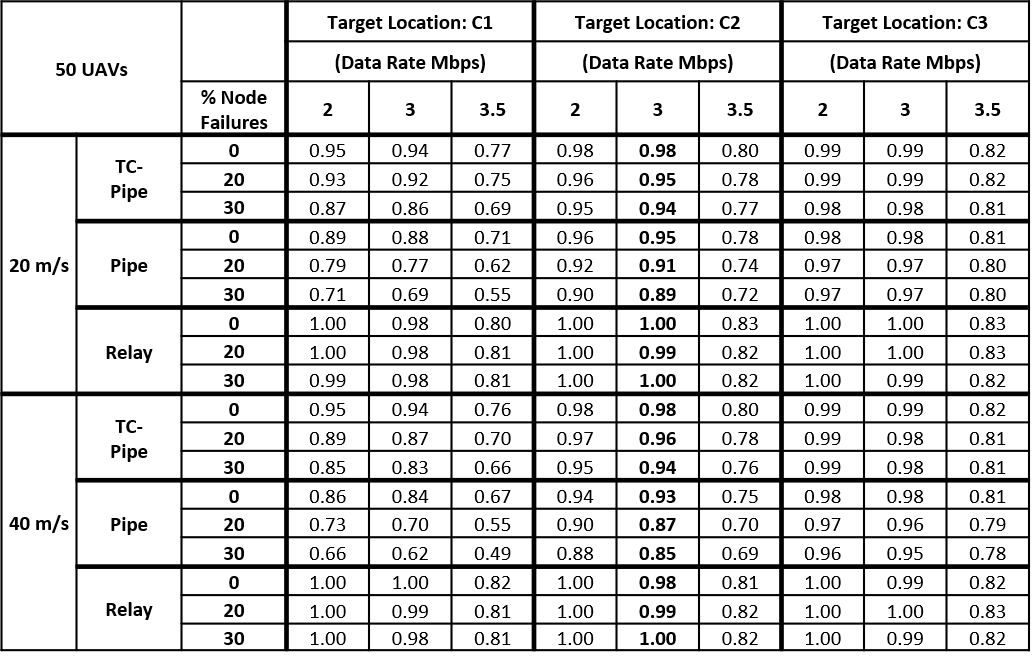}
\caption[Impact of node failures on PDR for 50 UAVs for single-target settings]{PDR values for 50 UAVs for single-target settings $C_1$, $C_2$ and $C_3$.}
\label{fig:1T-PDR-50n}
\end{figure*}

\begin{figure*}[htbp]
\centering
\includegraphics[width=0.8\textwidth]{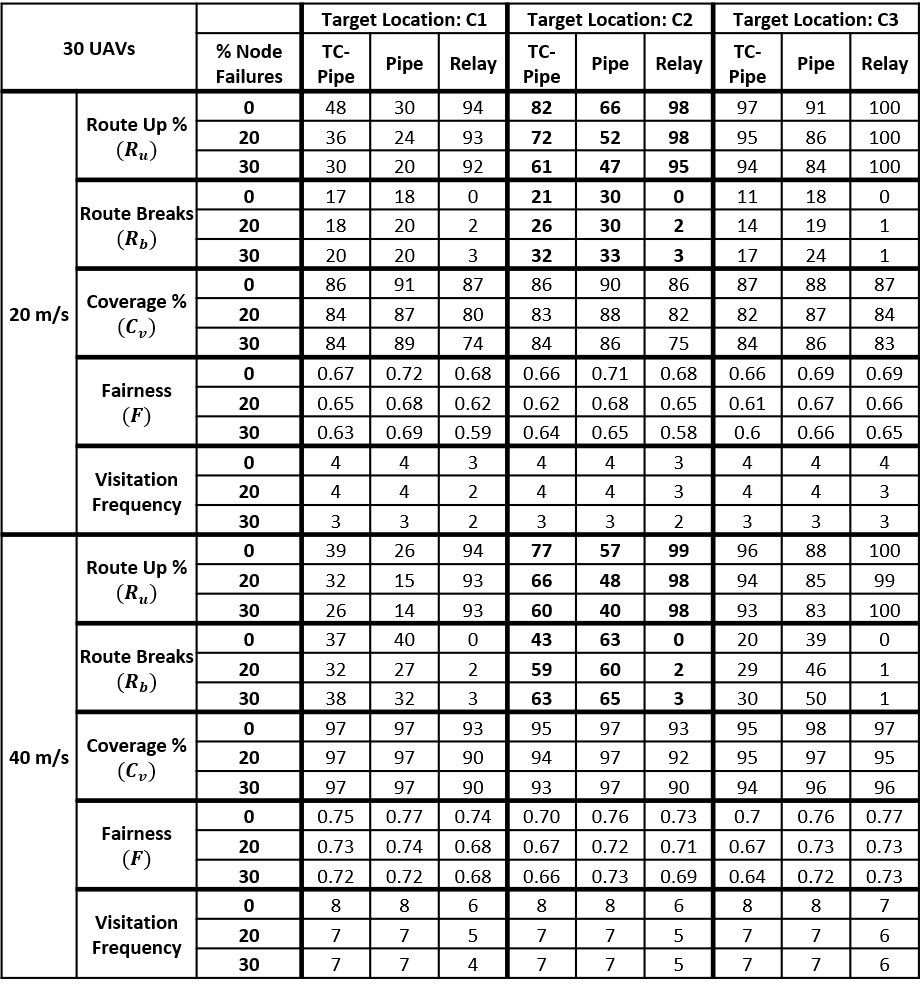}
\caption[Impact of node failures on routing and coverage metrics: 30 UAVs for single target settings]{Routing and Coverage metrics: 30 UAVs for single target settings $C_1$, $C_2$ and $C_3$.}
\label{fig:1T-ALL-30n}
\end{figure*}

\begin{figure*}[htbp]
\centering
\includegraphics[width=0.8\textwidth]{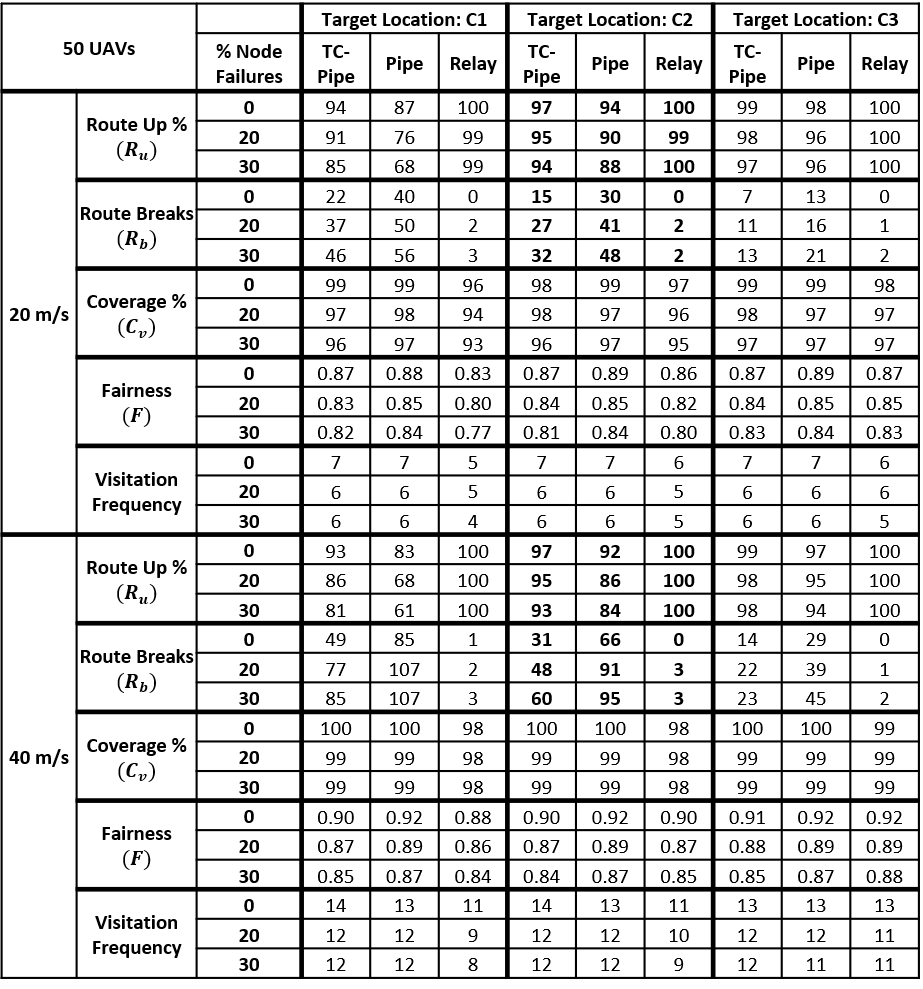}
\caption[Impact of node failures on routing and coverage metrics: 50 UAVs for single settings]{Routing and Coverage metrics: 50 UAVs for single settings $C_1$, $C_2$ and $C_3$.}
\label{fig:1T-ALL-50n}
\end{figure*}

For single target settings ($C_1$, $C_2$ and $C_3$), Figures \ref{fig:1T-PDR-30n} and \ref{fig:1T-PDR-50n} show the PDR performance of TC-Pipe, Pipe, and Relay schemes for 30 and 50 UAVs at both speeds, respectively, whereas Figures \ref{fig:1T-ALL-30n} and \ref{fig:1T-ALL-50n} show other routing and coverage metrics.

At low UAV densities and both speeds, fewer UAVs are left after node failures which are required to maintain the pipe while also performing area coverage. For 30 UAVs, the routing performance, including PDR (see Figure \ref{fig:1T-PDR-30n}, see highlighted column for $C_2$ at 3 Mbps data rate for illustration) and route up and route breaks (see Figure \ref{fig:1T-ALL-30n}, highlighted values), of TC-Pipe and Pipe schemes decrease with increase in \% node failures. The impact of node failures is more for longer route lengths. The loss of nodes reduces the rate at which the cells are covered. Since the coverage ($C_v$) represents the \% cells covered at least once during the observation period of 2000 s, the impact of node failures on $C_v$ is not visible; however, fairness value $F$ decreases with increase in \% node failures.
For a higher node density (50 UAVs) at both speeds, the effect of node failures on routing and coverage performances of Pipe and TC-Pipe schemes is relatively low (see Figures \ref{fig:1T-PDR-50n} and \ref{fig:1T-ALL-50n}, highlighted values).
%the PDR, routing metrics ($R_u$, $R_b$) and coverage metrics ($C_v$, $F$) performances of the TC-Pipe scheme is almost negligible as there are enough nodes in the network to support the robustness of the short pipe and its route     
% 
%However, for longer route lengths in target settings $C_1$ and $C_2$, the effect of node failures leading to decrease in performance increases slightly because more nodes are needed to support the route and make the pipe more robust (in TC-Pipe scheme).
% On the other hand, for 30 UAVs, we observe a noticeable decrease in the performance for setting $C_3$ (shorter route) which becomes worse for setting $C_1$ (longer route). At low UAV densities, there are fewer UAVs left after failure to maintain the node density along the pipe while also performing area coverage (see results under settings $C_3$ and $C_1$ in Figures \ref{fig:1T-PDR-30n} and \ref{fig:1T-ALL-30n}).
When a relay UAV fails in the Relay scheme, a new shortest route is re-established using AODV routing. Therefore, its routing performance remains almost unaffected by node failures; however, its coverage performance decreases considerably compared to the TC-Pipe and Pipe schemes.  

%The columns highlighted in Figures \ref{fig:1T-PDR-30n} and \ref{fig:1T-PDR-50n} illustrate the impact of node failures on PDR performance of routing schemes for both node densities. 
%A similar trend for the impact of node failures 
%
%Since the routes are more likely to break at a higher node speed, the PDR for a given node density is lower for a higher node speed. We observe that PDR decreases with node failures for both node speeds in Figures \ref{fig:1T-PDR-30n} and \ref{fig:1T-PDR-50n}.
% Thus, we see only a small drop in routing performance due to node failures.  

% 1Target-PDR

% 1Target-ALL

% ---------------------------------
% 3 target cases
% ---------------------------------
\subsubsection{Effect of Node Failures for Three Targets:} \label{Effects_node_failure-3T}

\begin{figure*}[htbp]
\centering
\includegraphics[width=0.8\textwidth]{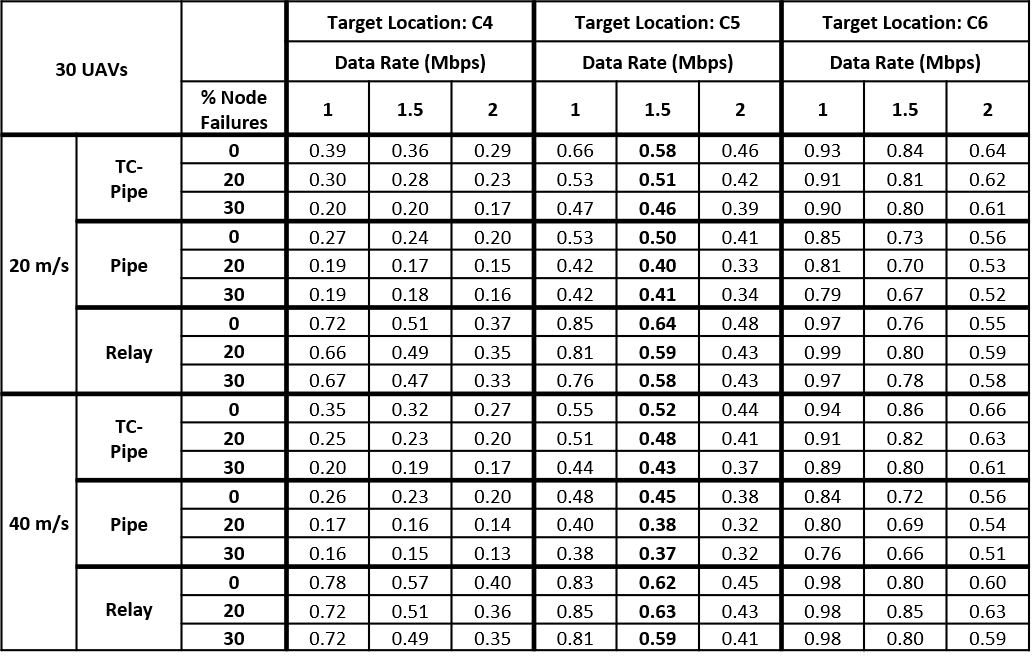}
\caption[Impact of node failures on PDR for 30 UAVs for 3-targets settings]{PDR values for 30 UAVs for 3-targets settings $C_4$, $C_5$ and $C_6$.}
\label{fig:3T-PDR-30n}
\end{figure*}

\begin{figure*}[htbp]
\centering
\includegraphics[width=0.8\textwidth]{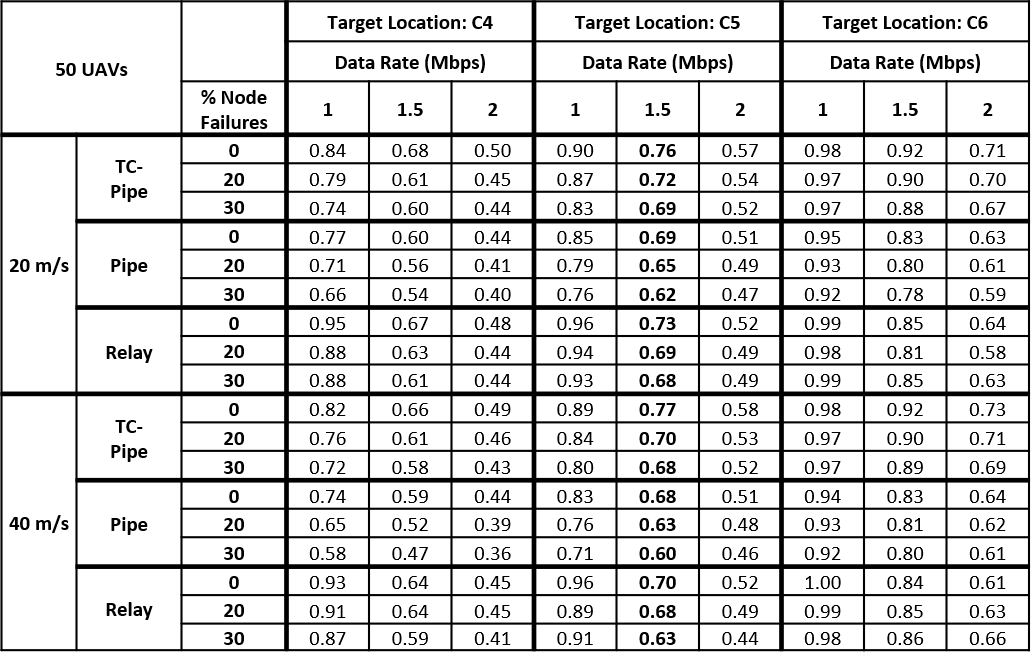}
\caption[Impact of node failures on PDR for 50 UAVs for 3-targets settings]{PDR values for 50 UAVs for 3-targets settings $C_4$, $C_5$ and $C_6$.}
\label{fig:3T-PDR-50n}
\end{figure*}

\begin{figure*}[htbp]
\centering
\includegraphics[width=0.8\textwidth]{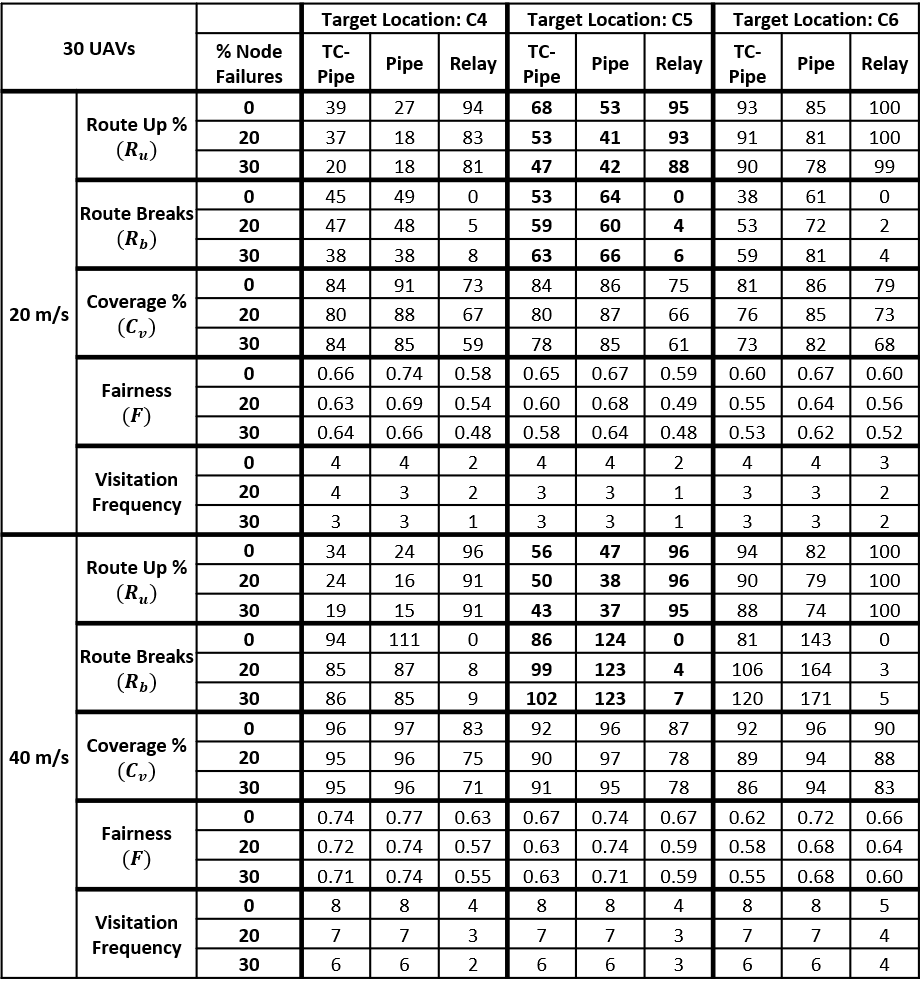}
\caption[Impact of node failures on routing and coverage metrics: 30 UAVs for 3-targets settings]{Routing and Coverage metrics: 30 UAVs for 3-targets settings $C_4$, $C_5$ and $C_6$.}
\label{fig:3T-ALL-30n}
\end{figure*}

\begin{figure*}[htbp]
\centering
\includegraphics[width=0.8\textwidth]{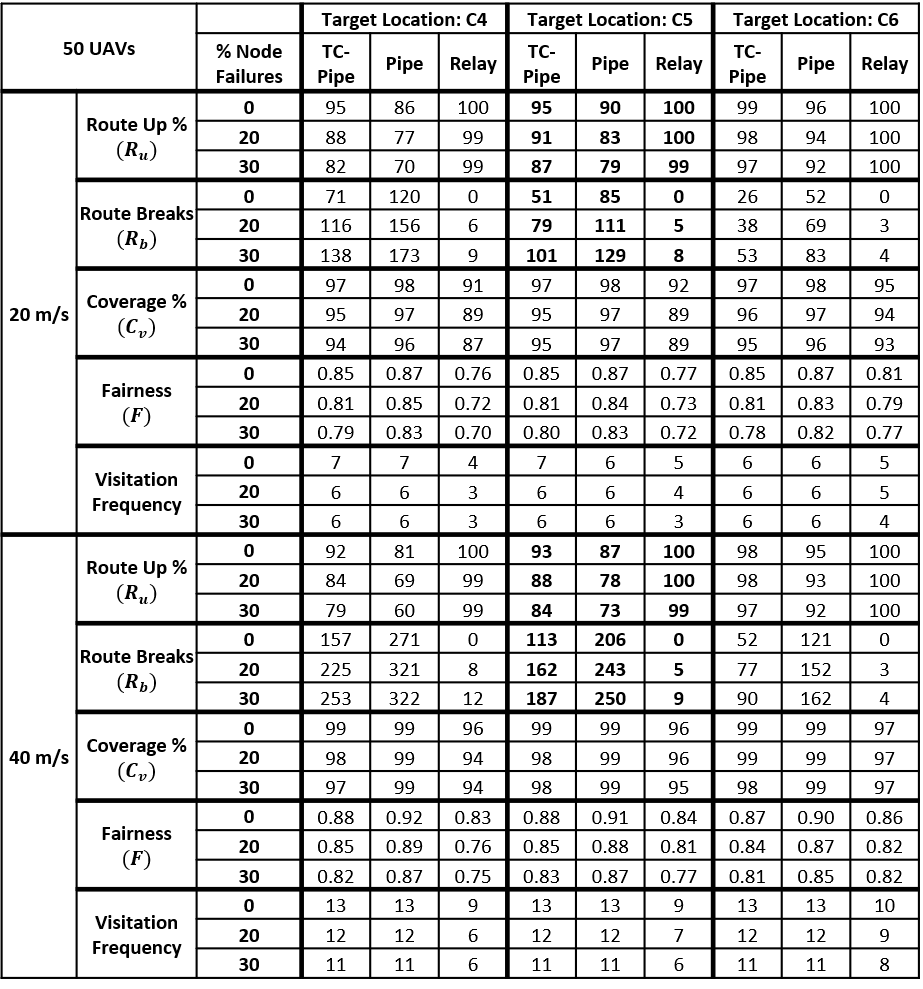}
\caption[Impact of node failures on routing and coverage metrics: 50 UAVs for 3-targets settings]{Routing and Coverage metrics: 50 UAVs for 3-targets settings $C_4$, $C_5$ and $C_6$.}
\label{fig:3T-ALL-50n}
\end{figure*}

In 3-target settings ($C_4$, $C_5$ and $C_5$ shown in Figure \ref{fig:Targets_loc}), a separate route is established to serve the data flow from each of three targets. Figures \ref{fig:3T-PDR-30n} and \ref{fig:3T-PDR-50n} show the PDR performance of the TC-Pipe, Pipe and Relay schemes for 30 and 50 UAVs at both speeds, respectively, whereas Figures \ref{fig:3T-ALL-30n} and \ref{fig:3T-ALL-50n} show other routing and coverage metrics. 

% Both PDR, route, and coverage performances decrease gracefully in proportion to the\% UAV failing in the network. The performance decrease is mainly attributed to the decreasing number of UAVs in the network. 

The overall trends representing the impact of node failures on different routing and coverage metrics are similar to the single target settings discussed above. 
Since more nodes are required to maintain the pipe region for routes for flows from three targets, the node failures result in a relatively greater impact on the routing and coverage performances for both node densities and at both speeds in the TC-Pipe and Pipe schemes.
%
%For 30 UAVs in setting $C_5$ with long routes (average route length of around 7), we see considerable decrease in PDR, coverage and other routing metrics due to node failures (see setting $C_4$ in Figures \ref{fig:3T-PDR-30n}  and  \ref{fig:3T-ALL-30n}).
%
For 30 UAVs in setting $C_5$ with long routes, we observe decrease in PDR, coverage and other routing metrics due to node failures (see Figures \ref{fig:3T-PDR-30n}  and  \ref{fig:3T-ALL-30n}). For 50 UAVs at both speeds, we observe a better overall performance (due to a higher node density) but the impact of node failures is still visible, especially for a longer route length and higher \% node failures (see setting $C_5$ in Figures \ref{fig:3T-PDR-50n}  and  \ref{fig:3T-ALL-50n}).
% 
%Also, the loss of nodes leads to fewer congestion-free routes, which further decreases PDR.
%
In the Relay scheme, the relay UAVs establish stable routes, so the PDR and other routing performance see a small decrease due to node failures. But its coverage performance becomes much worse, especially for the flows with longer route lengths. Since more UAVs are reassigned as relay nodes to form routes, fewer UAVs are available for covering the map (see $C_v$ performance in Figure \ref{fig:3T-ALL-30n}).  

% \textcolor{red}{PDR plots for different node failures, for the 30 and 50 UAVs for the target location settings $C_2$ and $C_5$, are shown in Figures \ref{fig:C2-allpdrvsfail} and \ref{fig:C5-allpdr-50n}, respectively.}

% 3Target-PDR

% 3Target-ALL

\subsection{Summary} \label{Results_summary}
We see that both pipe routing and topology control contribute to maintaining a robust set of alternative routes and lead to better routing performance (higher PDR, longer route-up time, and fewer route breaks) at all settings, with TC-Pipe (which uses both) outperforming Pipe, AODV, and ConCov-Pipe.  With no pipe region, AODV performs worst; the pipe region allows the other methods (TC-Pipe, Pipe, and ConCov-Pipe) to proactively switch routes within the pipe, leading to fewer route breaks and thus fewer route discoveries.
In general, the routing and coverage performance is a trade-off \cite{BS-CAP_ref}. TC-Pipe sees a slight decrease in coverage ($C_v$, $F$) compared to Pipe, AODV and ConCov-Pipe at lower node density (30 UAVs), but at higher density, TC-Pipe achieves comparable coverage performance with better PDR.

The Relay scheme achieves higher PDR than TC-Pipe in single target settings and some 3-targets settings (longer routes and lower data rates), because it uses dedicated relay UAVs to establish stable routes. 
However, it provides the worst coverage performance among the tested schemes, especially for multiple targets with long routes at low UAV density, because significantly fewer UAVs are available to perform area coverage.
% 
%For three targets, the TC-Pipe scheme achieves higher PDR at higher data rates for 50 UAVs and at higher data rates for shorter routes (for 30 UAVs) compared to the Relay scheme due to its congestion-aware routing.  
%
In both Pipe and TC-Pipe, the PDR and coverage performances decrease gracefully in proportion to the \% of UAVs failing in the network. At higher density, this decrease is less than at lower density, as there are enough UAVs to maintain a robust pipe. 
%Thus, both TC-Pipe and Pipe are well-suited for decentralized dynamic UAV networks performing target monitoring and area coverage. 

%\textcolor{red}{See \cite{openarch-TC_PIPE} for more extensive simulation results and descriptions of all single and 3-target settings.}

% \textbf{ALL RESULTS in Table Form}
% \include{Tables/1T-30} \label{alltable-30n}
% \include{Tables/1T-50} \label{alltable-50n}

% % 1Target-PDR
% \begin{figure*}[htbp]
% \centering
% \includegraphics[width=\textwidth]{RESULTS/1target/PDR-TEST-TABLE.png}
% \caption{PDR values for UAV speeds of 20 m/s (black) and 40 m/s (red), with 30 and 50 UAVs, for single-target settings $C_1$, $C_2$ and $C_3$.}
% \label{fig:1T-PDR-TEST}
% \end{figure*}

\section{Conclusion} \label{conclusion_section-TC}

We considered a decentralized, multi-hop, dynamic UAV network
consisting of low SWaP UAVs, where UAVs continuously monitor the map area to find new targets and maintain routes between UAVs monitoring targets to BS for data communication. 
We designed a mobility, congestion, and energy-aware hybrid reactive routing protocol, called the Pipe routing (Pipe), based on the MCA-AODV \cite{mca-aodv-shivam}, followed by a novel pipe routing with topology control scheme (TC-Pipe). 

In TC-Pipe routing, the 2-hop region around the active routes (the ``pipe'') is made robust using topology control; this helps to  establish stable routes and improves routing performance.
Both TC-Pipe and Pipe routing schemes saw fewer route breaks and better PDR than the AODV and ConCov-Pipe routing schemes. TC-Pipe achieved better routing performance than Pipe but saw a slight decrease in coverage performance at low UAV density. This is due to the inherent trade-off between connectivity and coverage performance. 
Though the Relay scheme establishes stable routes between the target UAVs and BS, the TC-Pipe scheme achieved better PDR for certain 3-target settings at higher data rates.
In both the TC-Pipe and Pipe schemes, the PDR and coverage performances decreased gracefully in proportion to the number of failures in the network. 
Thus, both TC-Pipe and Pipe routing schemes are well suited for decentralized dynamic UAV networks performing target monitoring and area coverage applications.

%% =========================== References ============================
\bibliographystyle{IEEEtran}
\bibliography{ref}

%% ===================================================================

\end{document}